\documentclass[preprint,amsmath,amssymb,aps,prd,nofootinbib]{revtex4}
\usepackage{epsfig,graphicx,color,appendix}
\begin{document}
\begin{flushright}
IBS-CTPU-17-07
\end{flushright}
\def\CP{{\it CP}~}
\def\cp{{\it CP}}
\title{\mbox{}\\[10pt]
Axion and Neutrino physics\\ in a $U(1)$-enhanced supersymmetric model}

\author{Y. H. Ahn}
\affiliation{Center for Theoretical Physics of the Universe, Institute for Basic Science (IBS), Daejeon, 34051, Korea}
\email{yhahn@ibs.re.kr}

%\thanks{}


%\date{\today}

%%%%%%%%%%%%%%%%%%%%%%%%%%%%%%%%%%%%%%%%%%%%%%%%%%%%%%%%%%%%%%%%%%%%%%%%%%%%%%%%%%%%%%%%%%
\begin{abstract}
\noindent
Motivated by the flavored Peccei-Quinn symmetry for unifying the flavor physics and string theory, we construct an explicit model by introducing a $U(1)$ symmetry such that the $U(1)_X$-$[gravity]^2$ anomaly-free condition together with the standard model flavor structure demands additional sterile neutrinos as well as no axionic domain-wall problem. Such additional sterile neutrinos play the role of a realization of baryogenesis via a new Affleck-Dine leptogenesis. We provide grounds for that the $U(1)_X$ symmetry could be interpreted as a fundamental symmetry of nature. The model will resolve rather recent, but fast-growing issues in astro-particle physics, including leptonic mixings and CP violation in neutrino oscillation, high-energy neutrinos, QCD axion, and axion cooling of stars.
The QCD axion decay constant, through its connection to the astrophysical constraints of stellar evolution and the SM fermion masses, is shown to be fixed at $F_A=1.30^{+0.66}_{-0.54}\times10^{9}$ GeV (consequently, its mass is $m_a=4.34^{+3.37}_{-1.49}$ meV and axion-photon coupling is $|g_{a\gamma\gamma}|=1.30^{+1.01}_{-0.45}\times10^{-12}\,{\rm GeV}^{-1}$). Interestingly enough, we show that neutrino oscillations at low energies could be connected to astronomical-scale baseline neutrino oscillations. The model predicts non-observational neutrinoless double beta ($0\nu\beta\beta$) decay rate as well as a remarkable pattern between leptonic Dirac CP phase ($\delta_{CP}$) and atmospheric mixing angle ($\theta_{23}$); {\it e.g.} $\delta_{CP}\simeq220^{\circ}-240^{\circ}$, $120^{\circ}-140^{\circ}$ for $\theta_{23}=42.3^{\circ}$ for normal mass ordering, and $\delta_{CP}\simeq283^{\circ},250^{\circ},100^{\circ},70^{\circ}$ for $\theta_{23}=49.5^{\circ}$ for inverted one. We stress that future measurements on $\theta_{23}$, $0\nu\beta\beta$ decay rate, sum of active neutrino masses, track-to-shower ratio of a cosmic neutrino, astrophysical constraints on axions, QCD axion mass, and its axion-photon coupling are of importance to test the model in the near future.
\end{abstract}

\maketitle %
%%%%%%%%%%%%%%%%%%%%%%%%%%%%%%%%%%%%%%%%%%%%%%%%%%%%%%%%%%%%%%%%%%%%%%%%%%%%%%%%%%%%%%%%%%
\section{Introduction}
The standard model (SM) of particle physics has been successful in describing properties of known matter and forces to a great precision until now, but we are far from satisfied since it suffers from some problems or theoretical arguments that have not been solved yet, which follows: inclusion of gravity in gauge theory, instability of the Higgs potential, cosmological puzzles of matter-antimatter asymmetry, dark matter, dark energy, and inflation, and flavor puzzle associated with the SM fermion mass hierarchies, their mixing patterns with the CP violating phases, and the strong CP problem. The SM therefore cannot be the final answer. It is widely believed that the SM should be extended to a more fundamental underlying theory. If nature is stringy, string theory should give insight into all such fundamental problems or theoretical arguments\,\footnote{In Ref.\,\cite{Ahn:2016typ} a concrete model is designed to bridge between string theory as a fundamental theory and low energy flavor physics.}. As indicated in Refs.\,\cite{Ahn:2016typ,Ahn:2014gva}\,\footnote{Ref.\,\cite{Ahn:2014gva} introduces a superpotential for unifying flavor and strong CP problems, the so-called flavored
PQ symmetry model in a way that no axionic domain wall problem occurs.}, such several fundamental challenges strongly hint that a supersymmetric framework with new gauge symmetries as well as higher dimensional operators responsible for the SM flavor puzzles may be a promising way to proceed.
In favor of such a new extension of the SM, axions and neutrinos could be powerful sources for the arguments, in that they stand out as their convincing physics and the variety of experimental probes. Many of the outstanding mysteries of astrophysics may be hidden from our sight at all wavelengths of the electromagnetic spectrum because of absorption by matter and radiation between us and the source. So, data from a variety of observational windows, especially, through direct observations with neutrinos and axions, may be crucial. Thus, the axions and neutrinos in astrophysics and cosmology could provide a natural laboratory
for a new extension of SM particle physics\,\footnote{See Ref.\,\cite{Ballesteros:2016xej} for a new extension of SM particle physics, and Ref.\,\cite{Quevedo:2016tbh} for a landscape of new physics.}.

%The SM doest not provide a reason for the absence of strong CP violation. Until now, the most elegant solution to the strong CP problem is the PQ mechanism, which yields a light Nambu-Goldstone (NG) boson, called the QCD axion\,\cite{Peccei-Quinn, axion}. Moreover, 
Axions in stars available at low energies are well suited for very sensitive tests. If the axion exists, it solves the strong CP problem of QCD through the Peccei-Quinn (PQ) mechanism\,\cite{Peccei-Quinn, axion}, fits easily into a string theoretic framework, and appears cosmologically as a form of cold dark matter. 
The axion lies at the intersection of elementary particle physics, astrophysics, cosmology and string theory, potentially playing a crucial role in each.
There are being discussed two prototype axion models\,\footnote{There are good reviews Ref.\,\cite{Kim:1986ax}, Ref.\,\cite{Cheng:1987gp} and Ref.\,\cite{Peccei} on the axion.}, Kim-Shifman-Vainshtein-Zakharov (KSVZ)\,\cite{KSVZ} and Dine-Fischler-Srednicki-Zhitnitsky (DFSZ) models. And another new type model so-called flavored PQ symmetry is appeared\,\cite{Ahn:2014gva}.
These minimal models, commonly introducing SM gauge singlet scalar fields carrying PQ charges, are categorized by what couples to $U(1)_{\rm PQ}$ with domain-wall number $N_{\rm DW}$\,\footnote{At the QCD phase transition, each axionic string becomes the edge to $N_{\rm DW}$ domain-walls, and the process of axion radiation stops. 
If $N_{\rm DW}>1$ separating the various domains (like in the DFSZ model) the string-wall network is stable and has a sizable surface energy density $\sigma\approx m_a F^2_A\approx6.3\times10^{9}\,{\rm GeV}^3(F_A/10^{12}{\rm GeV})$, which is enormously bigger than the critical density of the Universe today $\rho_{c}\simeq10^{-26}{\rm kg/m^3}\sim10^{-47}\,{\rm GeV}^4$, where $m_a$ is an axion mass and $F_A$ is an axion decay constant. And since the energy density in these walls, $\rho_{\rm wall}=\sigma\,T$, dissipates slowly as the Universe expand and $\rho_{\rm wall}$ now would vastly exceed the closure density of the Universe, this is a serious problem\,\cite{Zeldovich:1974uw}. This disaster is avoided if $N_{\rm DW}=1$ or if the PQ phase transition occurred during (or before) inflation.}: 
(i) the KSVZ model\,\cite{KSVZ} couples to hadrons and photons with $N_{\rm DW}=1$, where only new heavy quarks are charged under $U(1)_{\rm PQ}$, and (ii) the DFSZ model\,\cite{DFSZ} couples to hadrons, photons and charged-leptons with $N_{\rm DW}=6$, where only known quarks and Higgs doublets carry PQ charge.
(iii) The flavored PQ symmetry model\,\cite{Ahn:2014gva} couples to hadrons, photons and leptons with $N_{\rm DW}=1$, in which the SM fermion fields as well as SM gauge singlet fields carry PQ charges but electroweak Higgs doublet fields do not. We refer to the model as flavored-Axion (FA) model.

In the case of neutrinos, the neutrino oscillations at low energies are quite well-studied from the experiments available in nuclear power plants, particle accelerators, nuclear bombs, and general atmospheric phenomena. And, after the observation of a non-zero mixing angle $\theta_{13}$ in the Daya Bay\,\cite{An:2012eh} and RENO\,\cite{Ahn:2012nd} experiments, the Dirac CP-violating phase $\delta_{CP}$ and a precise measurement of the atmospheric mixing angle $\theta_{23}$ are the next observables on the agenda of neutrino oscillation experiments. %Even nothing is known on the physics related to the leptonic CP violation, the measurements of non-vanishing $\theta_{13}$ have opened up the possibilities for searching for CP violation in neutrino oscillation experiments. 
Meanwhile, the very different structure of leptonic mixings compared to the quark ones indicates an unexpected texture of the mass matrix and may provide important clues to our understanding of the physics of fundamental constituents of matter. In some sense, our understanding of the SM fermion masses and mixing angles remains at a very primitive level.
On the other hand, high energy neutrinos are available in the most violent astrophysical sources: events like during the births, collisions, and deaths of stars, especially the explosion of supernovae, gamma-ray bursts, and cataclysmic phenomena involving black holes and neutron stars.
The SM weakly interacting neutrinos, known as three different flavors $\nu_e,\nu_\mu,\nu_\tau$, can deliver astrophysical information ({\it e.g.} IceCube detector\,\cite{ice-cube}, etc) from the edge of the Universe and from deep inside the most cataclysmic high-energy processes. %A majority of the neutrinos floating around were born, soon after the birth of the Universe. After inflation the growth of the Universe has continued, but at a slow rate, and thereby cooled, and neutrinos have just kept on going. 
%Theoretically, there are now so many of neutrinos that constitute a cosmic background radiation whose temperature is 1.9 K. 
Moreover, the observations of cosmic structures ({\it e.g.} cosmic microwave background (CMB), galaxy surveys, etc.) can give the information on the neutrino masses and the effective number of species of neutrino $N^{\rm eff}_{\nu}$\,\cite{BBN,deGouvea:2009fp}. 
Neutrino oscillation, a CP property not yet fully understood, may play a role in the decoupling process and therefore can affect $N^{\rm eff}_{\nu}$. Additional neutrinos, if existed in nature, should be sterile with respect to the SM gauge interactions because the $Z$-boson decay $Z\rightarrow\nu\bar{\nu}$ showed that there are only 3 species of active neutrinos with ordinary weak interactions. Such sterile neutrinos are light or heavy and do not participate in the weak interaction. However, the latest results\,\cite{Planck2014} from Planck and Baryon Acoustic Oscillations (BAO) show that the contribution of light sterile neutrinos to $N^{\rm eff}_{\nu}$ at the Big-Bang Nucleosynthesis (BBN) era is negligible\,\footnote{See the arguments related with Eq.\,(\ref{D_bound0}).}%and indicate that BBN favors just 3 known species of light neutrinos
; such light sterile neutrinos can play the role of a realization of baryogenesis via a new Affleck-Dine leptogensis\,\cite{Ahn:2016hbn}. Such additional sterile neutrinos could be further constrained by the mass orderings of active neutrinos,  the BBN constraints\,\cite{BBN}, the solar neutrino oscillations\,\cite{deGouvea:2009fp}, and the inflationary and leptogenesis scenarios\,\footnote{The inflationary and leptogenesis scenarios on Ref.\,\cite{Ahn:2016hbn} will be separated and appear in more detail soon.}.
Hence it needs a new paradigm to explain the peculiar structure of lepton sector compared to the quark one as well as the astrophysical and cosmological observations on neutrinos. 

Since astrophysical and cosmological observations have increasingly placed tight constraints on parameters for axion and neutrino, it is in time for a new scenario on axion and neutrino to mount an interesting challenge.
In a theoretical point of view axion physics together with neutrino physics\,\footnote{There are certainly models of neutrino masses without new gauge interactions.} requires new gauge interactions and a set of new fields that are SM singlets. %Even though sterile neutrinos, as discussed above, are not protected by SM gauge interactions from acquiring masses, their masses of order sub-eV are unnatural as well as BBN favors just three species of light neutrinos. 
Thus in extensions of the SM, sterile neutrinos and axions could be naturally introduced, {\it e.g.}, in view of $U(1)$ symmetry.
Motivated by the aforementioned fundamental challenges, we investigate a minimal and economic supersymmetric extension of SM realized within the framework of $G\equiv SM\times U(1)_X\times A_4$. The non-Abelian discrete symmetry $A_4$ as a symmetry of geometrical solid could be originated from superstring theory; indeed, orbifolds have certain geometrical symmetries, and thus field theories in orbifold can realize $A_4$\,\cite{Altarelli:2006kg}. All renormalizable and nonrenormalizable operators allowed by such gauge symmetries, non-Abelian discrete symmetry, and $R$-parity exist in the superpotential.  
We assign the $U(1)_X$ quantum numbers in the following ways, see TABLE\,\ref{reps}: in a way that 
\begin{description}
\item[(i)] the mixed $U(1)_X$-$[{gravity}]^2$ anomaly is free in the presence of gravity, so that additional sterile neutrinos are introduced. 
\item[(ii)] the $U(1)_X$ quantum numbers of the SM quarks do not give rise to axionic domain-wall problem, implying that flavor structure of the SM may be correlated to axionic domain-wall.
\item[(iii)] the $U(1)_X$ symmetry is responsible for vacuum configuration as well as for describing mass hierarchies of leptons and quarks in the SM.
\end{description}
Then the flavored PQ symmetry $U(1)_X$ embedded in the non-Abelian $A_4$ finite group\,\footnote{E. Ma and G. Rajasekaran\,\cite{Ma:2001dn} have introduced to leptonic sector $A_4$ symmetry which is the smallest group for three families.} could economically explain the mass hierarchies of quarks and leptons including their peculiar mixing patterns as well as provide a neat solution to the strong CP problem and its resulting axion.
Here if we assume that the non-Abelian discrete $A_4$ symmetry is a subgroup of a gauge symmetry, it can be protected from quantum-gravitational effects\,\cite{Krauss:1988zc}. Moreover, in the model since such $A_4$ symmetry is broken completely by higher order effects, there is no residual symmetry; so there is no room for a spontaneously broken discrete symmetry to give rise to domain-wall problem. Differently from Ref.\,\cite{Ahn:2014gva}, in the present model we impose $U(1)_X$-$[gravity]^2$ anomaly-free condition together with the SM flavor structure in a way that  no axionic domain-wall problem occurs, which in turn demands additional sterile neutrinos. Such additional neutrinos may play a crucial role as a bridge between leptogenesis and new neutrino oscillations along with high energy cosmic events. In addition, in order to fix the QCD axion decay constant appropriately we impose several astrophysical constraints, see Sec.\,\ref{astro_ax_nu}.

The rest of this paper is organized as follows. In Sec.\,II we construct a minimalistic SUSY model for quarks, leptons, and axions based on $A_4\times U(1)_X$ symmetry in a way that the mixed $U(1)_X$-$[gravity]^2$ anomaly-free condition together with the SM flavor structure demands additional sterile neutrinos as well as no axionic domain-wall problem. In detail, In Sec.\,\ref{vacCon} the vacuum configuration is described to explain the peculiar mixing patterns of the SM.
In Sec.\,\ref{qla} we describe the Yukawa superpotential for leptons, quarks, and Nambu-Goldstone (NG) modes.
In Sec.\,\ref{qla1} we show that the global $U(1)_X$ is the remnant of the broken $U(1)_X$ gauge symmetry by the Green-Schwarz (GS) mechanism\,\cite{GS}, so it can be protected from  quantum-gravitational effects. Along this line, we provide a reason that the $U(1)_X$ symmetry could be interpreted as a fundamental symmetry of nature. And we show, through the GS anomaly cancellation mechanism, how the $U(1)_X$ gauge bosons acquire masses leaving behind the corresponding global symmetries and how the QCD axion could be derived from string theory.
In Sec.\,III we describe how the QCD axion could be realized in the model under the two global $U(1)_X$ symmetry. And we show explicitly symmetry breaking scales by considering the astrophysical constraints on star coolings, and provide model predictions on the axion mass and axion-photon coupling. 
In Sec.\,IV we investigate how neutrino oscillations at low energies could be connected to new oscillations available on high energy neutrinos. In turn, we explore what values of CP phase and atmospheric mixing angle in the low energy neutrino oscillation can be predicted, depending on mass hierarchies of the active neutrinos and mass splittings responsible for new oscillations. In addition, we examine a possibility to observe the effects of the pseudo-Dirac property of neutrinos by performing astronomical-scale baseline experiments to uncover the oscillation effects of tiny mass splitting, and such possibility has the ability to distinguish between normal mass ordering and inverted one of the active neutrino mass.
What we have done is summarized in Sec.\,V, and we provide our conclusions.

%%%%%%%%%%%%%%%%%%%%%%%%%%%%%%%%%%%%%%%%%%%%%%%%%%%%%%%%%%%%%%%%%%%%%%%%%%%%%%%%%%%%%%%%%%
\section{The Model setup: flavored $A_{4}\times U(1)_{X}$ symmetry}
Unless flavor symmetries are assumed, particle masses and mixings are generally undetermined in the SM gauge theory. In order to describe the present SM flavor puzzles associated with the fermion mass hierarchies including the large leptonic mixing angles and small quark mixing angles, we introduce the non-Abelian discrete $A_{4}$ flavor symmetry\,\cite{Altarelli:2005yp, nonAbelian} which is mainly responsible for the peculiar mixing patterns with an additional continuous global symmetry $U(1)_{X}$ which is mainly for vacuum configuration as well as for  describing mass hierarchies of leptons and quarks. Moreover, the spontaneous breaking of $U(1)_{X}$ realizes the existence of the NG mode (called axion) and provides an elegant solution of the strong CP problem. Along with Ref.\,\cite{Ahn:2014gva} in a way that no axionic domain wall problem occurs, this global $U(1)$ symmetry is referred to as ``flavored-PQ symmetry".  Then the symmetry group for matter fields (leptons and quarks), flavon fields and driving fields is $A_{4}\times U(1)_{X}$, whose quantum numbers are assigned in TABLE\,\ref{DrivingRef} and \ref{reps}.

To impose the $A_{4}$ flavor symmetry on our model properly, apart from the usual two Higgs doublets $H_{u,d}$ responsible for electroweak symmetry breaking, which are invariant under $A_{4}$ ({\it i.e.} flavor singlets $\mathbf{1}$ with no $T$-flavor), the scalar sector is extended by introducing two types of new scalar multiplets, flavon fields\,\footnote{The flavon fields are responsible for the spontaneous breaking of the flavor symmetry, while the driving fields are introduced to break the flavor group along required vacuum expectation value (VEV) directions and to allow the flavons to get VEVs, which couple only to the flavons.} $\Phi_{T},\Phi_{S},\Theta,\tilde{\Theta}, \Psi, \tilde{\Psi}$ that are $SU(2)$-singlets and driving fields $\Phi^{T}_{0},\Phi^S_{0},\Theta_{0},\Psi_{0}$ that are associated to a nontrivial scalar potential in the symmetry breaking sector: we take the flavon fields $\Phi_{T},\Phi_{S}$ to be $A_{4}$ triplets, and $\Theta,\tilde{\Theta},\Psi,\tilde{\Psi}$ to be $A_{4}$ singlets with no $T$-flavor ($\mathbf{1}$ representation), respectively, that are $SU(2)$-singlets, and driving fields $\Phi_{0}^{T},\Phi_{0}^{S}$ to be $A_{4}$ triplets and $\Theta_{0}, \Psi_{0}$ to be an $A_{4}$ singlet. In addition, the superpotential $W$ in the model (see, Eqs.\,(\ref{potential},\ref{lagrangian1}) and (\ref{lagrangian2})) is uniquely determined by the $U(1)_R$ symmetry, containing the usual $R$-parity as a subgroup: $\{matter\,fields\rightarrow e^{i\xi/2}\,matter\,fields\}$ and $\{driving\,fields\rightarrow e^{i\xi}\,driving\,fields\}$, with $W\rightarrow e^{i\xi}W$, whereas flavon and Higgs fields remain invariant under an $U(1)_R$ symmetry. As a consequence of the $R$ symmetry, the other superpotential term $\kappa_{\alpha}L_{\alpha}H_{u}$ and the terms violating the lepton and baryon number symmetries are not allowed\,\footnote{In addition, higher-dimensional supersymmetric operators like $Q_{i}Q_{j}Q_{k}L_{l}$ ($i,j,k$ must not all be the same) are not allowed either, and stabilizing proton.}.

In the lepton sector the $A_{4}$ model giving non-zero $\theta_{13}$ as well as bi-large mixings, $\theta_{23}, \theta_{12}$, works as follows. According to the $\mu$--$\tau$ power law in Ref.\,\cite{Ahn:2014gva}, one can assign charged-leptons to the three inequivalent singlet representations of $A_{4}$: we assign the left-handed charged leptons denoted as $L_e,\,L_\mu,\,L_\tau$, the electron flavor to the ${\bf 1}$ ($T$-flavor 0), the muon flavor to the ${\bf 1}'$  ($T$-flavor $+$1), and the tau flavor to the ${\bf 1}''$ ($T$-flavor $-$1), while the right-handed charged leptons denoted as $e^c,\,\mu^c,\,\tau^c$, the electron flavor to the ${\bf 1}$ ($T$-flavor 0), the muon flavor to the ${\bf 1}''$  ($T$-flavor $-$1), and the tau flavor to the ${\bf 1}'$ ($T$-flavor $+$1). In addition, we assign the right-handed neutrinos $SU(2)_L$ singlets denoted as $N^c$ to the ${\bf 3}$, while the right-handed neutrinos $SU(2)_L$ singlets denoted as $S^c_e$, $S^c_\mu$ and $S^c_\tau$ to the ${\bf 1}$, ${\bf 1}''$ and ${\bf 1}'$, respectively. On the other hand, for the quark flavors we assign the left-handed quark $SU(2)_{L}$ doublets denoted as $Q_{1}$, $Q_{2}$ and $Q_{3}$ to the ${\bf 1}$, ${\bf 1}''$ and ${\bf 1}'$, respectively, while the right-handed up-type quarks are assigned as $u^{c}$, $c^{c}$ and $t^{c}$ to the ${\bf 1}$, ${\bf 1}'$ and ${\bf 1}''$ under $A_{4}$, respectively, and the right-handed down-type quark $SU(2)_L$ gauge singlet $D^{c}=\{d^{c}, s^{c}, b^{c}\}$ to the ${\bf 3}$ under $A_{4}$.

Finally, the additional symmetry $U(1)_{X}$ is imposed\,\footnote{It is likely that an exact continuous global symmetry is violated by quantum gravitational effects\,\cite{Krauss:1988zc}. Here the global $U(1)_X$ symmetry is a remnant of the broken $U(1)_X$ gauge symmetry which connects string theory with flavor physics\,\cite{Ahn:2016typ}, see Sec.\,\ref{qla1}.}, which is an anomalous symmetry and under which matter fields, flavon fields, and driving fields carry their own $X$-charges. 
The $U(1)_{X}$ invariance forbids renormalizable Yukawa couplings for the light families, but would allow them through effective nonrenormalizable couplings suppressed by $({\cal F}/\Lambda)^n$ with $n$ being positive integers. Then, the gauge singlet flavon field ${\cal F}$ is activated to dimension-4(3) operators with different orders\,\cite{Ahn:2014zja, Froggatt:1978nt}
 \begin{eqnarray}
  c_{0}\,{\cal OP}_{4}\,({\cal F})^{0}+c'_{1}\,{\cal OP}_{3}\,({\cal F})^{1}+c_{1}\,{\cal OP}_{4}\,\left(\frac{{\cal F}}{\Lambda}\right)^1+c_{2}\,{\cal OP}_{4}\,\left(\frac{{\cal F}}{\Lambda}\right)^{2}+c_{3}\,{\cal OP}_{4}\,\left(\frac{{\cal F}}{\Lambda}\right)^{3}+...
 \label{flavon}
 \end{eqnarray}
where ${\cal OP}_{4(3)}$ is a dimension-$4(3)$ operator, and all the coefficients $c_{i}$ and $c'_{i}$ are complex numbers with absolute value of order unity. Even with all couplings being of order unity, hierarchical masses for different flavors can be naturally realized. 
The flavon field ${\cal F}$ is a scalar field which acquires a VEV and breaks spontaneously the flavored-PQ symmetry $U(1)_{X}$. Here $\Lambda$, above which there exists unknown physics, is the scale of flavor dynamics, and is associated with heavy states which are integrated out. The effective theory below $\Lambda$ is rather simple, while the full theory will have many heavy states. We assume that the cut-off scale $\Lambda$  in the superpotentials\,(\ref{lagrangian1}) and (\ref{lagrangian2}) is a scale where the complex structure and axio-dilaton moduli are stabilized through fluxes.
So, in our framework, the hierarchy $\langle H_{u,d}\rangle=v_{u,d}\ll\Lambda$ is maintained, and below the scale $\Lambda$ the higher dimensional operators express the effects from the unknown physics.
Since the Yukawa couplings are eventually responsible for the fermion masses they must be related in a very simple way at a large scale in order for intermediate scale physics to produce all the interesting structure in the fermion mass matrices. %On the other hand, cosmological observables, such as power spectrum of curvature perturbations and spectral index, do not generically receive significant contributions from possible higher-dimensional non-renormalizable operators, as these are suppressed by the Planck mass $M_{\rm P}$. So inflationary dynamics is mainly governed by a few renormalizable operators which might have observable implications for laboratory experiments.  

Here we recall that $A_{4}$ is the symmetry group of the tetrahedron and the finite groups of the even permutation of four objects having four irreducible representations: its irreducible representations are ${\bf 3}, {\bf 1}, {\bf 1}' , {\bf 1}''$ with ${\bf 3}\otimes{\bf 3}={\bf 3}_{s}\oplus{\bf 3}_{a}\oplus{\bf 1}\oplus{\bf 1}'\oplus{\bf 1}''$, and ${\bf 1}'\otimes{\bf 1}'={\bf 1}''$. The details of the $A_{4}$ group are shown in Appendix\,\ref{A4group}.
Let $(a_{1}, a_{2}, a_{3})$ and $(b_{1}, b_{2}, b_{3})$ denote the basis vectors for two ${\bf 3}$'s. Then, we have
 \begin{eqnarray}
  (a\otimes b)_{{\bf 3}_{\rm s}} &=& \frac{1}{\sqrt{3}}(2a_{1}b_{1}-a_{2}b_{3}-a_{3}b_{2}, 2a_{3}b_{3}-a_{2}b_{1}-a_{1}b_{2}, 2a_{2}b_{2}-a_{3}b_{1}-a_{1}b_{3})~,\nonumber\\
  (a\otimes b_c)_{{\bf 3}_{\rm a}} &=& i(a_{3}b_{2}-a_{2}b_{3}, a_{2}b_{1}-a_{1}b_{2}, a_{1}b_{3}-a_{3}b_{1})~,\nonumber\\
  (a\otimes b)_{{\bf 1}} &=& a_{1}b_{1}+a_{2}b_{3}+a_{3}b_{2}~,\nonumber\\
  (a\otimes b)_{{\bf 1}'} &=& a_{1}b_{2}+a_{2}b_{1}+a_{3}b_{3}~,\nonumber\\
  (a\otimes b)_{{\bf 1}''} &=& a_{1}b_{3}+a_{2}b_{2}+a_{3}b_{1}~.
  \label{A4reps}
 \end{eqnarray}
Under $A_{4}\times U(1)_{X}\times U(1)_{R}$, the driving, flavon, and Higgs fields are assigned as in TABLE\,\ref{DrivingRef}.
%\begin{center}
\begin{table}[h]
\caption{\label{DrivingRef} Representations of the driving, flavon, and Higgs fields under $A_4 \times U(1)_{X}$. Here $U(1)_X\equiv U(1)_{X_1}\times U(1)_{X_2}$ symmetries which are generated by the charges $X_1=-2p$ and $X_2=-q$.}
\begin{ruledtabular}
\begin{tabular}{cccccccccccccccc}
Field &$\Phi^{T}_{0}$&$\Phi^{S}_{0}$&$\Theta_{0}$&$\Psi_{0}$&\vline\vline&$\Phi_{S}$&$\Phi_{T}$&$\Theta$&$\tilde{\Theta}$&$\Psi$&$\tilde{\Psi}$&\vline\vline&$H_{d}$&$H_{u}$\\
\hline
$A_4$&$\mathbf{3}$&$\mathbf{3}$&$\mathbf{1}$&$\mathbf{1}$&\vline\vline&$\mathbf{3}$&$\mathbf{3}$&$\mathbf{1}$&$\mathbf{1}$&$\mathbf{1}$&$\mathbf{1}$&\vline\vline&$\mathbf{1}$&$\mathbf{1}$\\
$U(1)_{X}$&$0$&$4p$&$4p$&$0$&\vline\vline&$-2p$&$0$&$-2p$&$-2p$&$-q$&$q$&\vline\vline&$0$&$0$\\
$U(1)_R$&$2$&$2$&$2$&$2$&\vline\vline&$0$&$0$&$0$&$0$&$0$&$0$&\vline\vline&$0$&$0$\\
%$SU(2)\times U(1)_Y$&$1_{0}$&$1_{0}$&$1_{0}$&$1_{0}$&$1_{0}$&$1_{0}$&$1_{0}$\\
\end{tabular}
\end{ruledtabular}
\end{table}
%\end{center}
%%%%%%%%%%%%%%%%%%%%%%%%%%%%%%%%%%%%%%%%%%%%%%%%%%%%%%%%%%%%%%%%%%%%%%%%%%%%%%%
\subsection{Vacuum configuration}
\label{vacCon}
\noindent 
The superpotential dependent on the driving fields, which is invariant under  $SU(3)_c\times SU(2)_L\times U(1)_{Y}\times U(1)_{X}\times A_{4}$, is given at leading order by
 \begin{eqnarray}
W_{v} &=& \Phi^{T}_{0}\left(\tilde{\mu}\,\Phi_{T}+\tilde{g}\,\Phi_{T}\Phi_{T}\right)+\Phi^{S}_{0}\left(g_{1}\,\Phi_{S}\Phi_{S}+g_{2}\,\tilde{\Theta}\Phi_{S}\right)\nonumber\\
 &+& \Theta_{0}\left(g_{3}\,\Phi_{S}\Phi_{S}+g_{4}\,\Theta\Theta+g_{5}\,\Theta\tilde{\Theta}+g_{6}\,\tilde{\Theta}\tilde{\Theta}\right)+g_{7}\,\Psi_{0}\left(\Psi\tilde{\Psi}-\mu^2_\Psi\right)\,,
 \label{potential}
 \end{eqnarray}
where the fields $\Psi$ and $\tilde{\Psi}$ charged by $-q,q$, respectively, are ensured by the $U(1)_{X}$ symmetry extended to a complex $U(1)$ due to the holomorphy of the supepotential. 
Note here that the PQ scale $\mu_\Psi\equiv\sqrt{v_{\Psi}v_{\tilde{\Psi}}/2}$ corresponds to the scale of the spontaneous symmetry breaking scale, see Eqs.\,(\ref{vev_ph}) and (\ref{PQ_scale}).
Recalling that the model\,\footnote{In the model there are three $U(1)$ symmetries, $U(1)_L$ (lepton number), $U(1)_{\rm PQ}$ and $U(1)_{Y}$ except for $U(1)_{R}$ and $U(1)_B$ (baryon number). All of these threes are finally broken. $U(1)_Y$ is broken by the electroweak symmetry breakdown. When flavon fields acquire VEVs, both $U(1)_L$(which is hidden) and $U(1)_{\rm PQ}$ appear to be broken. Actually, there are linear combinations of the two $U(1)_{X_i}$ symmetries, which are $U(1)_{\tilde{X}}\times U(1)_f$. Here the $U(1)_{\tilde{X}}$ symmetry as $U(1)_{\rm PQ}$ has anomaly, while the $U(1)_f$ is anomaly-free. Note that $U(1)_f$ is not identified with $U(1)_L$.} implicitly has two $U(1)_{X}\equiv U(1)_{X_1}\times U(1)_{X_2}$ symmetries which are generated by the charges $X_{1}=-2p$ and $X_{2}=-q$.
Since there is no fundamental distinction between the singlets $\Theta$ and $\tilde{\Theta}$ as indicated in TABLE\,\ref{DrivingRef}, we are free to define $\tilde{\Theta}$ as the combination that couples to $\Phi^{S}_{0}\Phi_{S}$ in the superpotential $W_{v}$\,\cite{Altarelli:2005yp}.
Due to the assignment of quantum numbers under $A_{4}\times U(1)_{X}\times U(1)_{R}$ the usual superpotential term $\mu H_{u}H_{d}$ is not allowed, while the following operators driven by $\Psi_0$ and $\Phi^T_0$ are allowed by
 \begin{eqnarray}
  g_{\Psi_0}\Psi_0\,H_uH_d+\frac{g_{T}}{\Lambda}(\Phi^{T}_{0}\Phi_{T})_{{\bf 1}}H_{u}H_{d}\,,
 \label{muterm}
 \end{eqnarray}
which is to promote the $\mu$-term $\mu_{\rm eff}\equiv g_{\Psi_0}\langle\Psi_{0}\rangle+g_{T}\langle\Phi^{T}_{0}\rangle\, v_{T}/(\sqrt{2}\Lambda)$ of the order of $m_S$ and/or $m_{S}\,v_{T}/\Lambda$ (here $\langle\Psi_{0}\rangle$ and $\langle\Phi^{T}_{0}\rangle$: the VEVs of the scalar components of the driving fields, $m_{S}$: soft SUSY breaking mass). Here\,\footnote{As discussed in Ref.\,\cite{Ahn:2016hbn}, the field $\Psi_0$ identified as inflaton can predominantly decay into Higgses (and Higgsinos) through the first term after inflation, which is important for inflation and Affleck-Dine leptogenesis, while the second term is crucial for relating the sizable $\mu$-term with the low energy flavor physics. The size of the renormalizable superpotential coupling of the inflaton to particles of the SM is severely restricted by the reheating temperature, $T^{\rm reh}_{\Psi_0}$, and in turn a successful leptogenesis. Consequently, we have $\mu_{\rm eff}\simeq g_{T}\langle\Phi^{T}_{0}\rangle\, v_{T}/\Lambda$ as in Ref.\,\cite{Ahn:2014gva}, which can describe the correct Cabibbo-Kobayashi-Maskawa (CKM) mixing matrix with $v_T/\Lambda\sim0.04\simeq\lambda^2/\sqrt{2}$. Since the field $\Phi_T$ is not charged under the $U(1)_X$, the non-trivial next-to-leading order operators in the down-type quark superpotential (\ref{lagrangian1}) could be generated via $\Phi_T$, see footnote 18.} we assume $g_{\Psi_0}\langle\Psi_{0}\rangle\ll g_{T}\langle\Phi^{T}_{0}\rangle\, v_{T}/(\sqrt{2}\Lambda)$. The supersymmetry of the model is assumed broken by all possible holomorphic soft terms which are invariant under $A_{4}\times U(1)_{X}\times U(1)_{R}$ symmetry, where the soft breaking terms are already present at the scale relevant to flavor dynamics.
And it is evident that at leading order the scalar supersymmetric $W(\Phi_{T}\Phi_{S})$ terms are absent due to different $U(1)_{X}$ quantum number, which is crucial for relevant vacuum alignments in the model to produce the present large leptonic mixing and small quark mixing. It is interesting that at the leading order the electroweak scale does not mix with the potentially large scales $v_{S},v_{T},v_{\Theta}$ and $v_{\Psi}$.
The $A_4$ flavor symmetry is broken by two triplets $\Phi_S$ and $\Phi_T$ and by a singlet $\Theta$. As demonstrated in Appendix\,\ref{flavon}, the fields develop a phenomenologically nontrivial VEV along the direction in Eq.\,(\ref{vev_ph}). Therefore, as we shall see later, such VEV direction is very crucial to realize the present experimental data of small quark mixing angles and leptonic tri-bimaximal mixing (TBM)-like angles. See also below Eq.\,(\ref{TBM1}).

We take the $U(1)_{X}$ breaking scale, which corresponds to the $A_{4}$ symmetry breaking scale, to be much above the electroweak scale in our scenario\,\footnote{See the symmetry breaking scales from the astrophysical constraints Eq.\,(\ref{a_nucleon03}).}, that is, 
\begin{eqnarray}
 \langle H_{u,d}\rangle\ll\langle\Theta\rangle,\langle\Phi_{T}\rangle,\langle\Phi_{S}\rangle<\langle\Psi\rangle,\langle\tilde{\Psi}\rangle\,.
 \label{hierarchy_vev}
\end{eqnarray}
Here we assume that the electroweak symmetry is broken by some mechanism, such as radiative effects when SUSY is broken. In supergravity SUSY is broken by the non-vanishing VEV of some auxiliary field.
Setting to zero from the beginning the matter fields $\{q^c,\ell,H_u,...\}$, with the almost vanishing cosmological constant for the remaining fields the gravitino mass $m_{3/2}$ is directly related to the scale of supersymmetry breaking, 
 \begin{eqnarray}
  |F|^2-3m^2_{3/2}M^2_P+\frac{1}{2}D^2_{X_i}\approx0\,,
 \end{eqnarray}
implying that the $F$- and $D$-term potentials should vanish in the limit $m_{3/2}=e^{\tilde{K}/2M^2_P}|W|/M^2_P$ (here $\tilde{K}$ is a Kahler potential in Eq.\,(\ref{Kahler0})) going to zero and some of them should scale as $m_{3/2}$ at the minimum.
In global SUSY limit, {\it i.e.} $M_{P}\rightarrow\infty$, the vacuum configurations are obtained by the $F$- and $D$-terms of all the fields being required to vanish. 
The relevant $F$-term potential is written as
\begin{eqnarray}
 V^{\rm global}_{F}&=&\left|\frac{2g_{1}}{\sqrt{3}}\left(\Phi_{S1}\Phi_{S1}-\Phi_{S2}\Phi_{S3}\right)+g_{2}\Phi_{S1}\tilde{\Theta}\right|^{2}\nonumber\\
  &+&\left|\frac{2g_{1}}{\sqrt{3}}\left(\Phi_{S2}\Phi_{S2}-\Phi_{S1}\Phi_{S3}\right)+g_{2}\Phi_{S3}\tilde{\Theta}\right|^{2}\nonumber\\
  &+&\left|\frac{2g_{1}}{\sqrt{3}}\left(\Phi_{S3}\Phi_{S3}-\Phi_{S1}\Phi_{S2}\right)+g_{2}\Phi_{S2}\tilde{\Theta}\right|^{2}\nonumber\\
  &+&\left|g_{3}\left(\Phi_{S1}\Phi_{S1}+2\Phi_{S2}\Phi_{S3}\right)+g_{4}\Theta^{2}+g_{5}\Theta\tilde{\Theta}+g_{6}\tilde{\Theta}^{2}\right|^{2}\nonumber\\
  &+&\left|g_{7}\left(\Psi\tilde{\Psi}-\mu^2_\Psi\right)\right|^2+|g_7|^2|\Psi_0|^2\left(|\Psi|^2+|\tilde{\Psi}|^2\right)+\sum_{i={\rm the~others}}\left|\frac{\partial W_{v}}{\partial z_{i}}\right|^{2}\,,
 \label{V_F}
\end{eqnarray}
where $g_i$ are dimensionless couplings.
The model contains two Fayet-Iliopolos (FI) $D$-terms, ${\cal L}_{\rm FI}=-\xi^{\rm FI}_i\int d^2\theta V_{X_i}=-\xi^{\rm FI}_i\,g_{X_i}\,D_{X_i}$, giving rise to the $D$-term potential. The $D$-term potential is given by
 \begin{eqnarray}
  V^{\rm global}_D&=&\frac{|X_1|^2g^2_{X_1}}{2}\Big(\frac{\xi^{\rm FI}_1}{|X_1|}-|\Phi_S|^2-|\Theta|^2-|\tilde{\Theta}|^2\Big)^2
  +\frac{|X_2|^2g^2_{X_2}}{2}\Big(\frac{\xi^{\rm FI}_2}{|X_2|}-|\Psi|^2+|\tilde{\Psi}|^2\Big)^2
 \label{grobal_V}
 \end{eqnarray}
with $D_{X_i}=g_{X_i}(\xi^{\rm FI}_i-\sum_i X_i|\Phi_i|^2)$, where $\Phi_1=\{\Phi_S,\Theta\}$ and $\Phi_2=\{\Psi,\tilde{\Psi}\}$, and $\xi^{\rm FI}_i=2E_i/\tau_i$ are constant parameters with dimensions of mass squared and here $E_i$ are measure of the strength of the fluxes for the gauge fields living on the D7 branes\,\cite{Burgess:2003ic}. In $V^{\rm global}_D$ the flavon fields charged under the $U(1)_X$ gauge group for which the fluxes provide FI factors.  
Since SUSY is preserved after the spontaneous symmetry breaking of $U(1)_X\times A_4$, the scalar potential in the limit $M_P\rightarrow\infty$ vanishes at its ground states, {\it i.e.}, $\langle V^{\rm global}_D\rangle=0$ and $\langle V^{\rm global}_F\rangle=0$ vanishing $F$-terms must have also vanishing $D$-terms. 
Consequently, the VEVs of the flavon fields are from the minimization conditions of the $F$-term scalar potential: from Appendix-\ref{flavon} the phenomenologically non-trivial solutions\,\cite{Ahn:2014gva}  
 \begin{eqnarray}
  \langle\Phi_S\rangle=\frac{1}{\sqrt{2}}(v_S,v_S,v_S)\,, \quad\langle\Phi_T\rangle=\frac{1}{\sqrt{2}}(v_T,0,0)\,, \quad\langle\Theta\rangle=\frac{v_\Theta}{\sqrt{2}}\,, \quad\langle\Psi\rangle=\langle\tilde{\Psi}\rangle=\frac{v_\Psi}{\sqrt{2}}\,,
 \label{vev_ph}
 \end{eqnarray}
with $v_{\Theta}=v_{S}\sqrt{-3\frac{g_{3}}{g_{4}}}$ and $v_{T}=-(\tilde{\mu}/\tilde{g})\,\sqrt{3/2}$ where $\tilde{g}$ is a dimensionless coupling, as well as a set of trivial solutions
 \begin{eqnarray}
  \langle\Phi_S\rangle=(0,0,0)\,, \qquad\langle\Phi_T\rangle=(0,0,0)\,, \qquad\langle\Theta\rangle=0\,, \qquad\langle\Psi\rangle=\langle\tilde{\Psi}\rangle=\frac{v_\Psi}{\sqrt{2}}\,,
 \label{vev_tr}
 \end{eqnarray}
in which the undetermined VEVs indicate that in the SUSY limit there exist flat directions in the flavon potential along which the scalar fields $\Phi_{S},\Theta$ and $\Psi,\tilde{\Psi}$ do not feel the potential. Even these VEVs could be slightly perturbed by higher dimensional operators contributing to the driving superpotential, their corrections to the lepton and quark mass matrices are absorbed into the leading order terms and redefined due to the same VEV directions, or can be kept small enough and negligible, as shown in Ref.\,\cite{Ahn:2014gva}.
The above two supersymmetric solutions are taken by the D-flatness conditions, respectively, for (i) phenomenologically viable case 
 \begin{eqnarray}
  \xi^{\rm FI}_1=|X_1|(\langle|\Phi_S|^2\rangle+\langle|\Theta|^2\rangle)\,, \qquad\xi^{\rm FI}_2=0\,, \qquad\langle\Psi\rangle=\langle\tilde{\Psi}\rangle\,,
 \end{eqnarray}
 and (ii) phenomenologically trivial case 
 \begin{eqnarray}
  \xi^{\rm FI}_1=\langle\Phi_S\rangle=\langle\Theta\rangle=0\,, \qquad\xi^{\rm FI}_2=0\,, \qquad\langle\Psi\rangle=\langle\tilde{\Psi}\rangle\,,
 \label{phtriv}
 \end{eqnarray}
both of which indicate that the VEVs of the flavon fields strictly depend on the moduli stabilization, particularly on the VEVs of the fluxes $E_i$ in the FI terms\,\cite{Burgess:2003ic}. 
So it seems hard for the first case (i) to stabilize $|\Phi_i|$ at large VEVs$\sim{\cal O}(10^{9-10})$ GeV. And there is a tension between $\langle\Phi_i\rangle=0$ and $\langle\xi^{\rm FI}_i\rangle\neq0$ which are possible as long as $E_i$ are below the string scale. Therefore it is imperative that, in order for the $D$-terms to act as uplifting potential, the $F$-terms have to necessarily break SUSY.

In order for the solution in Eq.\,(\ref{vev_tr})  to be phenomenologically non-trivial, by taking $m^{2}_{\Phi_{S}}, \,m^{2}_{\Theta}, \,m^{2}_{\Psi}, \,m^{2}_{\tilde{\Psi}}<0$, $\Phi_1$ and $\Phi_2$ roll down toward its true minimum from a large scale, which we assume to be stabilized far away from the origin by Planck-suppressed higher dimensional corrections in the SUSY broken phase. And by adding a soft SUSY breaking mass term to the scalar potential one can execute $\langle\tilde{\Theta}\rangle=0$ for the scalar field $\tilde{\Theta}$ with $m^{2}_{\tilde{\Theta}}>0$. Then, the vacuum alignment is taken as the absolute minimum.
The phenomenologically viable VEVs of the flavon fields can be determined by considering both the SUSY breaking effect which lift up the flat directions and supersymmetric next-to-leading order Planck-suppressed terms\,\cite{Nanopoulos:1983sp, Murayama:1992dj} invariant under $A_4\times U(1)_X$. 
The supersymmetric next-to-leading order terms are given by
 \begin{eqnarray}
  \Delta W_v&\simeq&\frac{\alpha}{M_{P}}\Psi\tilde{\Psi}(\Phi_T\Phi^T_0)_{\bf 1}+\frac{\beta}{M_P}(\Phi^S_0\Phi_T)_{\bf 1}\Theta\Theta\nonumber\\
  &+&\frac{1}{M_P}\left\{\gamma_1(\Phi_S\Phi_S)_{\bf 1}(\Phi_T\Phi^S_0)_{\bf 1}+\gamma_2(\Phi_S\Phi_S)_{\bf 1'}(\Phi_T\Phi^S_0)_{\bf 1''}+\gamma_3(\Phi_S\Phi_S)_{\bf 1''}(\Phi_T\Phi^S_0)_{\bf 1'}\right\}\,,
 \end{eqnarray}
where $\alpha$, $\beta$, and $\gamma_{1,2,3}$ are real-valued constants being of order ${\cal O}(0.1)={\cal O}(1)/\sqrt{8\pi}$. Note that here we have neglected irrelevant operators including $\tilde{\Theta}$, $(\Phi_S\Phi_S)_{\bf 3s}$, $(\Phi_S\Phi_T)_{\bf 3s}$, and $(\Phi_S\Phi_T)_{\bf 3a}$ in $\Delta W_v$ since we are considering the phenomenologically non-trivial solutions as in Eq.\,(\ref{vev_ph}).
Since soft SUSY-breaking terms are already present at the scale relevant to flavor dynamics, the scalar potentials for $\Psi(\tilde{\Psi})$ and $\Phi_S(\Theta)$ at leading order read
\begin{eqnarray}
  V(\Phi_S,\Theta)&\simeq&\beta_1m^2_{3/2}|\Phi_S|^2+\beta_2m^2_{3/2}|\Theta|^2+\frac{v_T^2|\beta\Theta^2+\gamma\Phi^2_S|^2}{2M^2_{P}},\nonumber\\
  V(\Psi,\tilde{\Psi})&\simeq&\alpha_1m^2_{3/2}|\Psi|^2+\alpha_2m^2_{3/2}|\tilde{\Psi}|^2+|\alpha|^2\frac{v_T^2|\Psi|^2|\tilde{\Psi}|^2}{2M^2_{P}},
 \label{V_susy}
 \end{eqnarray}
 leading to the PQ breaking scales
 \begin{eqnarray}
  \mu^2_{\Psi}&=&\frac{v_{\Psi}v_{\tilde{\Psi}}}{2}=\frac{2\sqrt{\alpha_1\alpha_2}}{|\alpha|^2}\,\left(\frac{m_{3/2}}{v_T}\,M_{P}\right)^{2}\,,\\
 v^2_S&=& \frac{2\,\beta_1\,\kappa^2}{\gamma\,(\beta+\gamma)}\,\left(\frac{m_{3/2}}{v_T}\,M_{P}\right)^2=\kappa^2\,v^2_\Theta\,,
 \label{PQ_scale}
\end{eqnarray}
where $\gamma=3(\gamma_1+\gamma_2+\gamma_3)$, $\beta_1\beta=\gamma\beta_2$, and $\kappa=(-3g_3/g_4)^{-\frac{1}{2}}$.  It indicates that the gravitino mass (or soft SUSY breaking mass, $m_S=m_{3/2}$, see Ref.\,\cite{Ahn:2016typ}) strongly depends on the scales of PQ fields and $\Phi_T$ as well as the ratios $\sqrt{\alpha_1\,\alpha_2}/|\alpha|^2$ and $\beta_1/\gamma(\beta+\gamma)$; for example, for $\mu_\Psi\sim10^{10}$ GeV and $v_T\sim10^{9}$ GeV satisfying the SM fermion mass hierarchies\,\cite{Ahn:2014gva} one can obtain $m_{3/2}\sim{\cal O}(10)$ TeV, and/or subsequently $v_S\sim v_\Theta\sim10^{9}$ GeV with $\sqrt{\alpha_1\,\alpha_2}/|\alpha|^2\sim\beta_1/\gamma(\beta+\gamma)\sim{\cal O}(10^{-6})$ which is comparable with the axion decay constants (for example, as in Eqs.\,(\ref{axion-electron01}) and (\ref{a_nucleon02})).
With the soft SUSY-breaking potential, the radial components of the fields $\Psi$ and $\tilde{\Psi}$ are stabilized at
\begin{eqnarray}
  v_{\Psi}\simeq\mu_{\Psi}\sqrt{2}\left(\frac{\alpha_2}{\alpha_1}\right)^{1/4}\,,\qquad v_{\tilde{\Psi}}\simeq\mu_{\Psi}\sqrt{2}\left(\frac{\alpha_1}{\alpha_2}\right)^{1/4}\,,
 \label{PQ_scale1}
\end{eqnarray}
respectively.
The saxion field $h_{\Psi}$ is defined in Eq.\,(\ref{NGboson}) which is the deviation of $|\Psi|$ from the VEV Eq.\,(\ref{PQ_scale1}) along the flat direction.
And in the SUSY limit the driving fields $\Phi^{T}_0,\Phi^{S}_0,\Theta_0$ and $\Psi_0$ develop VEVs along the directions
 \begin{eqnarray}
  \qquad\langle\Phi^T_0\rangle=\left(0,0,0\right),\qquad\langle\Phi_{S}\rangle=\left(0,0,0\right),\qquad
  \langle\Theta_0\rangle=0,\qquad\langle\Psi_0\rangle=0\,,
 \label{vevd}
 \end{eqnarray}
in which the vacuum structures are corrected being of order $m_{S}$ when the SUSY breaking effect lifts up the flat directions.
%The massive modes of flavons and driving fields in the SUSY limit are either proportional to the mass parameter of the superpotential $W_v$ or to the undetermined VEVs.

As mentioned before, the model has two $U(1)$ symmetries which are generated by the charges $X_{1}\equiv-2p$ and $X_{2}\equiv-q$.
The $A_{4}$ flavor symmetry along with the flavored PQ symmetry $U(1)_{X_1}$ is spontaneously broken by two $A_{4}$-triplets $\Phi_{T},\Phi_{S}$ and by a singlet $\Theta$ in TABLE\,\ref{DrivingRef}.
And the $U(1)_{X_2}$ symmetry is spontaneously broken by $\Psi,\tilde{\Psi}$, whose scales are denoted as $v_{\Psi}$ and $v_{\bar{\Psi}}$, respectively, and the VEV of $\Psi$ (scaled by the cutoff $\Lambda$) is assumed as
 \begin{eqnarray}
 \frac{\langle\Psi\rangle}{\Lambda}=\frac{\langle\tilde{\Psi}\rangle}{\Lambda}\equiv\frac{\lambda}{\sqrt{2}}~.
 \label{Cabbibo}
 \end{eqnarray}
Here the parameter $\lambda\approx0.225$ stands for the Cabbibo parameter\,\cite{PDG}. After getting VEVs $\langle\Theta\rangle,\langle\Phi_{S}\rangle\neq0$ (which generates the heavy neutrino masses given by Eq.\,(\ref{MR1})) and $\langle\Psi\rangle\neq0$, the flavored PQ symmetry $U(1)_{X}$ is spontaneously broken at a scale much higher than the electroweak scale and is realized by the existence of the NG modes $A_{1,2}$ that couples to ordinary quarks and leptons at the tree level through the Yukawa couplings as in Eq.\,(\ref{AxionLag2}) (see also Eqs.\,(\ref{AxionLag14}), (\ref{AxionLag15}) and (\ref{Axion_nu_La}), and one of linear combinations of NG bosons becomes the QCD axion\,\footnote{The VEV configurations in Eq.\,(\ref{vev_ph}) break the $U(1)_X$ spontaneously and the superpotential dependent on the driving field $\Theta_0$ in Eq.\,(\ref{potential}) becomes, for simplicity, if we let $\Phi_{S1}=\Phi_{S2}=\Phi_{S3}$,
 $W_{\Theta_0} = \Theta_{0}\left(g_{3}\,\Phi_{S}\Phi_{S}+g_{4}\,\Theta\Theta+6\kappa\, g_3\left\{v_{\Theta}\Phi_{Si}-v_S\Theta\right\}+g_{5}\,(\Theta+2\frac{v_S}{\kappa})\tilde{\Theta}+g_{6}\,\tilde{\Theta}\tilde{\Theta}\right)$ after shifting by $v_{\Theta},v_S$. This shows clearly that the linear combination $(v_{\Theta}\Theta+v_{S}\Phi_{Si})/\sqrt{v^2_{\Theta}+v^2_{S}}$ is a massless superfield.}. Through triangle anomalies, the axion mixes with mesons (leading to a non-zero mass), and thus couples to photons, and nucleons. The explicit breaking of the $U(1)_X$ by the chiral anomaly effect further breaks it down to $Z_{N_{\rm DW}}$ discrete symmetry, where $N_{\rm DW}$ is the domain-wall number. At the QCD phase transition, the $Z_{N_{\rm DW}}$ symmetry is spontaneously broken, and which gives rise to a domain wall problem\,\cite{Sikivie:1982qv}.
Such domain wall problem can be overcome because the model has two anomalous axial $U(1)$ symmetries which are generated by the charges $X_1$ and $X_2$, $U(1)_{X}\equiv U(1)_{X_1}\times U(1)_{X_2}$.

%%%%%%%%%%%%%%%%%%%%%%%%%%%%%%%%%%%%%%%%%%%%%%%%%%%%%%%%%%%%%%%%%%%%%%%%%%%%%
\subsection{Quarks, Leptons, and Axions}
\label{qla}
\noindent Under $A_{4}\times U(1)_{X}$, the matter fields are assigned as in TABLE\,\ref{reps}. Because of the chiral structure of weak interactions, bare fermion masses are not allowed in the SM. Fermion masses arise through Yukawa interactions\,\footnote{Since the right-handed neutrinos $N^c$ ($S^c$) having a mass scale much above (below) the weak interaction scale are complete singlets of the SM gauge symmetry, they can possess bare SM invariant mass terms. However, the flavored-PQ symmetry $U(1)_{X}$ guarantees the absence of bare mass terms $M\,N^{c}N^{c}$ and $\mu_s\,S^cS^c$.}.  Recalling that $v_{\Psi}/\Lambda=v_{\tilde{\Psi}}/\Lambda\equiv\lambda$  in Eq.\,(\ref{Cabbibo}) is used when the $U(1)_X$ quantum numbers of the SM charged fermions are assigned.
%\begin{center}
\begin{table}[h]
\caption{\label{reps} Representations of the matter fields under $A_4 \times U(1)_{X}$.}
\begin{ruledtabular}
\begin{tabular}{ccccccccccccc}
Field &$Q_{1},Q_{2},Q_{3}$&$D^c$&$u^c,c^c,t^c$&$L_{e},L_{\mu},L_{\tau}$&$e^c,\mu^c,\tau^c$&$N^{c}$&$S_e^c,S_\mu^c,S_\tau^c$\\
\hline
$A_4$&$\mathbf{1}$, $\mathbf{1}''$, $\mathbf{1^{\prime}}$&$\mathbf{3}$&$\mathbf{1}$, $\mathbf{1}'$, $\mathbf{1^{\prime\prime}}$&$\mathbf{1}$, $\mathbf{1^{\prime}}$, $\mathbf{1^{\prime\prime}}$&$\mathbf{1}$, $\mathbf{1^{\prime\prime}}$, $\mathbf{1^\prime}$&$\mathbf{3}$&$\mathbf{1}$, $\mathbf{1^{\prime\prime}}$, $\mathbf{1^{\prime}}$\\
$U(1)_{X}$&$(-3q-r,-2q-r,-r)$ &$2p+r$&$r-3q,r,r$& $ -9q-p $ & $p+15q, p+13q, p+11q$& $p$&$p+25q$\\
$U(1)_R$&$1$ &~$1$~&~$1$& $ 1 $ & $1$~~~& $1$& $1$\\
%$SU(2)\times U(1)_Y$&$2_{-1}$&$2_\frac{1}{3}$&$1_{\frac{4}{3}}$&$1_{-\frac{2}{3}}$&$1_{-2}$&$1_{0}$&$2_{1}$&$2_{-1}$\\
\end{tabular}
\end{ruledtabular}
\end{table}
%\end{center}
The superpotential for Yukawa interactions in the quark sector, which are invariant under  $SU(3)_c\times SU(2)_L\times U(1)_{Y}\times U(1)_{X}\times A_{4}$, is given at leading order by
 \begin{eqnarray}
 W_{q} &=&  y_{u}\,Q_{1}u^{c}\,H_{u}+y_{c}\,Q_{2}\,c^{c}\,H_{u}+y_{t}\,Q_{3}\,t^{c}\,H_{u}\,,\nonumber\\
 &+& y_{d}\,Q_{1}(D^c\Phi_{S})_{{\bf 1}}\,\frac{H_{d}}{\Lambda}+y_{s}\,Q_{2}(D^c\Phi_{S})_{{\bf 1}'}\,\frac{H_{d}}{\Lambda}+y_{b}\,Q_{3}(D^c\Phi_{S})_{{\bf 1}''}\,\frac{H_{d}}{\Lambda}\,.
 \label{lagrangian1} 
 \end{eqnarray}
In the above superpotential, $W_q$, each quark sector has three independent Yukawa terms at the leading: apart from the Yukawa couplings, each up-type quark sector does not involve flavon fields, while the down-type quark sector involves\,\footnote{The operators including the field $\Phi_T$ appear in the next-to-leading order superpotential, {\it i.e.}, $\Delta W_d= x_{d}\,Q_{1}(D^c\Phi_{T})_{{\bf 1}}\,\frac{\Theta}{\Lambda^2}\,H_{d}+x_{s}\,Q_{2}(D^c\Phi_{T})_{{\bf 1}'}\,\frac{\Theta}{\Lambda^2}\,H_{d}+x_{b}\,Q_{3}(D^c\Phi_{T})_{{\bf 1}''}\,\frac{\Theta}{\Lambda^2}\,H_{d}
 +x^{as}_{d}\,Q_{1}(D^c\Phi_{T}\Phi_{S})_{{\bf 1}}\,\frac{H_{d}}{\Lambda^2}+x^{as}_{s}\,Q_{2}(D^c\Phi_{T}\Phi_{S})_{{\bf 1}'}\,\frac{H_{d}}{\Lambda^2}+x^{as}_{b}\,Q_{3}(D^c\Phi_{T}\Phi_{S})_{{\bf 1}''}\,\frac{H_{d}}{\Lambda^2}$ where $x_{d,s,b}$ and $x^{as}_{d,s,b}$ are Yukawa coupling constants, which plays crucial roles for the CKM mixing angles to be correctly fitted. See also Ref.\,\cite{Ahn:2014gva}.} the $A_{4}$-triplet flavon fields $\Phi_{T}$ and $\Phi_{S}$. The left-handed quark doublets $Q_1,Q_2,Q_3$ transform as ${\bf 1}, {\bf 1}''$, and ${\bf 1}'$, respectively; the right-handed quarks $u^{c}\sim{\bf 1}, c^{c}\sim{\bf 1}', t^{c}\sim {\bf 1}''$ and $D^{c}\equiv\{d^{c}, s^{c}, b^{c}\}\sim {\bf 3}$. Since the right-handed down-type quark transforms as ${\bf 3}$, in contrast with the up-type quark sector, the down-type quark sector can have non-trivial next-to-leading order terms as shown in Ref.\,\cite{Ahn:2014gva}, and which in turn explains the CKM matrix. The up-type quark superpotential in (\ref{lagrangian1}) does not contribute to the CKM matrix due to the diagonal form of mass matrix, while the down-type quark superpotential does contribute the CKM matrix. Naively speaking, since the leading order operators in the down-type quark superpotential has six physical parameters, they could not explain the four CKM parameters and three down-type quark masses. Thus, one can consider the next-to-leading order corrections as in footnote 17 to account for the correct CKM matrix.

In the lepton sector, based on the field contents in TABLE\,\ref{DrivingRef} and \ref{reps} the superpotential for Yukawa interactions under $SU(3)_c\times SU(2)_L\times U(1)_{Y}\times U(1)_{X}\times A_{4}$ reads at leading order
 \begin{eqnarray}
W_{\ell\nu} &=&y^s_{1}\,L_{e}\,S^{c}_e\,H_{u}+y^s_{2}\,L_{\mu}\,S^{c}_\mu\,H_{u}+y^s_{3}\,L_{\tau}\,S^{c}_\tau\,H_{u}\nonumber\\
&+&\frac{1}{2}\left(y^{ss}_{1}\,S^c_{e}S^{c}_e+y^{ss}_{2}\,S^c_{\mu}\,S^{c}_\tau+y^{ss}_{2}\,S^c_{\tau}\,S^{c}_\mu\right)\tilde{\Psi}\nonumber\\
&+&y^{\nu}_{1}\,L_{e}(N^{c}\Phi_{T})_{{\bf 1}}\frac{H_{u}}{\Lambda}+y^{\nu}_{2}\,L_{\mu}(N^{c}\Phi_{T})_{{\bf 1}''}\frac{H_{u}}{\Lambda}+y^{\nu}_{3}\,L_{\tau}(N^{c}\Phi_{T})_{{\bf 1}'}\frac{H_{u}}{\Lambda}\nonumber\\
&+& \frac{1}{2}(\hat{y}_{\Theta}\,\Theta+\hat{y}_{\tilde{\Theta}}\,\tilde{\Theta}) (N^{c}N^{c})_{{\bf 1}}+\frac{\hat{y}_{R}}{2} (N^{c}N^{c})_{{\bf 3}_{s}}\Phi_{S}\nonumber\\
&+&y_{e}\,L_{e}\,e^{c}\,H_{d}+y_{\mu}\,L_{\mu}\,\mu^{c}\,H_{d}+y_{\tau}\,L_{\tau}\,\tau^{c}\,H_{d}\,.
 \label{lagrangian2}
 \end{eqnarray}
In the above leptonic Yukawa superpotential, $W_{\ell\nu}$, charged lepton sector has three independent Yukawa terms at the leading: apart from the Yukawa couplings, each term does not involve flavon fields. The left-handed lepton doublets $L_e, L_\mu, L_\tau$ transform as ${\bf 1}$, ${\bf 1}'$, and ${\bf 1}''$, respectively; the right-handed leptons $e^{c}\sim{\bf 1}, \mu^{c}\sim{\bf 1}''$, and $\tau^{c}\sim {\bf 1}'$.
In neutrino sector, 
two right-handed Majorana neutrinos $S$ and $N$ are introduced to make light neutrinos pseudo-Dirac particles and to realize TBM pattern\,\footnote{See Eq.\,(\ref{TBM1}) the exact TBM mixing\,\cite{Harrison:2002er}.}, respectively; $S^c_e, S^c_\mu, S^c_\tau$ and $N$ transform as ${\bf 1}$, ${\bf 1}''$, ${\bf 1}'$, and ${\bf 3}$ under $A_4$ symmetry, respectively. They compose two Majorana mass terms; one is associated with an $A_{4}$ singlet $\tilde{\Psi}$, while the other one is associated with an $A_{4}$ singlet $\Theta$ and an $A_{4}$ triplet $\Phi_{S}$, in which all flavon fields associated with the Majorana mass terms are the SM gauge singlets. The two different assignments of $A_4$ quantum number to Majorana neutrinos guarantee the absence of the Yukawa terms $S^cN^c\times{flavon\,fields}$. %Below the cutoff scale $\Lambda$, the mass term of the Majorana neutrinos $N$ comprises an exact TBM pattern, which will be shown later. 
Correspondingly, two Dirac neutrino mass terms are generated; one is associated with $S^c$, and the other is $N^c$.
Imposing the continuous global $U(1)_{X}$ symmetry in TABLE\,\ref{reps} explains the absence of the Yukawa terms $LN^{c}\Phi_{S}$ and $N^{c}N^{c}\Phi_{T}$ as well as does not allow the interchange between $\Phi_{T}$ and $\Phi_{S}$, both of which transform differently under $U(1)_{X}$, so that bi-large $\theta_{12},\theta_{23}$ mixings with a non-zero $\theta_{13}$ mixing for the leptonic mixing matrix could be obtained after seesawing\,\cite{Minkowski:1977sc} (as will be shown later, the effective mass matrix achieved by seesawing contributes to TBM mixing pattern and pseudo-Dirac mass splittings, except for active neutrino masses. Such pseudo-Dirac mass splittings are responsible for very long wavelength, which in turn connect to an axion decay constant, see Eqs.\,(\ref{AxioAhn1}) and (\ref{scaleLambda}).).

Since the $U(1)_{X}$ quantum numbers are assigned appropriately to the matter fields content as in TABLE\,\ref{reps}, it is expected that the SM gauge singlet flavon fields derives higher-dimensional operators, which are eventually visualized into the Yukawa couplings of charged fermions as a function of flavon fields $\Psi(\tilde{\Psi})$, {\it i.e.}, $y_{u,c}=y_{u,c}(\tilde{\Psi})$, $y_{d,s}=y_{d,s}(\tilde{\Psi})$, $y_{e,\mu,\tau}=y_{e,\mu,\tau}(\Psi)$, except for the top and bottom Yukawa couplings :
\begin{eqnarray}
y_{u}&=&\hat{y}_{u}\left(\frac{\tilde{\Psi}}{\Lambda}\right)^{6},\qquad\qquad\, y_{c}=\hat{y}_{c}\left(\frac{\tilde{\Psi}}{\Lambda}\right)^{2},\qquad\qquad y_{t}=\hat{y}_{t}\,
 \nonumber\\
y_{d}&=&\hat{y}_{d}\left(\frac{\tilde{\Psi}}{\Lambda}\right)^{3},\qquad\qquad\, y_{s}=\hat{y}_{s}\left(\frac{\tilde{\Psi}}{\Lambda}\right)^2,\qquad\qquad y_{b}=\hat{y}_{b}\,.
 \nonumber\\
y_{e}&=&\hat{y}_{e}\left(\frac{\Psi}{\Lambda}\right)^{6}\,,\qquad\qquad
y_{\mu}=\hat{y}_{\mu}\left(\frac{\Psi}{\Lambda}\right)^{4}\,,~\,\quad\qquad
y_{\tau}=\hat{y}_{\tau}\left(\frac{\Psi}{\Lambda}\right)^{2}\,.
 \label{YukawaW}
\end{eqnarray}
From the top Yukawa coupling and pole mass ($\hat{y}_{t}$ and $m_{t}$) and the neutral Higgs VEV ratio ($\tan\beta=v_{u}/v_{d}$),
by requiring $\hat{y}_{t}$ to be order of unity, $1/\sqrt{10}\lesssim|\hat{y}_{t}|\lesssim\sqrt{10}$, we have the allowed range for $\tan\beta$: $1.7\lesssim\tan\beta<10$ where\,\footnote{We take a lower bound of $\tan\beta$ preferred in the Minimal Supersymmetric Standard Model (MSSM). For $\tan\beta<1.7$ the top quark Yukawa coupling blows up before the momentum scale $\mu\approx2\times10^{16}$ GeV.} we have used $m_{t}=173.07\pm0.52\pm0.72$ GeV\,\cite{PDG}. Especially, the value of $\tan\beta=2$ with the above Yukawa couplings is preferred because of the mixed $U(1)_X$-$[gravity]^2$ anomaly free condition together with the observed mass hierarchies of the SM charged fermions.
On the other hand, the neutrino Yukawa couplings in terms of the flavons $\Psi(\tilde{\Psi})$ and $\Theta$ are given as
\begin{eqnarray}
y^s_{i}&=& \hat{y}^s_{i}\left(\frac{\Psi}{\Lambda}\right)^{16}\,,\qquad \qquad y^{ss}_{i}= \hat{y}^{ss}_{i}\left(\frac{\Psi}{\Lambda}\right)^{51}\frac{\Theta}{\Lambda}\,,\nonumber\\
y^{\nu}_{i}&=& \hat{y}^{\nu}_{i}\left(\frac{\tilde{\Psi}}{\Lambda}\right)^{9}\,,\qquad \qquad\hat{y}_{\Theta}\approx \hat{y}_{\tilde{\Theta}}\approx \hat{y}_{R}\approx{\cal O}(1)\,.
 \label{YukawaWnu}
\end{eqnarray}
Here the hat Yukawa couplings $\hat{y}$ are complex numbers and of order unity, {\it i.e.} $1/\sqrt{10}\lesssim|\hat{y}|\lesssim\sqrt{10}$. The above Yukawa superpotentials (\ref{lagrangian1}) and (\ref{lagrangian2}) with Eqs.\,(\ref{YukawaW}) and (\ref{YukawaWnu}) indicate that, since the flavon fields charged under $U(1)_X$ are the SM gauge singlets, a direct NG mode coupling to ordinary quarks and leptons is possible through Yukawa interactions.
Since the fields associated with the superpotentials (\ref{lagrangian1}) and (\ref{lagrangian2}) are charged under $U(1)_{X}$, it is expected that the top quark and hat neutrino Yukawa couplings appearing in the superpotentials are of order unity and complex numbers.
We note that the flavon fields $\Phi_{S}$ and $\Phi_{T}$ derive dimension-5 operators in the down-type quark sector and Dirac neutrino sector, respectively, apart from the Yukawa couplings, while the flavon fields $\Psi$ and $\tilde{\Psi}$ derives higher dimensional operators through the Yukawa couplings with the $U(1)_{X}$ flavor symmetry responsible for the hierarchical charged lepton masses as shown by Eqs.\,(\ref{YukawaW}) and (\ref{YukawaWnu}).
The model is assumed to be broken by all possible holomorphic soft-terms, where the soft breaking terms are already present at the scale relevant to flavor dynamics.

The model incorporates the SM gauge singlet flavon fields ${\cal F}_A=\Phi_{S},\Theta,\Psi,\tilde{\Psi}$ with the following interactions invariant under the $U(1)_{X}\times A_{4}$ and the resulting chiral symmetry, {\it i.e.}, the kinetic and Yukawa terms, and the scalar potential $V_{\rm SUSY}$ in SUSY limit\,\footnote{In our superpotential, the superfields $\Phi_S, \Theta$ and $\Psi(\tilde{\Psi})$ are gauge singlets and have $-2p$ and $-q(q)$ $X$-charges, respectively.
Given soft SUSY-breaking potential, the radial components of the $X$-fields $|\Phi_S|$, $|\Theta|$ $|\Psi|$ and $|\tilde{\Psi}|$ are stabilized. The $X$-fields contain the axion, saxion (the scalar partner of the axion), and axino (the fermionic superpartner of the axion).} are of the form
 \begin{eqnarray}
  {\cal L} &\supset& \partial_{\mu}{\cal F}^{\dag}_A\,\partial^{\mu}{\cal F}_A+{\cal L}_{Y}-V_{\rm SUSY}+{\cal L}_{\vartheta}+\overline{\psi}\,i\! \! \not\!\partial\psi+\frac{1}{2}\overline{N}\,i\! \! \not\!\partial N +\frac{1}{2}\overline{S}\,i\! \! \not\!\partial S \,.
  \label{AxionLag}
 \end{eqnarray}
Here the $V_{\rm SUSY}$ term is written in terms of Eqs.\,(\ref{V_F}) and (\ref{grobal_V}), which is replaced by $V_{total}$ including soft SUSY breaking term when SUSY breaking effects are considered, and $\psi$ stands for all Dirac fermions.
The kinetic terms $\frac{\partial^2K}{\partial{\cal F}^{\dag}_A\partial{\cal F}_A}\partial_\mu{\cal F}^\dag_A\partial^{\mu}{\cal F}_A$ with Kahler potential $\tilde{K}\supset|{\cal F}_A|^2+$higher order terms (c.f. Eq.\,(\ref{kinetic})) for canonically normalized fields are written as
 \begin{eqnarray} 
 \partial_{\mu}\Phi^{\dag}_{S}\partial^{\mu}\Phi_{S}+\partial_{\mu}\Theta^{\dag}\partial^{\mu}\Theta+\partial_{\mu}\Psi^{\dag}\partial^{\mu}\Psi+\partial_{\mu}\tilde{\Psi}^{\dag}\partial^{\mu}\tilde{\Psi}\,.
 \label{AxionLag1}
 \end{eqnarray}
The scalar fields $\Phi_{S},\Theta$ and $\Psi(\tilde{\Psi})$ have $X$-charges $X_{1}=-2p$ and $X_{2}=-q(q)$, respectively, that is
 \begin{eqnarray}
 \Phi_{S_i}\rightarrow e^{i\xi_1 X_1}\Phi_{S_i},\quad\Theta\rightarrow e^{i\xi_1 X_1}\Theta\,;\quad\Psi\rightarrow e^{i\xi_2 X_2}\Psi,\quad\tilde{\Psi}\rightarrow e^{-i\xi_2 X_2}\tilde{\Psi}
 \label{X_sclar}
 \end{eqnarray}
where $\xi_k$ ($k=1,2$) are constants. So, the potential $V_{\rm SUSY}$ has $U(1)_X$ global symmetry.
In order to extract NG bosons resulting from spontaneous breaking of $U(1)_{X}$ symmetry, we set the decomposition of complex scalar fields as follows\,\footnote{Note that the massless modes are not contained in the fields $\tilde{\Theta},\Phi_{T},\Phi^{T}_{0},\Phi^{S}_{0},\Theta_{0},\Psi_{0}$.}
 \begin{eqnarray}
  &&\Phi_{Si}=\frac{e^{i\frac{\phi_{S}}{v_{S}}}}{\sqrt{2}}\left(v_{S}+h_{S}\right)\,,\qquad\qquad\quad\,\Theta=\frac{e^{i\frac{\phi_{\theta}}{v_{\Theta}}}}{\sqrt{2}}\left(v_{\Theta}+h_{\Theta}\right)\,,\nonumber\\
&&\Psi=\frac{v_{\Psi}}{\sqrt{2}}e^{i\frac{\phi_{\Psi}}{v_{g}}}\left(1+\frac{h_{\Psi}}{v_{g}}\right)\,,\qquad\qquad\,\tilde{\Psi}=\frac{v_{\tilde{\Psi}}}{\sqrt{2}}e^{-i\frac{\phi_{\Psi}}{v_{g}}}\left(1+\frac{h_{\Psi}}{v_{g}}\right)\,,
  \label{NGboson}
 \end{eqnarray}
in which we have set $\Phi_{S1}=\Phi_{S2}=\Phi_{S3}\equiv\Phi_{Si}$ in the SUSY limit, and $v_{g}=\sqrt{v^2_{\Psi}+v^2_{\tilde{\Psi}}}$. And the NG modes $A_1$ and $A_2$ are expressed as
 \begin{eqnarray}
  A_1=\frac{v_{S}\,\phi_{S}+v_{\Theta}\,\phi_{\theta}}{\sqrt{v^{2}_{S}+v^{2}_{\Theta}}}\,,\qquad A_{2}=\phi_{\Psi}
 \end{eqnarray}
with the angular fields $\phi_{S}$, $\phi_{\theta}$ and $\phi_{\Psi}$.
With Eqs.\,(\ref{AxionLag1}) and (\ref{NGboson}), the derivative couplings of $A_k$ arise from the kinetic terms
 \begin{eqnarray}
  \partial_{\mu}{\cal F}^{\ast}_{k}\,\partial^{\mu}{\cal F}_{k}
  &=&\frac{1}{2}\left(\partial_{\mu}A_1\right)^2\left(1+\frac{h_{\cal F}}{v_{\cal F}}\right)^2+\frac{1}{2}\left(\partial_{\mu}A_2\right)^2\left(1+\frac{h_{\Psi}}{v_{g}}\right)^2+\frac{1}{2}\left(\partial_{\mu}h_{\cal F}\right)^2+\frac{1}{2}\left(\partial_{\mu}h_{\Psi}\right)^2\nonumber\\
  &+&...
  \label{derivative}
 \end{eqnarray}
where $v_{\cal F}=v_{\Theta}(1+\kappa^2)^{1/2}$ and $h_{\cal F}=(\kappa h_{S}+h_{\Theta})/(1+\kappa^2)^{1/2}$, and the dots stand for the orthogonal components $h^{\bot}_{\cal F}$ and $A^{\bot}_{1}$. Recalling that $\kappa\equiv v_{S}/v_{\Theta}$. Clearly, the derivative interactions of $A_k$ ($k=1,2$) are suppressed by the VEVs $v_{\cal F}$ and $v_{\Psi}$. From Eq.\,(\ref{derivative}), performing $v_{\cal F}, v_{\Psi}\rightarrow\infty$, the NG modes $A_{1,2}$, whose interactions are determined by symmetry, are invariant under the symmetry and distinguished from the radial modes, $h_{\cal F}$ and $h_{\Psi}$.

In Eq.\,(\ref{AxionLag}) the Yukawa Lagrangian is given as follows. 
Once the scalar fields $\Phi_{S}, \Theta, \tilde{\Theta},\Psi$ and $\tilde{\Psi}$ get VEVs, the flavor symmetry $U(1)_{X}\times A_{4}$ is spontaneously broken\,\footnote{If the symmetry $U(1)_{X}$ is broken spontaneously, the massless modes $A_1$ of the scalar $\Phi_{S}$ (or $\Theta$) and $A_{2}$ of the scalar $\Psi(\tilde{\Psi})$ appear as phases.}. 
And at energies below the electroweak scale, all quarks and leptons obtain masses.
The relevant Yukawa interaction terms with chiral fermions $\psi$ charged under the flavored $ U(1)_X$ symmetry is given by 
 \begin{eqnarray}
  -{\cal L}_{Y} &=&
  \overline{q^{u}_{R}}\,\mathcal{M}_{u}\,q^{u}_{L}+e^{i\frac{A_{1}}{v_{\cal F}}}\,\overline{q^{d}_{R}}\,\mathcal{M}_{d}\,q^{d}_{L}
   + \overline{\ell_{R}}\,{\cal M}_{\ell}\,\ell_{L}\nonumber\\
 &+& \frac{1}{2} \begin{pmatrix} \overline{\nu^c_L} & \overline{S_R} & \overline{N_R} \end{pmatrix} \begin{pmatrix} 0 & e^{16i\frac{A_{2}}{v_{g}}}\,m^T_{DS} & e^{-9i\frac{A_{2}}{v_{g}}}\,m^T_D  \\ e^{16i\frac{A_{2}}{v_{g}}}\, m_{DS} & e^{i(50\frac{A_{2}}{v_{g}}+\frac{A_1}{v_{\cal F}})}\, M_{S} & 0  \\ e^{-9i\frac{A_{2}}{v_{g}}}\,m_D & 0 & e^{i\frac{A_{1}}{v_{\cal F}}}\,M_R \end{pmatrix} \begin{pmatrix} \nu_L \\ S^c_R \\ N^c_R \end{pmatrix} +\text{h.c.}\,,
  \label{AxionLag2}
 \end{eqnarray}
where $q^{u}=(u,c,t)$ and $q^{d}=(d,s,b)$.
And in the above Lagrangian\,(\ref{AxionLag2}) the Dirac and Majorana neutrino mass terms read
 \begin{eqnarray}
m_{DS}&=&{\left(\begin{array}{ccc}
 \hat{y}^s_{1} &  0 &  0 \\
 0 &  \hat{y}^s_{2} &  0   \\
 0 &  0  &  \hat{y}^s_{3}
 \end{array}\right)}\left(\frac{v_{\Psi}}{\sqrt{2}\Lambda}\right)^{16}v_{u}, 
 \label{YDS1}\\
 M_{S}&=&{\left(\begin{array}{ccc}
 \hat{y}^{ss}_{1} &  0 &  0 \\
 0 &  0 &  \hat{y}^{ss}_{2} \\
 0 &  \hat{y}^{ss}_{2} &  0 \end{array}\right)}\frac{v_{\tilde{\Psi}}}{\sqrt{2}}\left(\frac{v_{\Psi}}{\sqrt{2}\Lambda}\right)^{51}\frac{v_\Theta}{\sqrt{2}\Lambda}\,,\label{YS1}\\
 m_{D}&=&{\left(\begin{array}{ccc}
 \hat{y}^{\nu}_{1} &  0 &  0 \\
 0 &  0 &  \hat{y}^{\nu}_{2}   \\
 0 &  \hat{y}^{\nu}_{3}  &  0
 \end{array}\right)}\frac{v_{T}}{\sqrt{2}\Lambda}\left(\frac{v_{\tilde{\Psi}}}{\sqrt{2}\Lambda}\right)^{9}v_{u}
 =\hat{y}^{\nu}_{1}{\left(\begin{array}{ccc}
 1 &  0 &  0 \\
 0 &  0 &  y_{2}   \\
 0 &  y_{3}  &  0
 \end{array}\right)}\frac{v_{T}}{\sqrt{2}\Lambda}\left(\frac{v_{\tilde{\Psi}}}{\sqrt{2}\Lambda}\right)^{9}v_{u}, \label{Ynu1}\\
% \end{eqnarray}
% \begin{eqnarray}
 M_{R}&=&{\left(\begin{array}{ccc}
 1+\frac{2}{3}\tilde{\kappa}\,e^{i\phi} &  -\frac{1}{3}\tilde{\kappa}\,e^{i\phi} &  -\frac{1}{3}\tilde{\kappa}\,e^{i\phi} \\
 -\frac{1}{3}\tilde{\kappa}\,e^{i\phi} &  \frac{2}{3}\tilde{\kappa}\,e^{i\phi} &  1-\frac{1}{3}\tilde{\kappa}\,e^{i\phi}\\
 -\frac{1}{3}\tilde{\kappa}\,e^{i\phi} &  1-\frac{1}{3}\tilde{\kappa}\,e^{i\phi} &  \frac{2}{3}\tilde{\kappa}\,e^{i\phi}
 \end{array}\right)}M~,
 \label{MR1}
 \end{eqnarray}
where
 \begin{eqnarray}
 y_{2}\equiv\frac{\hat{y}^{\nu}_{2}}{\hat{y}^{\nu}_{1}}~,\quad y_{3}\equiv\frac{\hat{y}^{\nu}_{3}}{\hat{y}^{\nu}_{1}}~,\quad\tilde{\kappa}\equiv\sqrt{\frac{3}{2}}\left|\hat{y}_{R}\frac{v_{S}}{M}\right|~,\quad\phi\equiv\arg\left(\frac{\hat{y}_{R}}{\hat{y}_{\Theta}}\right)~\,\text{with}~M\equiv \left|\hat{y}_{\Theta}\,\frac{v_{\Theta}}{\sqrt{2}}\right|~.
 \label{MR2}
 \end{eqnarray}
Recalling that the hat Yukawa couplings in Eqs.\,(\ref{ChL1}-\ref{MR2}) are all of order unity and complex numbers. We will discuss the neutrino physics in detail in Sec.\,\ref{low_nut}. Now, we move to discussion on the charged-fermion sector, in which the physical mass hierarchies are directly responsible for the assignment of $U(1)_X$ quantum numbers.
The axion coupling matrices to the up-type quarks, charged leptons, and down-type quarks, respectively, are diagonalized through biunitary transformations : $V^{\psi}_R{\cal M}_{\psi}V^{\psi\dag}_L=\hat{{\cal M}}_\psi$ (diagonal form), and the mass eigenstates $\psi'_R=V^\psi_R\,\psi_R$ and $\psi'_L=V^\psi_L\,\psi_L$. These transformation include, in particular, the chiral transformation necessary to make ${\cal M}_u$ and ${\cal M}_d$ real and positive. This induces a contribution to the QCD vacuum angle as in Eq.\,(\ref{QCDlag1}). Note here that under the chiral rotation of the quark field given by Eq.\,(\ref{chiralR}) the effective QCD angle $\vartheta_{\rm eff}$ is invariant.
In the above Lagrangian\,(\ref{AxionLag2}) the mass matrices ${\cal M}_{u}, {\cal M}_{d}$ and ${\cal M}_{\ell}$ for up-, down-type quarks and charged leptons, respectively, are expressed as
 \begin{eqnarray}
 {\cal M}_{u}&=& {\left(\begin{array}{ccc}
 y_{u} \,e^{-6i\frac{A_{2}}{v_{g}}} &  0 &  0 \\
 0 &  y_{c}\,e^{-2i\frac{A_{2}}{v_{g}}} &  0   \\
 0 &  0  &  y_{t}
 \end{array}\right)}v_{u}\,,  \label{ChL1}\\
 {\cal M}_{d}&=&{\left(\begin{array}{ccc}
 \tilde{y}_{d} &  y_{s} &  y_{b} \\
 y_{d}  &  \tilde{y}_{s} &  y_{b}   \\
 y_{d}  &  y_{s}  & \tilde{y}_{b}
 \end{array}\right)}{\left(\begin{array}{ccc}
 e^{-3i\frac{A_{2}}{v_{g}}} &  0 &  0 \\
 0 &  e^{-2i\frac{A_{2}}{v_{g}}} &  0   \\
 0 &  0  &  1
 \end{array}\right)}\frac{v_{S}}{\sqrt{2}\Lambda}v_{d}\,,
 \label{ChL2}\\
 {\cal M}_{\ell}&=& {\left(\begin{array}{ccc}
 y_{e}\,e^{6i\frac{A_{2}}{v_{g}}} & 0 &  0 \\
 0 & y_{\mu}\,e^{4i\frac{A_{2}}{v_{g}}} & 0 \\
 0 & 0 & y_{\tau}\,e^{2i\frac{A_{2}}{v_{g}}}
 \end{array}\right)}v_{d}\,,
 \label{ChL3}
 \end{eqnarray}
where $v_{d}\equiv\langle H_{d}\rangle=v\cos\beta/\sqrt{2}$, and $v_{u}\equiv\langle H_{u}\rangle =v\sin\beta/\sqrt{2}$ with $v\simeq246$ GeV, $\tilde{y}_{f}=y_{f}+x_{f}\,\frac{1}{\kappa}\frac{v_{T}}{\Lambda}$ with $f=d,s,b$ (in which the Yukawa couplings $x_f$ come from higher-dimensional operators driven by the flavon field $\Phi_T$ in Ref.\,\cite{Ahn:2014gva}), and the corresponding Yukawa terms for charged leptons and up-type quarks are given by
 \begin{eqnarray}
 y_{\tau}&=&\Big(\frac{\lambda}{\sqrt{2}}\Big)^2\,\hat{y}_{\tau}\,,\quad
 y_{\mu}=\Big(\frac{\lambda}{\sqrt{2}}\Big)^4\,\hat{y}_{\mu}\,,\qquad
 y_{e}=\Big(\frac{\lambda}{\sqrt{2}}\Big)^6\,\hat{y}_{e}\,,\nonumber\\
 y_{t}&=& \hat{y}_{t}\,,\qquad\qquad~~
 y_{c}=\Big(\frac{\lambda}{\sqrt{2}}\Big)^2\,\hat{y}_{c}\,,\qquad
 y_{u}=\Big(\frac{\lambda}{\sqrt{2}}\Big)^6\,\hat{y}_{u}\,.
 \label{Top1}
 \end{eqnarray}
The physical structure of the charged-fermion Lagrangian given by Eqs.\,(\ref{AxionLag14}) and (\ref{AxionLag15}) may be examined, and these results are in a good agreement with the empirical charged lepton and up-type quarks mass ratios calculated from the measured values\,\cite{PDG}:
 \begin{eqnarray}
 \frac{m_{e}}{m_{\tau}}&\simeq&2.9\times10^{-4}\,,\qquad\quad \frac{m_{\mu}}{m_{\tau}}\simeq5.9\times10^{-2}\,.
 \label{MlRatio}\\
  \frac{m_{u}}{m_{t}} &\simeq& 1.4\times 10^{-5}\,,\qquad\quad
  \frac{m_{c}}{m_{t}}\simeq7.4\times10^{-3}\,.
\label{MuRatio}
 \end{eqnarray}
On the other hand,  ${\cal M}_{d}$ in Eq.\,(\ref{ChL2}) generates the down-type quark masses :
 \begin{eqnarray}
 \widehat{\mathcal{M}}_{d}=V^{d\dag}_{R}\,{\cal M}_{d}\,V^{d}_{L}
 ={\rm diag}(m_{d},m_{s},m_{b})\,,
 \label{Quark21}
 \end{eqnarray}
where $V^{d}_{L}$ and $V^{d}_{R}$ can be determined by diagonalizing the matrices for ${\cal M}^{\dag}_{d}{\cal M}_{d}$ and ${\cal M}_{d}{\cal M}^{\dag}_{d}$, respectively. Especially, the mixing matrix $V^{d}_{L}$ becomes one of the matrices composing the CKM mixing matrix.
The Hermitian matrix ${\cal M}^\dag_{d}{\cal M}_{d}$ is diagonalized by the mixing matrix $V^{d}_{L}$ :
 \begin{eqnarray}
 V^{d\dag}_{L}{\cal M}^\dag_{d}{\cal M}_{d}V^{d}_{L}&=&
 v^2_{d}\,3\left(\frac{v_{S}}{\sqrt{2}\Lambda}\right)^2\,V^{d\dag}_{L}{\left(\begin{array}{ccc}
 (\frac{\lambda}{\sqrt{2}})^{6}|\hat{y}_{d}|^2 & (\frac{\lambda}{\sqrt{2}})^5\hat{y}^{\ast}_{d}\hat{y}_{s} & (\frac{\lambda}{\sqrt{2}})^3\hat{y}^{\ast}_{d}\hat{y}_{b}  \\
 (\frac{\lambda}{\sqrt{2}})^5\hat{y}_{d}\hat{y}^{\ast}_{s} & (\frac{\lambda}{\sqrt{2}})^4|\hat{y}_{s}|^2 & (\frac{\lambda}{\sqrt{2}})^2\hat{y}^{\ast}_{s}\hat{y}_{b}   \\
 (\frac{\lambda}{\sqrt{2}})^3\hat{y}_{d}\hat{y}^{\ast}_{b} & (\frac{\lambda}{\sqrt{2}})^2\hat{y}_{s}\hat{y}^{\ast}_{b}  & |\hat{y}_{b}|^2
 \end{array}\right)}V^{d}_{L}\nonumber\\
 &=&{\rm diag}(|m_{d}|^{2}, |m_s|^{2}, |m_{b}|^{2})\,.
 \label{MDMD0}
 \end{eqnarray}
Due to the strong hierarchal structure of the Hermitian matrix, one can fit the results calculated from the measured values\,\cite{PDG} :
 \begin{eqnarray}
  \frac{m_{d}}{m_{b}} &\simeq&  1.2\times 10^{-3}\,,\qquad
  \frac{m_{s}}{m_{b}} \simeq 2.4\times10^{-2}\,.
 \label{massRatio}
 \end{eqnarray}
Naively speaking, since the leading matrix ${\cal M}_d$ has 6 physical parameters, while observables are seven (CKM parameters: 4, down-type quark masses: 3), its alone may not generate the correct CKM matrix in the standard parameterization in Ref.\,\cite{PDG}.
Therefore, in order to achieve the correct CKM mixing matrix, we should include non-trivial next leading order corrections which are driven by the field $\Phi_T$ neutral under $SU(2)_L\times U(1)_Y\times U(1)_X$, see more details in Ref.\,\cite{Ahn:2014gva}.

%%%%%%%%%%%%%%%%%%%%%%%%%%%%%%%%%%%%%%%%%%%%%%%%%%%%%%%%%%%%%%%%%%%%%%%%%%%%%
\subsection{The mixed  $U(1)_X$ anomalies and a bridge between string theory and flavor physics}
\label{qla1}
It is well known that any discrete or continuous global symmetry is not protected from violations by quantum gravity effects\,\cite{Krauss:1988zc}. Here we discuss that the global $U(1)_X$ is the remnant of the broken $U(1)_X$ gauge symmetry by the GS mechanism, and so it can be protected from quantum-gravitational effects, similar to Ref.\,\cite{Ahn:2016typ}.
String theory when compactified to four dimensions generically contains an anomalous $U(1)$ with anomaly cancellation. The model group $SU(3)_C\times SU(2)_L\times U(1)_Y\times U(1)_R\times U(1)_X$ we are interested may be realized in a four-stack model $U(3)\times U(2)\times U(1)\times U(1)$ on D-branes where the gauged $U(1)$s are generically anomalous\,\cite{string_book}. Hypercharge $U(1)_Y$ is the unique anomaly-free linear combination of the four $U(1)$s. The other combinations contribute to $U(1)_{X}$ and a gauged $U(1)_R$\,\cite{Villadoro:2005yq} which contains an $R$-symmetry as a subgroup. In addition, a non-Abelian discrete flavor symmetry, $A_4$, has been introduced to describe flavor mixing pattern, which can be realized in field theories on orbifolds\,\cite{Altarelli:2006kg}.
Here if we assume that the non-Abelian discrete symmetry $A_4$ is a subgroup of a gauge symmetry, it can be protected from quantum-gravitational effects. Moreover, in the model since such non-Abelian discrete symmetry is broken completely by higher order effects, there is no residual symmetry; so there is no room for a spontaneously broken discrete symmetry to give rise to domain-wall problem. 

We assume throughout that the model can be derived as consistent type IIB string vacuum. In such a vacuum, as will be shown later, the $U(1)_X$-mixed anomalies such as $U(1)_X[U(1)_Y]^2$, $U(1)_X[SU(2)_L]^2$, $U(1)_X[SU(3)_C]^2$, and $U(1)_Y[U(1)_X]^2$ should be cancelled by appropriate shifts of Ramond-Ramond axions in the bulk\,\cite{Sagnotti:1992qw}. 
On the other hand, non-perturbative quantum gravitational anomaly effects\,\cite{Kamionkowski:1992mf, Dvali:2005an} lead to a non-conservation of the corresponding current, 
 \begin{eqnarray}
 \partial_\mu J^\mu_X\propto R\tilde{R}
 \label{gravi}
 \end{eqnarray}
where $R$ is the Riemann tensor and $\tilde{R}$ is its dual, which spoils the axion solution to the strong CP problem.
Therefore, in order to eliminate the breaking effects of the axionic shift symmetry by gravity we impose an $U(1)_X$-$[{gravity}]^2$ anomaly cancellation condition.
Since the $U(1)_X$ charges in TABLE\,\ref{reps} are flavor-dependent, the $U(1)_X$ symmetry serves as a natural flavor symmetry, and helps explanation of the pattern of quark and lepton mixings as seen in Sec.-\ref{qla}.
Thus the choices of $U(1)_X$ charges for ordinary quarks and leptons are strongly restricted by the $U(1)_X$-$[{gravity}]^2$ anomaly cancellation condition:
 \begin{eqnarray}
  0&=&\{3\cdot2(-5q-3r)+3(6p+3r)+3(-3q+3r)\}_{\rm quark}\nonumber\\
  &&+\{-2(27q+3p)+(3p+39q)+3p+3p+75q\}_{\rm lepton}\,.
  \label{anomal}
 \end{eqnarray}
This indicates that the $U(1)_X$ symmetry could be interpreted as a fundamental symmetry of nature when $p=-q$. Clearly, the $U(1)_X$ quantum numbers of quark flavors in Eq.\,(\ref{anomal}) are arranged in a way that no axionic domain-wall problem occurs, which plays a crucial role in cosmology when the $X$-symmetry breaking occurs after inflation. 
With the quantum numbers of SM quarks and charged leptons satisfying the observed mass spectra based on the framework of $A_4\times U(1)_X$, if the SM quark quantum numbers are arranged in a way that no domain-wall problem occurs, one can find an available arrangement of quantum numbers to satisfy the neutrino phenomenology (see Sec.\,\ref{low_nut}).

We work in a supergravity framework based on type IIB string theory, and assume that the dilaton and complex structure moduli are fixed at semi-classical level by turning on background fluxes\,\cite{Gukov:1999ya}. 
Below the scale where the complex structure and the axio-dilaton moduli are stabilized through fluxes as in Refs.\,\cite{Giddings:2001yu,Dasgupta:1999ss}, the low-energy Kahler potential $K$ and superpotential $W$ for the Kahler moduli and matter superfields, invariant under $U(1)_X$ gauged symmetry, are given in type IIB string theory by\,\cite{Ahn:2016typ}
 \begin{eqnarray}
  K&=&-M^2_P\ln\big\{(T+\bar{T})\prod^2_{i=1}\big(T_i+\bar{T}_i-\frac{\delta^{\rm GS}_i}{16\pi^2}V_{X_i}\big)\big\}+\tilde{K}+...\label{Kahler0}\\
  &&\qquad\text{with}\quad\tilde{K}=\sum^2_{i=1}Z_i\Phi^\dag_i e^{-X_iV_{X_i}}\Phi_i+\sum_k Z_k|\varphi_k|^2\,,\nonumber\\
  W&=& W_{Y}+W_v+W_0+W(T)\label{Kahler}\,,
 \end{eqnarray}
in which $\Phi_1=\{\Phi_S, \Theta,\tilde{\Theta}\}$, $\Phi_2=\{\Psi,\tilde{\Psi}\}$,  $\varphi_{i}=\{\Psi_0, \Phi^{T}_{0},\Phi_{T}\}$,  dots represent higher-order terms, and $M_{P}=(8\pi G)^{-1/2}=2.436\times10^{18}$ GeV is the reduced Planck mass with the Newton's gravitational constant $G$. $W_0$ stands for the constant value of the flux superpotential at its minimum. Since the Kahler moduli do not appear in the superpotential $W$ at leading order, they are not fixed by the fluxes. So a non-perturbative superpotential $W(T)$ is introduced to stabilize the Kahler moduli. Although $W(T)$ in Eq.\,(\ref{Kahler}) is absent at tree level, 
the source of this non-perturbative term could be either D3-brane instantons or gagino condensation from the non-Abelian gauge sector of the $N$-wrapped D7-branes\,\cite{Derendinger:1985kk}. 
The Kahler moduli in $K$ of Eq.\,(\ref{Kahler0}) control the overall size of the compact space, 
 \begin{eqnarray}
  T=\frac{\tau}{2}+i\theta,\qquad T_i=\frac{\tau_i}{2}+i\theta_i \quad\text{with}~i=1,2\,,
 \end{eqnarray}
where $\tau/2(\tau_i/2)$ are the size moduli of the internal manifold and $\theta(\theta_i)$ are the axionic parts.
As can be seen from the Kahler potential above, the relevant fields participating in the four-dimensional GS mechanism are the $U(1)_{X_i}$ charged chiral matter superfields $\Phi_i$, the vector superfields $V_{X_i}$ of the gauged $U(1)_{X_i}$ which is anomalous, and the Kahler moduli $T_i$. The matter superfields in $K$ consist of all the scalar fields $\Phi_i$ that are not moduli and do not have Planck sized VEVs, and the chiral matter fields $\varphi_k$ are neutral under the $U(1)_{X_i}$ symmetry. We take, for simplicity, the normalization factors $Z_i=Z_k=1$, and the holomorphic gauge kinetic function on the Kahler moduli in the 4-dimensional effective SUGRA
 \begin{eqnarray}
 T_i=\frac{1}{g^2_{X_i}}+i\frac{a_{T_i}}{8\pi^2}
 \end{eqnarray}
where $g_{X_i}$ are the four-dimensional gauge couplings of $U(1)_{X_i}$. Actually, gaugino masses require a nontrivial dependence of the holomorphic gauge kinetic function on the Kahler moduli. This dependence is generic in most of the models of ${\cal N}=1$ SUGRA derived from extended supergravity and string theory\,\cite{Ferrara:2011dz}. And vector multiplets $V_{X_i}$ in Eq.\,(\ref{Kahler0}) are the $U(1)_{X_i}$ gauge superfields including gauge bosons $A^{\mu}_i$. The GS parameter $\delta^{\rm GS}_i$ characterizes the coupling of the anomalous gauge boson to the axion.
The kinetic terms of the Kahler moduli and scalar sectors in the flat space limit of the 4 dimensional ${\cal N} = 1$ supergravity are expressed as
  \begin{eqnarray}
   {\cal L}_{\rm kinetic}=K_{T\bar{T}}\,\partial_\mu{T}\partial^{\mu}\bar{T}+K_{T_i\bar{T}_i}\,\partial_\mu{T}_i\partial^{\mu}\bar{T}_i+K_{\Phi_i\bar{\Phi}_i}\,\partial_\mu{\Phi}_i\partial^{\mu}{\Phi}^\dag_i\,.
  \label{kinetic}
 \end{eqnarray}
Here we set $K_{\Phi_i\bar{\Phi}_i}=1$ for canonically normalized scalar fields, as Eq.\,(\ref{AxionLag1}).
In addition to the superpotential in Eq.\,(\ref{Kahler}) the Kahler potential in Eq.\,(\ref{Kahler0}) deviates from the canonical form due to the contributions of non-renormalizable terms scaled by an ultra violate cutoff $M_P$, invariant under the both gauge and the flavor symmetries. Here the kinetic terms for the axionic and size moduli do not mix in perturbation theory, due to the axionic shift symmetry, where any non-perturbative violations are small enough to be irrelevant.

The theory is invariant under the $U(1)_X$ gauge transformation $V_{X_i}\rightarrow V_{X_i}+i(\Lambda_i-\bar{\Lambda}_i)$, together with the matter and Kahler moduli superfields transform as\,\cite{Ahn:2016typ}
 \begin{eqnarray}
  \Phi_i\rightarrow e^{iX_i\Lambda_i}\Phi_i\,,\quad T_i\rightarrow T_i+i\frac{\delta^{\rm GS}_i}{16\pi^2}\Lambda_i
 \label{transf}
 \end{eqnarray}
where $\Lambda(\bar{\Lambda}_i)$ are (anti-)chiral superfields parameterizing $U(1)_{X_i}$ transformations on the superspace. Recalling that the scalar fields $\Phi_S$, $\Theta$ and $\Psi(\tilde{\Psi})$ have $X$-charges $X_1=-2p$ and $X_2=-q(q)$, respectively. So the axionic moduli $\theta_i$ and matter axions $A_i$ have shift symmetries
 \begin{eqnarray}
  \theta_i\rightarrow \theta_i-\frac{\delta^{\rm GS}_i}{16\pi^2}\xi_i\,,\quad A_i\rightarrow A_i+f_{a_i}\frac{\delta^{\rm GS}_i}{\delta^{\rm Q}_i}\xi_i
 \label{transf_axio}
 \end{eqnarray}
where the decay constants $f_{a_i}$ are defined in Eq.\,(\ref{AxioAhn1}), $\delta^Q_i$ is anomaly coefficient defined in Eq.\,(\ref{string_anomaly}), and $\xi_i=-{\rm Re}\Lambda_i|_{\theta=\bar{\theta}=0}$ and $\Phi_i|_{\theta=\bar{\theta}=0}=\frac{1}{\sqrt{2}}e^{i\frac{A_i}{v_i}}(v_i+h_i)$ (here $v_i$ and $h_i$ being the VEVs and Higgs bosons of scalar components, respectively, and the subscripts $\theta$ and $\bar{\theta}$ are the Grasmann variables.), with the gauge transformation 
 \begin{eqnarray}
  A^\mu_i\rightarrow A^\mu_i-\partial^\mu\xi_i\,.
 \label{transf_gauge}
 \end{eqnarray}

As discussed in Ref.\,\cite{Ahn:2016typ}, by introducing two gauged $U(1)$ symmetries author has stabilized the three size moduli $(\tau/2,\tau_i/2)$ and one axionic direction $\theta^{\rm st}$ with large masses, while the two axionic directions ($\theta^{\rm st}_1\equiv\theta-\theta_1$ and $\theta^{\rm st}_2\equiv\theta-\theta_2$) remain massless. The two massless axion directions are gauged by the $U(1)$ gauge interactions associated with D-branes, and the gauged flat directions of the $F$-term potential are removed through the Stuckelberg mechanism. Now we discuss how the corresponding massless NG modes could survive in the phase of scalars charged under the global continuous symmetry $U(1)_X$, as shown in Ref.\,\cite{Ahn:2016typ}.
Since we have two gauged anomalous $U(1)$ currents, there are two axions linear combinations of $A_i$ and $\theta^{\rm st}_i$ ($i=1,2$) that couple to the (non)-Abelian Chern-Pontryagin densities with coefficients by anomalies. And the two gauged anomalous $U(1)$ symmetries, $U(1)_{X_1}\times U(1)_{X_2}$, have the corresponding coefficients 
  \begin{eqnarray}
  \delta^{G}_i=2\,{\rm Tr}[X_iT^2_{SU(3)}]\,,\qquad \delta^{W}_i=2\,{\rm Tr}[X_iT^2_{SU(2)}]\,,\qquad\delta^{H}_i=2\,{\rm Tr}[X_iY^2]\,,
 \label{string_anomaly}
 \end{eqnarray}
respectively, which stand for the coefficients of the mixed $U(1)_{X_i}$-$[SU(3)_C]^2$, $U(1)_{X_i}$-$[SU(2)_L]^2$, and $U(1)_{X_i}$-$[U(1)_Y]^2$ anomalies which are cancelled by the GS mechanism. Here $U(n)$ generators $(n>2)$ are normalized according to ${\rm Tr}[T^aT^b]=\delta^{ab}/2$, and for convenience $\delta^{H}_i=2\,{\rm Tr}[X_iY^2]$ is defined for hypercharge. Then the anomaly generated by the triangle graph is cancelled by diagram in which the gauged anomalous $U(1)_X$ mixes with the axionic moduli, which in turn couples to  the Chern-Pontryagin density ${\rm Tr}(Q^{\mu\nu}\tilde{Q}_{\mu\nu})$ for the corresponding gauge group in the compactification. And so the axion decay constant depends on the Kahler metric, and in particular on where the moduli are stabilized, as shown in Ref.\,\cite{Ahn:2016typ}. Consider the four-dimensional effective action of the axions, $\theta^{\rm st}_i$ and $A_i$, and their corresponding gauge fields, $A^\mu_i$, which contains the following
%\begin{widetext}
  \begin{eqnarray}
  &&K_{T_i\bar{T}_i}\Big(\partial^{\mu}\theta^{\rm st}_i-\frac{\delta^{\rm GS}_i}{16\pi^2}A^{\mu}_i\Big)^2-\frac{1}{4g^2_{X_i}}F^{\mu\nu}_iF_{i\mu\nu}-g_{X_i}\xi^{\rm FI}_iD_{X_i}+|D_\mu\Phi_i|^2\nonumber\\
  &&+\theta^{\rm st}_i{\rm Tr}(Q^{\mu\nu}\tilde{Q}_{\mu\nu})+\frac{A_i}{f_{a_i}}\frac{\delta^{\rm Q}_i}{16\pi^2}{\rm Tr}(Q^{\mu\nu}\tilde{Q}_{\mu\nu})\,,
 \label{string_axion}
 \end{eqnarray}
 %\end{widetext}
where $\tilde{Q}\equiv\frac{1}{2}\epsilon^{\mu\nu\rho\sigma}Q_{\rho\sigma}$ with the gauge field strengths $Q=\{G, W, H\}$ for $SU(3)_C$, $SU(2)_L$, and $U(1)_Y$, respectively. $F^{\mu\nu}_i$ are the $U(1)_{X_i}$ gauge field strengths $F^{\mu\nu}_i=\partial^\mu A^\nu_i-\partial^\nu A^\mu_i$, and the $SU(3)_C$, $SU(2)_L$, $U(1)_Y$ gauge couplings are absorbed into their corresponding gauge field strengths.
In $|D_\mu\Phi_i|^2$ the scalar fields $\Phi_i$ couple to the $U(1)_{X_i}$ gauge bosons, where the gauge couplings $g_{X_i}$ are absorbed into the gauge bosons $A^{\mu}_i$ in the $U(1)_X$ gauge covariant derivative $D^\mu\equiv\partial^\mu+iX_iA^\mu_i$. As mentioned before, the introduction of FI terms leads to the D-term potentials in Eq.\,(\ref{grobal_V}) where the FI factors $\xi^{\rm FI}_i$ depend on the closed string moduli $\tau_i/2$. The first, third and fourth terms of Eq.\,(\ref{string_axion}) stem from expanding the Kahler potential of Eq.\,(\ref{Kahler0}).
Under the anomalous $U(1)_X$ gauge transformation in Eqs.\,(\ref{transf}) and (\ref{transf_axio}), the first and fifth terms together, and similarly the fourth and sixth terms in Eq.\,(\ref{string_axion}), are gauge invariant, that is, the interaction Lagrangians 
  \begin{eqnarray}
  {\cal L}^{\rm int}_{A_i\,\theta^{\rm st}_i}&=&-A^{\mu}_iJ^{X_i}_\mu+\frac{A_i}{f_{a_i}}\frac{\delta^{\rm Q}_i}{16\pi^2}\,{\rm Tr}(Q^{\mu\nu}\tilde{Q}_{\mu\nu})-A^{\mu}_iJ^{\theta_i}_\mu+\theta^{\rm st}_i\,{\rm Tr}(Q^{\mu\nu}\tilde{Q}_{\mu\nu})\,,
 \label{}
 \end{eqnarray}
are invariant. There are anomalous currents $J^{X_i}_\mu$ and $J^{\theta_i}_\mu$ coupling to the gauge bosons $A^\mu_{i}$, that is, $\partial_\mu J^{\mu}_{X_i}=\frac{\delta^{\rm GS}_i}{16\pi^2}\,{\rm Tr}(Q^{\mu\nu}\tilde{Q}_{\mu\nu})=-\partial_\mu J^{\mu}_{\theta_i}$:
 \begin{eqnarray}
  J^{\theta_i}_\mu=K_{T_i\bar{T}_i}\frac{\delta^{\rm GS}_i}{8\pi^2}\partial_\mu\theta^{\rm st}_i\,,\qquad J^{X_i}_\mu=-iX_i{\Phi}^\dag_i\overleftrightarrow{\partial_\mu}{\Phi}_i\,,
 \label{X_current}
 \end{eqnarray}
 leading to $\partial_\mu (J^{\mu}_{\theta_i}+J^{\mu}_{X_i})=0$.
 Expanding Lagrangian (\ref{string_axion}) and using $\theta^{\rm st}_{i}=a_{T_i}/(8\pi^2\,f^{\rm st}_i)$ with $f^{\rm st}_i=\sqrt{2K_{T_i\bar{T}_i}/(8\pi^2)^2}$ it reads
  \begin{eqnarray}
  &&\frac{1}{2}\left(\partial^{\mu}a_{T_i}\right)^2+\frac{a_{T_i}}{f^{\rm st}_i}\frac{1}{8\pi^2}{\rm Tr}(Q^{\mu\nu}\tilde{Q}_{\mu\nu})+\frac{1}{2}\left(\partial^{\mu}A_i\right)^2+\frac{A_i}{f_{a_i}}\frac{\delta^{\rm Q}_i}{16\pi^2}{\rm Tr}(Q^{\mu\nu}\tilde{Q}_{\mu\nu})\nonumber\\
  &&-J^{X_i}_\mu A^\mu_i-J^{\theta_i}_\mu A^\mu_i+\frac{1}{2g^2_{X_i}}m^2_{X_i}A^\mu_iA_{i\mu}-\frac{1}{4g^2_{X_i}}F^{\mu\nu}_iF_{i\mu\nu}-\frac{g^2_{X_i}}{2}\Big(\xi^{\rm FI}_i-\sum_i X_i|\Phi_i|^2\Big)^2
 \label{string_axion1}
 \end{eqnarray}
 where $a_{T_i}$ is the canonically normalized Kahler axions. Clearly it indicates that the values of $f^{\rm st}_i$ depend on the Kahler metric and on where the moduli are stabilized. And the gauge boson masses obtained by the super-Higgs mechanism are given by
  \begin{eqnarray}
  m_{X_i}=g_{X_i}\sqrt{2K_{T_i\bar{T}_i}\left(\frac{\delta^{\rm GS}_i}{16\pi^2}\right)^2+2f^2_{a_i}} \,,
 \label{mX_mass}
 \end{eqnarray}
Then the open string axions $A_i$ are linearly mixed with the closed string axions $\tilde{a}_{T_i}$ with decay constants $f^{\rm st}_i$ and $f_{a_i}$
 \begin{eqnarray}
  \tilde{A}_i=\frac{A_i\frac{\delta^{\rm GS}_i}{2}f^{\rm st}_i-a_{T_i}\,f_{a_i}}{\sqrt{f^2_{a_i}+(\frac{\delta^{\rm GS}_i}{2}f^{\rm st}_i)^2}}\,,\qquad
  G_i=\frac{a_{T_i}\frac{\delta^{\rm GS}_i}{2}f^{\rm st}_i +A_i\,f_{a_i}}{\sqrt{f^2_{a_i}+(\frac{\delta^{\rm GS}_i}{2}f^{\rm st}_i )^2}}\,.
 \label{axion1}
 \end{eqnarray}
Since the $U(1)_X$ is gauged, two linear combinations $G_i$ of the $A_i$ and $a_{T_i}$ fields are eaten by the $U(1)_X$ gauge bosons and obtain string scale masses, while the other combinations $\tilde{A}_i$ survive to low energies and contribute to the QCD axion
 \begin{eqnarray}
  \tilde{A}_i\approx A_i\,.
 \label{axion_gauge}
 \end{eqnarray}
For $f^{\rm st}_i\gg f_{a_i}$, the axions $\tilde{A}_i$ as would-be QCD axion are approximated to $A_i$. Below the scale $m_{X_i}$ the gauge bosons decouple, leaving behind low-energy symmetries which are anomalous global $U(1)_{X_i}$ with the low energy effective Lagrangian
 \begin{eqnarray}
  {\cal L}\supset{\cal L}_{\rm SM}+\frac{1}{2}(\partial_\mu A_i)^2+\frac{\delta^{\rm Q}_i}{16\pi^2}\frac{A_i}{f_{a_i}}\,{\rm Tr}(Q^{\mu\nu}\tilde{Q}_{\mu\nu})+iX_i{\Phi}^\dag_i\overleftrightarrow{\partial_\mu}{\Phi}_i\frac{\partial^\mu A_i}{f_{a_i}}\frac{\delta^{\rm GS}_i}{\delta^{\rm Q}_i}\,.
 \label{la_Lag}
 \end{eqnarray}
The gauged $U(1)_{X_i}$ symmetries are broken, and only the SM gauge group remains.
Since the gauge fields $\xi_i$ are absorbed into gauge transformation, {\it i.e.}, into the longitudinal mode of the $U(1)_{X_i}$ gauge bosons to make them massive, the gauge fields $\xi_i$ in the axionic shift symmetry defined in Eq.\,(\ref{transf_axio}) become constant, $\xi$.
Under the axionic shift symmetry 
 \begin{eqnarray}
  \frac{A_i}{f_{a_i}}\rightarrow \frac{A_i}{f_{a_i}}+\frac{\delta^{\rm GS}_i}{\delta^{\rm Q}_i}\xi\,({\rm constant})\,,
 \label{tran0}
 \end{eqnarray}
 the operator $\frac{\delta^{\rm Qi}_i}{\delta^{\rm GS}_i}\,\partial_\mu J^{\mu}_{Qi}\frac{A_i}{f_{a_i}}$ (the third term in the right hand side in Eq.\,(\ref{la_Lag})) transforms
 \begin{eqnarray}
 \frac{\delta^{\rm Q}_i}{\delta^{\rm GS}_i}\,\partial_\mu J^{\mu}_{Qi}\frac{A_i}{f_{a_i}}\rightarrow\frac{\delta^{\rm Q}_i}{\delta^{\rm GS}_i}\,\partial_\mu J^{\mu}_{Qi}\frac{A_i+f_{a_i}\frac{\delta^{\rm GS}_i}{\delta^{\rm Q}_i}\xi}{f_{a_i}}=-\frac{\delta^{\rm Q}_i}{\delta^{\rm GS}_i}\,J^\mu_{Qi}\partial_\mu\frac{A_i}{f_{a_i}}+\xi\,\partial_\mu J^{\mu}_{Qi}\,.
 \label{tran}
 \end{eqnarray}
Since, in a $U(1)$ gauge theory, the resulting surface term in the action would vanish for finite energy configurations, the last term $\partial_\mu J^{\mu}_{Hi}$ in Eq.\,(\ref{tran}) does not lead to parity or time-reversal violation. And the coupling $\partial_\mu J^{\mu}_{Wi}$, the last term in Eq.\,(\ref{tran}) corresponding to the $SU(2)$ weak vacuum structure,  can be removed from the Lagrangian through a $B+L$ transformation {\it i.e.} $\partial_\mu(J^{\mu}_B+J^{\mu}_L)=\frac{N_f}{8\pi^2}{\rm Tr}(W_{\mu\nu}\tilde{W}^{\mu\nu})$ (where $J^{\mu}_B$ and $J^{\mu}_L$ are baryon-and lepton-number currents, and $N_f$ is the number of generations). Thus the last term in Eq.\,(\ref{tran}) $\xi\,\partial_\mu J^{\mu}_{Wi}$ and $\xi\,\partial_\mu J^{\mu}_{Hi}$ for the $SU(2)$ and $U(1)_Y$ gauge groups are not physical; it means the third term for $Q=\{W, H\}$ on the right hand side in Eq.\,(\ref{la_Lag}) are just axion-derivative couplings. Below the weak scale the third terms on the right side of Eq.\,(\ref{la_Lag}) for $Q=\{W,H\}$ merge to give the electromagnetic anomaly coefficient of $U(1)_{X}$-$[U(1)_{\rm EM}]^2$, see Eq.\,(\ref{EManomaly}). On the other hand, in the case of $SU(3)$ gauge group the Chern-Pontryagin density $\partial_\mu J^\mu_G$ has physical effects leading to CP violation due to the existence of instantonic configurations in the QCD Lagrangian. So the operator $G^{a\mu\nu}\tilde{G}^a_{\mu\nu}A_i/f_{a_i}$ is not invariant under the axionic shift symmetry. The fourth term in the right hand side in Eq.\,(\ref{la_Lag}) can be traded by Eq.\,(\ref{AxionCoupling}).
As will be discussed below Eq.\,(\ref{QCDtra}), one linear combination of the global $U(1)_{X_i}$ in Eq.\,(\ref{axion_gauge}) (see also Eq.\,(\ref{AxioAhn})) is broken explicitly by QCD instantons. 

A crucial property of the above GS anomaly cancellation mechanism is that the two $U(1)$ gauge bosons acquire masses leaving behind the corresponding global symmetries.
These global symmetries $U(1)_{X_i}$ are remain exact to all orders in type IIB string perturbations theory around the orientifold vacuum. On the other hand, we expects non-perturbative violation of global symmetries and consequently exponentially small in the string coupling, as long as the vacuum stays at the orientifold point.
This GS mechanism can be applied to show the cancellation of the other mixed $U(1)$ anomalies, such as $U(1)_Y$-$[U(1)_{X_1}]^2$, $U(1)_Y$-$[U(1)_{X_2}]^2$, and $U(1)_Y$-$U(1)_{X_1}$-$U(1)_{X_2}$, by including Chern-Simons terms in the effective Lagrangian. 
The anomalies coefficients of the mixed $U(1)_Y$-$[U(1)_{X_i}]^2$ and $U(1)_Y$-$U(1)_{X_i}$-$U(1)_{X_j}$ with $j\neq i=1,2$ are given by
  \begin{eqnarray}
  \delta^{X}_i=2\,{\rm Tr}[Y(X_i)^2]\,,\qquad \delta^{X}_{ij}=2\,{\rm Tr}[YX_iX_j]\,.
 \label{string_anomaly1}
 \end{eqnarray}
Actually, in order for the hypercharge gauge invariance of the SM not to be violated without giving mass to the hypercharge gauge field, these anomalies should be removed. Thus we include the following Chern-Simons terms to the effective action Eq.~(\ref{string_axion})
 \begin{eqnarray}
  &&\frac{\delta^X_i}{32\pi^2}A^{\mu}_iA^{\nu}_i\tilde{F}_{Y\mu\nu}+ \frac{\delta^X_{ij}}{32\pi^2}A^{\mu}_iA^{\nu}_j\tilde{F}_{Y\mu\nu}\nonumber\\
  &&-\frac{(\delta^X_i)^2}{\delta^{\rm GS}_i}\frac{A_i}{32\pi^2f_{a_i}}F^{\mu\nu}_i\tilde{F}_{Y\mu\nu}+\frac{1}{32\pi^2}\left(\frac{(\delta^X_{ij})^2}{\delta^{\rm GS}_j}\frac{A_j}{f_{a_j}}F^{\mu\nu}_i-\frac{(\delta^X_{ij})^2}{\delta^{\rm GS}_i}\frac{A_i}{f_{a_i}}F^{\mu\nu}_j\right)\tilde{F}_{Y\mu\nu}
 \label{string_axion2}
 \end{eqnarray}
with $j\neq i=1,2$, where $F_{Y\mu\nu}$ is the hyperchrge field strength and its dual $\tilde{F}_{Y\mu\nu}$.
Under the $U(1)_X$ gauge transformation in Eqs.\,(\ref{transf}) and (\ref{transf_axio}), the first and third terms together, and similarly the second and fourth terms in Eq.\,(\ref{string_axion2}), are gauge invariant.

(Hereafter, without loss of generality, at low energies we absorb $\delta^{\rm GS}_i$ into $\xi_i$ in Eq.\,(\ref{QCDtra}).)

%%%%%%%%%%%%%%%%%%%%%%%%%%%%%%%%%%%%%%%%%%%%%%%%%%%%%%%%%%%%%%%%%%%%%%
\section{QCD axion and Axions in astro-particle physics} 
 \label{astro_ax_nu}
The would-be axions play crucial role in evolution of stars and solving the strong CP problem, which will be discussed in detail here.
In Eq.\,(\ref{AxionLag}) the CP-violating term appearing in the QCD Lagrangian is expressed as
 \begin{eqnarray}
   {\cal L}_{\vartheta} &=&\frac{\vartheta_{\rm eff}}{32\pi^2}G^{a\mu\nu}\tilde{G}^{a}_{\mu\nu}
\label{QCDlag}
 \end{eqnarray}
where $-\pi\leq\vartheta_{\rm eff}\leq\pi$ is the effective $\vartheta$ parameter defined and the color gauge coupling is absorbed into the gauge field, in the basis where quark masses are real and positive, diagonal, and $\gamma_5$-free, as
 \begin{eqnarray}
  \vartheta_{\rm eff}=\vartheta+\arg\left\{\det(\mathcal{M}_{u})\det(\mathcal{M}_{d})\right\}\,.
\label{QCDlag1}
 \end{eqnarray}
Here the angle $\vartheta$ is given above the electroweak scale, which is the coefficient of $\vartheta\,G^{a\mu\nu}\tilde{G}^{a}_{\mu\nu}/32\pi^{2}$ where $G^a$ is the color field strength tensor and its dual $\tilde{G}^{a}_{\mu\nu}=\frac{1}{2}\varepsilon_{\mu\nu\rho\sigma}G^{a\mu\nu}$ (here $a$ is an $SU(3)$-adjoint index), coming from the strong interaction. And, the second term comes from a chiral transformation of weak interaction for diagonalization of the quark mass matrices by $\psi_q\rightarrow e^{-i\gamma_{5}\arg[\det m_{q}]/2}\psi_q$, directly indicating the CKM CP phase in Ref.\,\cite{PDG}, which is of order unity. However, experimental bounds on CP violation in strong interactions are very tight, the strongest ones coming from the limits on the electric dipole moment of the neutron $d_{n}<0.29\times10^{-25}~e$\,\cite{Beringer:1900zz} which implies $|\vartheta_{\rm eff}|<0.56\times10^{-10}$. $\vartheta_{\rm eff}$ should be very small to make a theory consistent with experimental bounds.
A huge cancellation between $\vartheta$ and $\arg\left\{\det(\mathcal{M}_{u})\det(\mathcal{M}_{d})\right\}$ suggests that there should be a physical process.

The model has two anomalous $U(1)$ symmetries, $U(1)_{X_1}\times U(1)_{X_2}$, with respective anomalies $\delta^{\rm G}_1$ and $\delta^{\rm G}_2$, both of which are the coefficients of the $U(1)_{X_k}$-$SU(3)_C$-$SU(3)_C$ anomaly, so there are two would-be axions $A_{1}$ and $A_{2}$, with the transformation of the phase fields
 \begin{eqnarray}
  A_{1}\rightarrow A_{1}+\frac{v_{\cal F}\,X_1}{\delta^{\rm G}_1}\,\xi_{1}\,,\qquad A_{2}\rightarrow A_{2}+\frac{v_{g}\,X_2}{\delta^{\rm G}_2}\,\xi_{2}\,,
\label{QCDtra}
 \end{eqnarray}
respectively\,\cite{Ahn:2014gva}.
Their charges $X_1$ and $X_2$ are linearly independent. And the color anomaly coefficients are obtained by letting $2\sum_{\psi_i}X_{k\psi_i}\,{\rm Tr}(t^at^b)=\delta^{\rm G}_k\delta^{ab}$, where the $t^a$ are the generators of the representation of $SU(3)$ to which $\psi$ belongs and the sum runs over all Dirac fermion $\psi$ with $X$-charge.
Since the two $U(1)$s are broken by two types of field attaining VEVs, a new PQ symmetry $U(1)_{\tilde{X}}$ which is a linear combination of the two $U(1)$s has anomaly, while another $U(1)$ is anomaly-free (it is the broken $U(1)_{f}$ symmetry by $\langle\Theta\rangle,\langle\Phi_S\rangle\neq0$ responsible for lepton number violation).
Under $U(1)_{\tilde{X}}\times U(1)_f$ the fields are transformed as
 \begin{eqnarray}
  &&{\cal F}_{1}=\frac{v_{\cal F}\,e^{i\frac{A_1}{v_{\cal F}}}}{\sqrt{2}}\left(1+\frac{h_{\cal F}}{v_{\cal F}}\right)\,;\qquad{\cal F}_{1}\rightarrow e^{iX_{1}\,\xi_1}{\cal F}_{1}\,,\quad\text{with}~\,\xi_1=\delta^{\rm G}_2\,\alpha\,,\nonumber\\
  &&{\cal F}_{2}=\frac{v_{g}\,e^{i\frac{A_2}{v_{g}}}}{\sqrt{2}}\left(1+\frac{h_{\Psi}}{v_{g}}\right)\,;\qquad {\cal F}_{2}\rightarrow e^{iX_{2}\,\xi_2}{\cal F}_{2}\,,\quad\text{with}~\,\xi_2=-\delta^{\rm G}_1\,\alpha\,.
  \label{Ahnaxion01}
 \end{eqnarray}
One linear combination of the phase fields $A_{1}$ and $A_{2}$ becomes the axion ($\equiv A$), and the other orthogonal combination corresponds to the Goldstone boson ($\equiv G$):
 \begin{eqnarray}
  {\left(\begin{array}{c}
 A \\
 G
 \end{array}\right)}={\left(\begin{array}{cc}
 \cos\vartheta &  \sin\vartheta \\
 -\sin\vartheta &  \cos\vartheta
 \end{array}\right)}{\left(\begin{array}{c}
 A_1 \\
 A_2
 \end{array}\right)}
 \end{eqnarray}
Here, the $G$ is the ``true" Goldstone boson\,\footnote{It could be a massless Majoron-like particle.} of the spontaneously broken $U(1)_{f}$. And since the Goldstone boson interactions arise only through the derivative couplings as Eq.\,(\ref{derivative}), we can have the nonlinearly realized global symmetry below the symmetry breaking scale, $U(1)_f:\qquad G\rightarrow G+\Upsilon \text{(constant)}$.
Then, the angle is obtained as $\cos\vartheta=-\frac{\tilde{X}_2\,v_{g}}{\sqrt{\left(\tilde{X}_1\,v_{\cal F}\right)^2+\left(-\tilde{X}_2\,v_{g}\right)^2}}$ and $\sin\vartheta=\frac{\tilde{X}_1\,v_{\cal F}}{\sqrt{\left(\tilde{X}_1\,v_{\cal F}\right)^2+\left(-\tilde{X}_2\,v_{g}\right)^2}}$ with $\tilde{X}_1\equiv \delta^{\rm G}_2\,X_1$ and $\tilde{X}_2\equiv-\delta^{\rm G}_1\,X_2$. Therefore,
the axion $A$ and the Goldstone boson $G$ can be expressed as
 \begin{eqnarray}
 A=\frac{A_1\,\delta^{\rm G}_1\,f_{a_2}+A_2\,\delta^{\rm G}_2\,f_{a_1}}{\sqrt{\left(\delta^{\rm G}_2\,f_{a_1}\right)^2+\left(\delta^{\rm G}_1\,f_{a_2}\right)^2}}\,,\qquad G=\frac{A_2\,\delta^{\rm G}_1\,f_{a_2}-A_1\,\delta^{\rm G}_2\,f_{a_1}}{\sqrt{\left(\delta^{\rm G}_2\,f_{a_1}\right)^2+\left(\delta^{\rm G}_1\,f_{a_2}\right)^2}}\,,
  \label{AxioAhn}
 \end{eqnarray}
where the decay constants are given by
 \begin{eqnarray}
 f_{a_1}=|X_1|\,v_{\cal F}\,,\qquad f_{a_2}=|X_2|\,v_{g}\,.
  \label{AxioAhn1}
 \end{eqnarray}
 
Meanwhile, the $X$-current for $U(1)_{\tilde{X}}$ with the condition\,(\ref{Ahnaxion01}) is given by
 \begin{eqnarray}
  J^{\tilde{X}}_{\mu}=i\tilde{X}_{1}{\cal F}^{\dag}_{1}\overleftrightarrow{\partial}_{\mu}{\cal F}_{1}-i\tilde{X}_{2}{\cal F}^{\dag}_{2}\overleftrightarrow{\partial}_{\mu}{\cal F}_{2}+\frac{1}{2}\sum_{\psi} \tilde{X}_{\psi}\bar{\psi}\gamma_{\mu}\gamma_{5}\psi\,\quad\text{with}\,\tilde{X}_{\psi}\equiv\tilde{X}_{1\psi}-\tilde{X}_{2\psi}
\label{AxialC}
 \end{eqnarray}
where $\psi=$ all $X$-charged Dirac fermions, which is conserved, $\partial^{\mu}J^{\tilde{X}}_{\mu}=0$, up to the triangle anomaly. This current creates a massless particle, the axion.
The $X$-current in Eq.\,(\ref{AxialC}) is now decoupled in the limit $v_{\cal F},v_{g}\rightarrow\infty$ as
 \begin{eqnarray}
  J^{\tilde{X}}_{\mu}&=&\tilde{X}_1\,v_{\cal F}\,\partial_{\mu}A_1+(-\tilde{X}_2\,v_{g})\,\partial_{\mu}A_2+\frac{1}{2}\sum_{\psi} \tilde{X}_{\psi}\bar{\psi}\gamma_{\mu}\gamma_{5}\psi\nonumber\\
  &=&\frac{\partial_{\mu}A}{\sqrt{\left(\frac{1}{2f_{a_1}\delta^{\rm G}_2}\right)^2+\left(\frac{1}{2f_{a_2}\delta^{\rm G}_1}\right)^2}}+\frac{1}{2}\sum_{\psi} \tilde{X}_{\psi}\bar{\psi}\gamma_{\mu}\gamma_{5}\psi\,,
  \label{Jx}
 \end{eqnarray}
which corresponds to the charge flow satisfying the current conservation equation if the symmetry is exact. Since the $J^{\tilde{X}}_{\mu}$ does not couple to the Goldstone boson $G$ in Eq.\,(\ref{AxioAhn}), requiring $J^{\tilde{X}}_{\mu}$ not to create $G$ from the vacuum $\langle0|J^{\tilde{X}}_{\mu}|G\rangle=0$, it follows
 \begin{eqnarray}
  \left(\tilde{X}_1\,v_{\cal F}\right)^2=\left(\tilde{X}_2\,v_{g}\right)^2\,.
  \label{AhnMass}
 \end{eqnarray}
This indicates that, if one of symmetry breaking scales is determined, the other one is automatically fixed.
The NG boson $A$ (which will be the QCD axion) possess the decay constant, $f_{A}$, defined by $\langle0|J^{\tilde{X}}_{\mu}(x)|A(p)\rangle=ip_{\mu}\,f_{A}\,e^{-ip\cdot x}$.
Then, from Eq.\,(\ref{Jx}) we obtain the spontaneous symmetry breaking scale
 \begin{eqnarray}
  f_{A}= \left\{\left(\frac{1}{2f_{a_1}\,\delta^{\rm G}_2}\right)^2+\left(\frac{1}{2f_{a_2}\,\delta^{\rm G}_1}\right)^2\right\}^{-\frac{1}{2}}\,,
  \label{fA1}
 \end{eqnarray}
which is more reduced to $f_{A}= \sqrt{2}\,\delta^{\rm G}_2\,f_{a_1}=\sqrt{2}\,\delta^{\rm G}_1\,f_{a_2}$ by using Eq.\,(\ref{AhnMass}).
Under the $U(1)_{\tilde{X}}$ transformation, the axion field $A$ translates with the axion decay constant $F_{A}$
 \begin{eqnarray}
  A\rightarrow A+F_{A}\,\alpha\,\qquad\text{with}\,\,F_{A}\equiv f_{A}/N\quad\text{and}\quad N=2\delta^{\rm G}_1\delta^{\rm G}_2
   \label{scale_relation}
 \end{eqnarray}
where $\alpha\equiv\sum_i\alpha_{i}$. Note here that if $N$ were large, then $F_A$ can be lowered significantly compared to the symmetry breaking scale.

However, the current $J^{\tilde{X}}_{\mu}$ is anomalous, that is, it is violated at one loop by the triangle anomaly\,\cite{anomaly, Ahn:2014gva} 
 \begin{eqnarray}
  \partial^{\mu}J^{\tilde{X}}_{\mu}=\frac{N}{16\pi^2}{\rm Tr}(G_{\mu\nu}\tilde{G}^{\mu\nu})\,.
 \end{eqnarray}
Then, after chiral rotation as in Eq.\,(\ref{chiralR}) the corresponding Lagrangian has the form
 \begin{eqnarray}
  {\cal L}_{\rm eff}\supset\frac{1}{32\pi^2}\left(\vartheta_{\rm eff}+\frac{A_1}{f_{a1}}\delta^{\rm G}_1+\frac{A_2}{f_{a2}}\delta^{\rm G}_2\right)G^{a}_{\mu\nu}\tilde{G}^{a\mu\nu}=\frac{1}{32\pi^2}\left(\vartheta_{\rm eff}+\frac{A}{F_{A}}\right)G^{a}_{\mu\nu}\tilde{G}^{a\mu\nu}\,.
 \end{eqnarray}
Since $\vartheta_{\rm eff}$ is an angle of mod $2\pi$, after chiral rotations on Dirac fermion charged under $U(1)_{X_1}\times U(1)_{X_2}$, the Lagrangian should be invariant under
 \begin{eqnarray}
  \frac{A_1}{f_{a1}}\rightarrow\frac{A_1}{f_{a1}}+\frac{2\pi}{\delta^{\rm G}_1}n_1\,,\qquad\frac{A_2}{f_{a2}}\rightarrow\frac{A_2}{f_{a2}}+\frac{2\pi}{\delta^{\rm G}_2}n_2\,,
 \end{eqnarray}
where $n_{1,2}$ are non-negative integers. So,
it is clear to see the following by replacing $n_{i}$ with $N_{\rm DW}\delta^{\rm G}_i$: if $\delta^{\rm G}_{1}$ and $\delta^{\rm G}_{2}$ are relative prime (so, the domain wall number $N_{\rm DW}=1$), there can be no $Z_{N_{\rm DW}}$ discrete symmetry and therefore no axionic domain wall problem. Our model ($\delta^{\rm G}_1=3X_1=6$, $\delta^{\rm G}_2=-13X_2=13$ for $q=-p=1$ in Eq.\,(\ref{anomal})) corresponds to the case\,\footnote{Note that these color anomaly coefficients, $\delta^{\rm G}_1=6$, $\delta^{\rm G}_2=13$, coming from the mixed $U(1)_X$-$[gravity]^2$ anomaly-free condition are different from those ($\delta^{\rm G}_1=3$, $\delta^{\rm G}_2=17$) in Ref.\,\cite{Ahn:2014gva}.}.

%%%%%%%%%%%%%%%%%%%%%%%%%%%%%%%%
\subsection{Quarks and charged leptons, and their interactions with axions} 
In order to obtain the axion interactions with the SM fermions, let us remove the NG modes $A_1$ and $A_2$ from the mass matrices in Eq.\,(\ref{AxionLag2}) by chiral-rotation of the charged fermion and neutrino fields as in Eq.\,(\ref{chiralR}).
The Yukawa Lagrangian of the charged fermions in Eq.\,(\ref{AxionLag2}) have the $\tilde{X}$-symmetry with the transformation parameter $\rho$ under
 \begin{eqnarray}
 U(1)_{\tilde{X}}: &&u\rightarrow e^{-6i\tilde{X}_2\,\frac{\gamma_5}{2}\rho}\,u,\qquad\qquad c\rightarrow e^{-2i\tilde{X}_2\,\frac{\gamma_5}{2}\rho}\,c,\quad\qquad t=\text{invariant}\nonumber\\
 &&
d\rightarrow e^{i(\tilde{X}_1-3\tilde{X}_2)\,\frac{\gamma_5}{2}\rho}\,d,\qquad s\rightarrow e^{i(\tilde{X}_1-2\tilde{X}_2)\,\frac{\gamma_5}{2}\rho}\,s,\qquad b\rightarrow e^{i\tilde{X}_1\,\frac{\gamma_5}{2}\rho}\,b,\nonumber\\
  &&e\rightarrow  e^{6i\tilde{X}_2\,\frac{\gamma_5}{2}\rho}\,e,~\,\qquad\qquad \mu\rightarrow  e^{4i\tilde{X}_2\,\frac{\gamma_5}{2}\rho}\,\mu,\quad\qquad \tau\rightarrow  e^{2i\tilde{X}_2\,\frac{\gamma_5}{2}\rho}\,\tau\,.
  \label{chiralR0}
 \end{eqnarray}
After diagonalization of the mass matrices for charged fermions, between $1$ GeV and $246$ GeV the axion-charged fermion Lagrangian are expressed as
 \begin{eqnarray}
  -{\cal L}^{a-q} &\simeq&\frac{\partial_{\mu}A_1}{2f_{a1}}\left\{X_{1d}\,\bar{d}\gamma^{\mu}\gamma_{5}d+X_{1s}\,\bar{s}\gamma^{\mu}\gamma_{5}s+X_{1b}\,\bar{b}\gamma^{\mu}\gamma_{5}b\right\}\nonumber\\
  &+&\frac{\partial_{\mu}A_2}{2f_{a2}}\big\{X_{u}\,\bar{u}\gamma^{\mu}\gamma_{5}u
  +X_{c}\,\bar{c}\gamma^{\mu}\gamma_{5}c+X_{2d}\,\bar{d}\gamma^{\mu}\gamma_{5}d+X_{2s}\,\bar{s}\gamma^{\mu}\gamma_{5}s\big\}\nonumber\\
  &+&m_u\,\bar{u}u+m_c\,\bar{c}c+m_t\,\bar{t}t+m_d\,\bar{d}d+m_s\,\bar{s}s+m_b\,\bar{b}b-\bar{q}\,i\! \! \not\!\partial\,q,\label{AxionLag14}\\
  -{\cal L}^{a-\ell} &\simeq&\frac{\partial_{\mu}A_2}{2f_{a2}}\left\{X_{e}\,\bar{e}\,\gamma^{\mu}\gamma_{5}\,e+X_{\mu}\,\bar{\mu}\,\gamma^{\mu}\gamma_{5}\,\mu+X_{\tau}\,\bar{\tau}\,\gamma^{\mu}\gamma_{5}\,\tau\right\}\nonumber\\
  &+&m_e\,\bar{e}e+m_\mu\,\bar{\mu}\mu+m_\tau\,\bar{\tau}\tau-\bar{\ell}\,i\! \! \not\!\partial\,\ell,
  \label{AxionLag15}
 \end{eqnarray}
in which $q=u,c,t,d,s,b$ and $\ell=e,\mu,\tau$ represent mass eigenstates.  
And the derivative interactions can also be simplified, and in turn which can be expressed in terms of the hadronic axion $A$, see Eq.\,(\ref{AxionLag01}), as
 \begin{eqnarray}
  \sum_\psi \left(\frac{\partial_{\mu}A_1}{f_{a1}}X_{1\psi}+\frac{\partial_{\mu}A_2}{f_{a2}}X_{2\psi}\right)\bar{\psi}\gamma^{\mu}\gamma_{5}\psi
  =\frac{\partial_{\mu}A}{f_A}\sum_{\psi}\tilde{X}_{\psi}\bar{\psi}\gamma^{\mu}\gamma_{5}\psi\,.
  \label{AxionCoupling}
 \end{eqnarray}
The axion couplings are model dependent with the elements of the matrices, so the $X$-charges of the fermions are given as $X_u=-6X_2$, $X_c=-2X_2$, $X_e=6X_2$, $X_\mu=4X_2$,  $X_\tau=2X_2$, $X_{1d}=X_{1s}=X_{1b}=X_1$, $X_{2d}=-3X_2$ and $X_{2s}=-2X_2$. Recalling that $X_1=-2p$ and $X_2=-q$ with $p=-q$. 
The above axion-SM fermion interactions are applicable above $1$ GeV such as in $J/\Psi$ and $\Upsilon$ decays. It is clear that the hadronic axion, $A$, does not couple to charged-leptons at tree level, whereas the new NG bosons, $A_1$ and/or $A_2$, interact with both quarks and leptons\,\footnote{The $A$ as a linear combination of $A_1$ and $A_2$ could play a role as a QCD axion to give a natural solution to the strong CP problem, while $A_2$ alone does not. However, since the $A_2$ is an admixture of the QCD axion, its coupling also is controlled by the QCD axion quantities. In addition,  the $A_2$ coupling to electron is constrained by astrophysical constraints.}. Such couplings, however, are suppressed by factors $v/f_{a1}$ or $v/f_{a2}$. Consequently, both the hadronic axion and the new NG modes are invisible. Below the QCD scale (1 GeV$\approx4\pi f_{\pi}$), the axion-hadron interactions are meaningful rather than the axion-quark interactions in Eq.\,(\ref{AxionLag14}), see below Eq.\,(\ref{a_nucleon}): the chiral symmetry is broken and $\pi, K$ and $\eta$ are produced as pseudo-Goldstone bosons. Since the weakly coupled NG bosons and the hadronic axion could carry away a large amount of energy from the interior of stars, according to the standard stellar evolution scenario their couplings should be bounded with electrons (because second and third generation particles are absent in almost all astrophysical objects) and nucleons\,\footnote{Axion interaction with nucleon will be discussed in Sec.\,\ref{axion_int}}, respectively.

As seen in superpotential\,(\ref{lagrangian2}) since the SM charged lepton fields which are nontrivially $X$-charged Dirac fermions have $U(1)_{\rm EM}$ charges, the axion $A_2$ coupling to electrons are added to the Lagrangian through a chiral rotation, as shown in Eq.\,(\ref{AxionLag15}).
And the axion $A_2$ couples directly to electrons, thereby the axion can be emitted by Compton scattering, atomic axio-recombination and axio-deexcitation, and axio-bremsstrahlung in electron-ion or electron-electron collisions\,\cite{Redondo:2013wwa}. The axion $A_2$ coupling to electron in the model reads
 \begin{eqnarray}
 g_{Aee}&=& \frac{X_e\,m_e}{f_{a_2}}\,, 
 \label{axion-electron0}
 \end{eqnarray}
 where $m_e=0.511$ MeV, and without loss of generality we set $X_e=-6$  for the given $X$-charges $X_2=-q=-1$.
On top of the FA model, in the two conventional models the hadronic axion coupling to electron has a very small model-independent coupling induced at one-loop via photon coupling for KSVZ, and a model-dependent contribution proportional to an ${\cal O}(1)$ coefficient for DFSZ,
 \begin{eqnarray}
 g_{Aee}&=&\left\{
             \begin{array}{ll}
              \frac{m_e}{f_{\rm K}}\frac{3\alpha^2_{\rm em}}{4\pi}\left(\frac{E}{N}\log\frac{f_{\rm K}}{m_e}-\frac{2}{3}\,\frac{4+z+w}{1+z+w}\log\frac{\Lambda_{\rm QCD}}{m_e}\right), & \hbox{KSVZ} \\
               \frac{m_e}{f_{\rm D}}\tan\beta\,, & \hbox{DFSZ}
             \end{array}
           \right.
 \label{axion-electron}
 \end{eqnarray}
where $z=m_u/m_d$, $w=m_u/m_s$, $\tan\beta=v_u/v_d$, $f_{\rm K(D)}$ are their corresponding decay constants, the electromagnetic anomaly coefficient $E$ vanishes for KSVZ, and $\Lambda_{\rm QCD}$ is an energy scale close to the QCD confinement scale.
There are several restrictive astrophysical limits\,\cite{PDG} on the axion models that couples to electrons, which arise from the above mentioned processes : among them, (i) from stars in the red giant branch of the color-magnitude diagram of globular clusters\,\cite{Redondo:2013wwa}, $\alpha_{Aee}<1.5\times10^{-26}$ ($95\%$ CL)\,\cite{Viaux:2013lha}, (ii) from white dwarfs (WDs) where bremsstrahlung is mainly efficient\,\cite{Raffelt:1985nj},  $\alpha_{Aee}<6\times10^{-27}$\,\cite{Bertolami:2014wua}, and recently (iii) from the Sun the XENON100 experiment provides the upper bound, $g_{Aee}<7.7\times10^{-12}$ ($90\%$ CL)\,\cite{Aprile:2014eoa}.
Here a fine-structure constant, $\alpha_{Aee}=g^{2}_{Aee}/4\pi$, is related to the axion-electron coupling constant $g_{Aee}$.
Then, the astrophysical lower bound of the PQ breaking scale is derived from the above mentioned upper limits
 \begin{eqnarray}
  && f_{a_2}\gtrsim(3.98\times10^{8}-1.23\times10^{10})\,{\rm GeV}\,,\qquad\qquad\text{FA}\nonumber\\
  && f_{\rm K}\gtrsim(1.02\times10^{4}-3.15\times10^{5})\,{\rm GeV}\,,\,\,\qquad\qquad\text{KSVZ}\nonumber\\
  && f_{\rm D}\gtrsim(6.64\times10^7-2.04\times10^{9})\,\tan\beta\,{\rm GeV}\,,\qquad\text{DFSZ}\,.
 \label{axion-electron01}
 \end{eqnarray}
Such weakly coupled axions have a wealth of interesting phenomenological implications in the context of astrophysics, like the formation of a cosmic diffuse background of axions from core collapse supernova explosions\,\cite{Raffelt:2011ft} or neutron star cooling\,\cite{Keller:2012yr}. 
Indeed, the longstanding anomaly in the cooling of WDs might be explained by axions with $\alpha_{Aee}=(0.29-2.30)\times10^{-27}$\,\cite{WD01}, which is recently improved in Refs.\,\cite{Bertolami:2014wua, wd_recent2}, implying axion decay constants
 \begin{eqnarray}
4.1\times10^{-28}\lesssim\alpha_{Aee}\lesssim3.7\times10^{-27}\,\Leftrightarrow
\left\{
             \begin{array}{lll}
             f_{a_2}=(1.4-4.3)\times10^{10}\,{\rm GeV}\,, & \hbox{FA}\\
             f_{\rm K}=(3.7-11.0)\times10^{5}\,{\rm GeV}\,, & \hbox{KSVZ} \\
             f_{\rm D}=(2.4-7.1)\times10^{9}\,\tan\beta\,{\rm GeV}\,, & \hbox{DFSZ}
             \end{array}
           \right.
 \label{wd_bound}
 \end{eqnarray}
As will be seen later, with the lower bounds of decay constants in Eq.\,(\ref{gagg_const}) derived from the upper limits of the axion-photon couplings, the KSVZ model could be excluded by the anomaly Eq.\,(\ref{wd_bound}) with Eq.\,(\ref{a_nucleon01}).
In addition, it may indicate that direct searches for axions and calculations of their effects on the cooling of stars and on the supernova SN1987A\,\cite{Raffelt:2006cw} exclude most values of $f_{a_2}\lesssim10^{9}$ GeV. Note that here, if the constraint from WDs cooling as in Eq.\,(\ref{wd_bound}) is not considered, the prototype KSVZ model is allowed; in addition, the model (FA) prediction for axion decay constants could have a little bit wider ranges constrained by the extra cooling from the neutron star as seen in Eq.\,(\ref{a_nucleon02}).  

%%%%%%%%%%%%%%%%%%%%%%%%%%%%%%%%%%%%%%%%%%%%%%%%%%%%%%%%%%%%%%%%%%%%%%%%%%%%%%%%%%%%
\subsection{Strong CP problem and QCD axion}
\label{sCP}
Through a chiral rotation on $\psi$ as in Eq.\,(\ref{chiralR0}), we can dispose of the $\vartheta_{\rm eff}$ angle in Eq.\,(\ref{QCDlag}). Let us chiral-rotate the $f$-th $\psi$ in the Fujikawa measure of the path integral
 \begin{eqnarray}
  \psi_f\rightarrow \text{exp}\left(i\frac{\alpha_{f}\gamma_5}{2}\right)\psi_f\qquad\text{with}\,\,\alpha_{f}\equiv\rho \tilde{X}_{\psi_f}=\rho(\tilde{X}_{1\psi_f}-\tilde{X}_{2\psi_f})
\label{chiralR}
 \end{eqnarray}
on Dirac spinors, which contributes
 \begin{eqnarray}
  {\cal L}\rightarrow{\cal L}+\frac{g^2_s}{16\pi^2}\sum_{\psi_f}\rho \tilde{X}_{\psi_f} G^{a}_{\mu\nu}\tilde{G}^{b\mu\nu}\,{\rm Tr}(t^at^b)={\cal L}+\frac{g^2_s}{32\pi^2}\,\rho N\,G^{a}_{\mu\nu}\tilde{G}^{a\mu\nu}
 \label{angleT}
 \end{eqnarray}
to the Lagrangian, where the $N$ is the axion color anomaly of the $U(1)_{\tilde{X}}$ symmetry. (Here we resurrect the color gauge coupling $g_s$.) And the second term in the right hand side of Eq.\,(\ref{angleT}) is obtained by letting $2\sum_{\psi_f}\tilde{X}_{1\psi_f}\,{\rm Tr}(t^at^b)-2\sum_{\psi_f}\tilde{X}_{2\psi_f}\,{\rm Tr}(t^at^b)=N\delta^{ab}$, where the sum runs over all $\psi$ with $\tilde{X}$-charge.
Through a rotation Eq.\,(\ref{chiralR}), {\it i.e.} $\psi_f\rightarrow \text{exp}\{i\frac{\tilde{X}_{\psi}}{N}\frac{A}{F_{A}}\frac{\gamma_5}{2}\}\psi_f$, we obtain
the vanishing anomaly terms by adding the QCD vacuum given in Lagrangian\,(\ref{QCDlag}) to the above Lagrangian
 \begin{eqnarray}
   {\cal L}_{\vartheta} &=& \left(\vartheta_{\rm eff}+\frac{A_1}{F_{a_1}}+\frac{A_2}{F_{a_2}}\right)\frac{\alpha_s}{8\pi}G^{a\mu\nu}\tilde{G}^{a}_{\mu\nu}\equiv\left(\vartheta_{\rm eff}+\frac{A}{F_{A}}\right)\frac{\alpha_s}{8\pi}G^{\mu\nu a}\tilde{G}^{a}_{\mu\nu}\,.
 \label{angleT1}
 \end{eqnarray}
Here $F_{a_i}=f_{a_i}/\delta^{\rm G}_i$ with $i=1,2$.
At low energies $A$ will get a VEV, $\langle A\rangle=-F_{A}\vartheta_{\rm eff}$, eliminating the constant $\vartheta_{\rm eff}$ term. The axion then is the excitation of the $A$ field, $a=A-\langle A\rangle$.
Since the SM fields $\psi$ have $U(1)_{\rm EM}$ charges, the axion coupling to photon will be added to the Lagrangian through a rotation Eq.\,(\ref{chiralR}), which survive to the QCD scale:
 \begin{eqnarray}
  {\cal L}\rightarrow{\cal L}+e^{2}\frac{2\rho\sum_\psi \tilde{X}_{\psi}(Q^{\rm em}_i)^2}{32\pi^2}F_{\mu\nu}\tilde{F}^{\mu\nu}={\cal L}+\frac{e^{2}}{32\pi^2}\left(\frac{E}{N}\right)\frac{A}{F_{A}}F_{\mu\nu}\tilde{F}^{\mu\nu}
 \label{EManomaly}
 \end{eqnarray}
with the axion electromagnetic anomaly $E=2\sum_{\psi} \tilde{X}_{1\psi_f}(Q^{\rm em}_{f})^2-2\sum_{\psi} \tilde{X}_{2\psi_f}(Q^{\rm em}_{f})^2$ for here $\psi=$ all $\tilde{X}$-charged Dirac fermions, where $F_{\mu\nu}$ is the electromagnetic field strength and its dual $\tilde{F}^{\mu\nu}$. Note that since the field $A$ is not a constant, this term is not a total derivative, and so can not be neglected.

At energies far below $f_{A}$, after integrating out the $X$-charge carrying heavy degree of freedoms, in terms of the physical axion field $``a"$ (which is the excitation with the vacuum expectation removed) we can obtain the following effective Lagrangian ${\cal L}$  including the SM Lagrangian ${\cal L}_{\rm SM}$:
 \begin{eqnarray}
  {\cal L} &\supset&\frac{1}{2}(\partial_{\mu}a)^2-\frac{\partial_{\mu}a}{2f_{A}}\sum_\psi \tilde{X}_{\psi}\bar{\psi}\gamma^{\mu}\gamma_{5}\psi
  +\frac{g^{2}_{s}}{32\pi^2}\frac{a}{F_A}G^{a}_{\mu\nu}\tilde{G}^{a\mu\nu}+\frac{e^{2}}{32\pi^2}\left(\frac{E}{N}\right)\frac{a}{F_A}F_{\mu\nu}\tilde{F}^{\mu\nu}.
  \label{AxionLag01}
 \end{eqnarray}
Below the $SU(2)\times U(1)$ breaking scale where all quarks and leptons obtain masses, the $X$-current given in Eq.\,(\ref{Jx}) is constructed from the axion, quark and lepton transformations under the $X$-symmetry. The reason that the axion gets a mass is that the $X$-current has the color anomaly. Then, we neglect the lepton current for the axion mass.

We integrate out the heavy quarks ($c, b, t$) to obtain the effective couplings just above QCD scale.
Now there are three light quarks ($u,d,s$). In order to obtain the axion mass and derive the axion coupling to photons, we eliminate the coupling of axions to gluons through rotation of the light quark fields
 \begin{eqnarray}
  q\rightarrow \text{exp}\left(-i\alpha_q\frac{\gamma_5}{2}\right)q\qquad\text{with}\,\,q=u,d,s\,.
\label{chiralR2}
 \end{eqnarray}
With the above chiral-rotation, such that $a/F_A-\sum_q\alpha_q=0$, the quark-axion sector of the Lagrangian (\ref{AxionLag01}) reads
 \begin{eqnarray}
  {\cal L}_{A} &=&i\bar{q}\gamma^{\mu}D_{\mu}q+\frac{1}{2}(\partial^{\mu}a)^2-\frac{1}{2}\sum_{q=u,d,s}\left(\frac{\partial^{\mu}a}{f_A}\tilde{X}_q-\partial^\mu\alpha_q\right)\,\bar{q}\gamma^{\mu}\gamma_5q\nonumber\\
  &-&\left(\sum_{q=u,d,s} m_{q}\bar{q}_Le^{-i\alpha_q}q_R+\text{h.c.}\right)+\frac{e^{2}}{32\pi^2}\left(\frac{E}{N}\frac{a}{F_A}-6\sum_{q}\alpha_q(Q^{\rm em}_q)^2\right)F_{\mu\nu}\tilde{F}^{\mu\nu}\,.
  \label{AxionLag02}
 \end{eqnarray}
As can be seen here, the CP violating $\vartheta_{\rm eff}$ term at the minimum is canceled out, which provides a dynamical solution to the CP problem\,\cite{Peccei-Quinn}, but there is a phase in $m_q$.
Clearly, we have some freedom in choosing the phase\,\footnote{In the case that $m_{u}, m_{d}$ and $m_{s}$ are equal, it is natural to choose these phase to be the same, {\it i.e.} $\alpha_{u}=\alpha_{d}=\alpha_{s}\equiv\alpha/3$\,\cite{Dine:2000cj}.}:
since the QCD vacuum is a flavor singlet, {\it i.e.} $\langle\bar{u}u\rangle=\langle\bar{d}d\rangle=\langle\bar{s}s\rangle$, the $\alpha_{q}$ is determined by the flavor singlet condition, that is, $\alpha_{u}m_u=\alpha_{d}m_d=\alpha_{s}m_s$. From $a/F_A-\sum_q\alpha_q=0$ we obtain
 \begin{eqnarray}
  \alpha_u&=&\frac{a}{F_A}\frac{1}{1+z+w}\,,\qquad
  \alpha_d=\frac{a}{F_A}\frac{z}{1+z+w}\,,\qquad
  \alpha_s=\frac{a}{F_A}\frac{w}{1+z+w}\,,
 \end{eqnarray}
where $z=m_{u}\langle\bar{u}u\rangle/m_{d}\langle\bar{d}d\rangle=m_{u}/m_{d}$ and $w=m_{u}\langle\bar{u}u\rangle/m_{s}\langle\bar{s}s\rangle=m_{u}/m_{s}$ in the $SU(3)_{\rm flavor}$ symmetric vacuum.
Considering $u$, $d$ and $s$ quarks, the chiral symmetry breaking effect due to the mixing between axion and light mesons is
 \begin{eqnarray}
  \sum_{q}\alpha_q(Q^{\rm em}_q)^2=\frac{4+z+w}{9(1+z+w)}\frac{a}{F_A}\,.
 \end{eqnarray}
And the value of $E/N$ is determined by the $X$-charge carrying quarks and leptons
 \begin{eqnarray}
  \frac{E}{N}&=&\frac{2\cdot[(\tilde{X}_e+\tilde{X}_\mu+\tilde{X}_\tau)(-1)^2+3(\tilde{X}_u+\tilde{X}_c)\left(\frac{2}{3}\right)^2+3(\tilde{X}_d+\tilde{X}_s+\tilde{X}_b)\left(-\frac{1}{3}\right)^2]}{2(X_{1d}+X_{1s}+X_{1b})(X_{u}+X_{c}+X_{2d}+X_{2s})}
 \end{eqnarray}
which corresponds to $14/39$, where $\delta^{\rm G}_1=6$, $\delta^{\rm G}_2=13$ for the given $X$-charges $X_1=2$, $X_2=-1$ (with $q=-p=1$ in Eq.\,(\ref{anomal})). Here the axion color anomaly $N$ and electromagnetic anomaly $E$ are given below Eq.\,(\ref{angleT}) and Eq.\,(\ref{EManomaly}), respectively.

%%%%%%%%%%%%%%%%%%%%%%%%%%%%%%%%%%%%%%%%%%%%%%%%%%%%%%%%%%%%%%%%%%%%%%%%%%%%%%%%%%%%%%%%%%
\subsubsection{Axion mass}
\noindent Now, at below the QCD scale where the quarks have hadronized into mesons, which will result in mixing between axions and NG mesons of the broken chiral $SU(3)_L\times SU(3)_R$, the kinetic terms vanish
 \begin{eqnarray}
  -{\cal L}_{A} &=&
 \left(\sum_{q=u,d,s} m_{q}\bar{q}_Le^{-i\alpha_q}q_R+\text{h.c.}\right)-\frac{e^{2}}{32\pi^2}\left(\frac{E}{N}-\frac{2}{3}\,\frac{4+z+w}{1+z+w}\right)\frac{a}{F_A}F_{\mu\nu}\tilde{F}^{\mu\nu}\nonumber\\
 &+&\frac{\partial^{\mu}a}{2f_A}\Big\{\Big(\tilde{X}_u-\frac{N}{1+z+\omega}\Big)\bar{u}\gamma^{\mu}\gamma_5u+\Big(\tilde{X}_d-\frac{z\,N}{1+z+\omega}\Big)\bar{d}\gamma^{\mu}\gamma_5d\nonumber\\
  &&+\Big(\tilde{X}_s-\frac{\omega\,N}{1+z+\omega}\Big)\bar{s}\gamma^{\mu}\gamma_5s\Big\}\,.
   \label{Nucleon_Lagran}
 \end{eqnarray}
From the effective Lagrangian\,(\ref{AxionLag14}) the interaction for the light quarks preserves the $X$-symmetry, while it does not preserve the chiral symmetry. So, we may include the effects of the Yukawa interactions in the effective Lagrangian by adding a term which explicitly breaks the symmetry. Let us consider the form of the chiral Lagrangian
 \begin{eqnarray}
  -{\cal L}_{\rm eff} &=& \frac{f^{2}_{\pi}}{4}{\rm Tr}\left[D_{\mu}\Sigma^{\dag}D^{\mu}\Sigma\right]+\frac{1}{2}\mu f^{2}_{\pi}{\rm Tr}\left[\Sigma{\cal A} M_q+(\Sigma{\cal A}M_q)^{\dag}\right]
  \label{ChLagran}
 \end{eqnarray}
where
$\Sigma\equiv{\rm exp}\left[2i\pi^aT^a/f_{\pi}\right]$ ($a=1,...,8$) is the meson field, $T^a$ are the generators of $SU(3)$, $D_{\mu}$ is the appropriate covariant derivatives which introduce the electroweak interactions, $f_{\pi}=92$ MeV, $\mu$ is an undetermined constant, which is related to explicit chiral symmetry breaking, $M_q={\rm diag}(m_{u},m_{d},m_{s})$ is the light quark mass matrix, and ${\cal A}={\rm diag}(e^{i\alpha_{u}}, e^{i\alpha_{d}}, e^{i\alpha_{s}})$ is the axion phase rotation. The first term in the above Lagrangian\,(\ref{ChLagran}) is invariant under global transformation $\Sigma\rightarrow g_{L}\Sigma g^{\dag}_{R}$ where $g_{L}=I$ (unit matrix) and $g_{R}={\rm diag}(e^{i\alpha_{1}}, e^{i\alpha_{2}}, e^{i\alpha_{3}})$, while the second term is not invariant. Thus, the axion and mesons acquire masses from the second term in the Lagrangian\,(\ref{ChLagran}). Note that the invariance of the above Lagrangian\,(\ref{ChLagran}) under $U(1)_{\tilde{X}}$ requires that $\Sigma$ transform as
 \begin{eqnarray}
 \Sigma\rightarrow\Sigma{\left(\begin{array}{ccc}
 e^{-i\alpha\,\tilde{X}_{u}} &  0 &  0 \\
 0 &  e^{-i\alpha\,\tilde{X}_{d}} &  0 \\
 0 &  0 &  e^{-i\alpha\,\tilde{X}_{s}}
 \end{array}\right)}\,;\qquad A\rightarrow A+F_A\alpha\,.
 \end{eqnarray}
Even the axion $A$ field is generated at the high energy, it develops a VEV below QCD scale. Expanding $\Sigma$ and considering the constant term corresponding to ground state energy, the $A$ potential is given as
 \begin{eqnarray}
 V(A)%&=&-\frac{1}{2}\mu f^{2}_{\pi}\left(m_{u}e^{i\alpha_u}+m_{d}e^{i\alpha_d}+m_{s}e^{i\alpha_s}\right)+\text{h.c.}\nonumber\\
% &=&-\mu f^{2}_{\pi}\left(m_{u}\cos\alpha_u+m_{d}\cos\alpha_d+m_{s}\cos\alpha_s\right)\nonumber\\
 &=&-\mu f^{2}_{\pi}\Big\{m_{u}\cos\frac{1}{1+z+w}\left(\frac{A}{F_A}+\vartheta_{\rm eff}\right)\nonumber\\
 &+&m_{d}\cos\frac{z}{1+z+w}\left(\frac{A}{F_A}+\vartheta_{\rm eff}\right)
 +m_{s}\cos\frac{w}{1+z+w}\left(\frac{A}{F_A}+\vartheta_{\rm eff}\right)\Big\}\,,
   \label{axi_pot}
 \end{eqnarray}
which is minimized when $\langle A\rangle=-\vartheta_{\rm eff}F_A$.
Then, the QCD axion mass is proportional to the curvature of the effective potential induced by the anomaly.
Expanding $V(A)$ at the minimum gives the axion mass
 \begin{eqnarray}
 m^2_{a}=\left\langle\frac{\partial^2V(A)}{\partial a^2}\right\rangle_{\langle A\rangle=-\vartheta_{\rm eff}F_A}=\frac{ f^{2}_{\pi}}{F^2_{A}}\frac{\mu m_u}{1+z+w}\,.
  \label{axiMass1}
 \end{eqnarray}

The physical axion/meson states and the mixing parameters may be determined from the axion/meson mass matrix which can be obtained by expanding the symmetry breaking part in Lagrangian\,(\ref{ChLagran}) and taking the terms quadratic in the fields (see Eq.\,(\ref{neut-A})). The axion mass in terms of the pion mass is obtained as
 \begin{eqnarray}
 m^{2}_{a}F^{2}_{A}=m^{2}_{\pi^0}f^{2}_{\pi}F(z,w)\,,
\label{axiMass2}
 \end{eqnarray}
where $m^2_{\pi^0}$ is the $\pi^{0}\pi^{0}$ entry of ${\cal M}^2$ in Eq.\,(\ref{neut-C}), and
 \begin{eqnarray}
 F(z,w)=\frac{z}{(1+z)(1+z+w)}\,, \qquad F_A=\left\{\left(\frac{1}{F_{a1}}\right)^2+\left(\frac{1}{F_{a2}}\right)^2\right\}^{-\frac{1}{2}}\,.
 \label{axipra}
 \end{eqnarray}
It is clear that the axion mass vanishes in the limit $m_u$ or $m_{d}\rightarrow0$. The QCD axion mass derived in Eq.\,(\ref{axiMass2}) is equivalent to Eq.\,(\ref{axiMass1}).
In order to estimate the QCD axion mass, first we determine the parameters $\mu m_{u}$ and $w$ as a function of $z$ from the physical masses of the mesons. In Eq.\,(\ref{neut-A}) they can be extracted as $\mu m_{u}= (108.3{\rm MeV})^2\,z, w=0.315\,z$.
Then, we can estimate the axion mass
 \begin{eqnarray}
 m_{a}=4.34\,{\rm meV}\left(\frac{1.3\times10^{9}\,{\rm GeV}}{F_{A}}\right)\,,
 \label{axiMass4}
 \end{eqnarray}
where the Weinberg value for $z\equiv m_{u}/m_{d}=0.56$\,\cite{WB} and Eq.\,(\ref{AhnMass}) are used.

%%%%%%%%%%%%%%%%%%%%%%%%%%%%%%%%%%%%%%%%%%%%%%%%%%%%%%%%%%%%%%%%%%%%%%%%%%%%%%%%%%%%%%%%%%
\subsubsection{Axion interactions with nucleon}
\label{axion_int}
\noindent Below the chiral symmetry breaking scale, the axion-hadron interactions are meaningful for the axion production rate in the core of a star where the temperature is not as high as 1 GeV, which is given by
 \begin{eqnarray}
  -{\cal L}^{a-\psi_N} &=& \frac{\partial_{\mu}a}{2F_{A}}\,X_{\psi_N}\overline{\psi}_N\,\gamma_\mu\gamma^5\,\psi_N
  \label{a_nucleon}
 \end{eqnarray}
where $\psi_N$ is the nucleon doublet $(p,n)^T$ (here $p$ and $n$ correspond to the proton field and neutron field, respectively).  The couplings of the axion to the nucleon can be derived from the last part in Lagrangian\,(\ref{Nucleon_Lagran})
 \begin{eqnarray}
  -{\cal L}_{A} &\supset&\frac{\partial^{\mu}a}{2F_A}\Big\{\Big(\frac{\tilde{X}_u}{N}-\frac{1}{1+z+\omega}\Big)\bar{u}\gamma^{\mu}\gamma_5u+\Big(\frac{\tilde{X}_d}{N}-\frac{z}{1+z+\omega}\Big)\bar{d}\gamma^{\mu}\gamma_5d\nonumber\\
  &&+\Big(\frac{\tilde{X}_s}{N}-\frac{\omega}{1+z+\omega}\Big)\bar{s}\gamma^{\mu}\gamma_5s\Big\}\,.
   \label{Nucleon_Lagran01}
 \end{eqnarray}
Then nucleon couplings, $X_{n,p}$, are related to axial-vector current matrix elements by Goldberger-Treiman relations\,\cite{PDG}, which are applied in the FA model as
 \begin{eqnarray}
  X_{p}&=&\Big(\frac{\tilde{X}_u}{N}-\eta\Big)\Delta u+\Big(\frac{\tilde{X}_d}{N}-\eta z\Big)\Delta d+\Big(\frac{\tilde{X}_s}{N}-\eta\omega\Big)\Delta s\,,\nonumber\\
  X_{n}&=&\Big(\frac{\tilde{X}_u}{N}-\eta\Big)\Delta d+\Big(\frac{\tilde{X}_d}{N}-\eta z\Big)\Delta u+\Big(\frac{\tilde{X}_s}{N}-\eta\omega\Big)\Delta s\,.
  \label{coupling_n}
 \end{eqnarray}
Here, $\eta=(1+z+\omega)^{-1}$ with $z=m_u/m_d$ and $\omega=m_u/m_s\ll z$ and the $\Delta q$ are given by the axial vector current matrix element $\Delta q\,S_\mu=\langle p|\bar{q}\gamma_\mu\gamma^5q|p\rangle$. 
From Eqs.\,(\ref{a_nucleon}-\ref{coupling_n}) the QCD axion coupling to the neutron can be obtained as
 \begin{eqnarray}
  g_{Ann}=\frac{X_n\,m_n}{F_A}\,.
  \label{coupling_n01}
 \end{eqnarray}
Here the neutron mass $m_n=939.6$ MeV, and the decay constant $F_A=f_A/N$ is replaced by $F_K=f_{\rm K}/N$ and $F_D=f_{\rm D}/N$  for KSVZ and DFSZ model, respectively, in which the color anomaly coefficients are model dependent, $N= 1, N_g(\tan\beta+1/\tan\beta)$, respectively. Now, for numerical estimations on Eq.\,(\ref{coupling_n01}) we adopt the central values of $\Delta u=0.84\pm0.02$, $\Delta d=-0.43\pm0.02$ and $\Delta s=-0.09\pm0.02$, and taken the Weinberg value for $0.38<z<0.58$\,\cite{PDG}. 
We obtain the axion-neutron coupling
 \begin{eqnarray}
  X_{n}&=&\Big(\frac{4}{\delta^{\rm G}_2}-\eta\Big)\Delta d+\Big(\frac{3}{2\delta^{\rm G}_2}+\frac{1}{2\delta^{\rm G}_1}-\eta z\Big)\Delta u+\Big(\frac{1}{\delta^{\rm G}_2}+\frac{1}{2\delta^{\rm G}_1}-\eta\omega\Big)\Delta s\,,
  \label{coupling_n02}
 \end{eqnarray}
 which gives a restrictive bound, whose value lies in ranges $0.174\gtrsim X_n\gtrsim0.070$.
Combining the measurement of axion-neutron coupling in Ref.\,\cite{Leinson:2014ioa} with that of axion-electron coupling in Eq.\,(\ref{wd_bound}),  the decay constants for the FA model are fixed as in Eq.\,(\ref{a_nucleon03}), while the decay constant for DFSZ model has a wide range once the unknown parameter $\tan\beta$ is determined\,\footnote{For example, if one takes $\tan\beta=10$, one obtains $0.022\lesssim X_n\lesssim0.327$ and $9.34\times10^{9}\lesssim f_{\rm D}/{\rm GeV}\lesssim 1.16\times10^{11}$ from the axion-neutron coupling. Combining this result with that of Eq.\,(\ref{wd_bound}) one gets a PQ symmetry breaking scale $f_{\rm D}=(2.4-7.1)\times10^{10}\,{\rm GeV}$  in DFSZ.}, and the decay constant for the KSVZ is not so tightly constrained. The reason is that
for KSVZ axions $\tilde{X}_u=\tilde{X}_d=\tilde{X}_s=0$ leads to $0.081\gtrsim X_n\gtrsim-0.023$ including $X_n=0$, and for DFSZ axions $\tilde{X}_u=\tan\beta, \tilde{X}_d=\tilde{X}_s=1/\tan\beta$ leading to $X_{n}=(\frac{\cos^2\beta}{N_g}-\eta)\Delta d+(\frac{\sin^2\beta}{N_g}-\eta z)\Delta u+(\frac{\sin^2\beta}{N_g}-\eta\omega)\Delta s$ with $N_g=3$ depends on the value of $\tan\beta$.
Interestingly enough, there is a hint for extra cooling from the neutron star in the supernova remnant ``Cassiopeia A" by axion neutron bremsstrahlung, requiring a coupling to the neutron of size\,\cite{Leinson:2014ioa}, which is translated into
 \begin{eqnarray}
  g_{Ann}=(3.8\pm3)\times10^{-10}\Leftrightarrow 7.66\times10^{7}\lesssim F_A/{\rm GeV}\lesssim1.95\times10^{9}\,,
  \label{a_nucleon01}
 \end{eqnarray}
which is compatible with the state-of-the-art upper limit on this coupling, $g_{Ann}<8\times10^{-10}$, from neutron star cooling\,\cite{Sedrakian:2015krq}.
From Eq.\,(\ref{a_nucleon01}) the coupling $g_{Ann}$ can be translated in terms of the scales of $X$-symmetry breakdown, $f_{a_i}$, into
 \begin{eqnarray}
  f_{a_1}&=&(0.65-16.54)\times10^{9}\,{\rm GeV}\,,\qquad\quad f_{a_2}=(0.14-3.58)\times10^{10}\,{\rm GeV}\,
  \label{a_nucleon02}
 \end{eqnarray}
  where we have used $f_{a_i}=\sqrt{2}\,\delta^{\rm G}_i\,F_A$ in Eq.\,(\ref{scale_relation}).
Combining the above result in Eq.\,(\ref{a_nucleon02}) with the axion-electron coupling in Eq.\,(\ref{wd_bound}) we obtain a more restrictive bound on the scale of $U(1)_X$ symmetry breakdown by using Eq.\,(\ref{axipra})
 \begin{eqnarray}
  f_{a_1}=1.1^{+0.6}_{-0.5}\times10^{10}\,{\rm GeV}\,,\qquad\quad f_{a_2}=2.4^{+1.2}_{-1.0}\times10^{10}\,{\rm GeV}\,,
  \label{a_nucleon03}
 \end{eqnarray}
which corresponds to  
 \begin{eqnarray}
  F_A=1.30^{+0.66}_{-0.54}\times10^{9}\, {\rm GeV}\,.
  \label{a_nucleon04}
 \end{eqnarray}
%Taking into account Eq.\,(\ref{hierarchy_vev}) we obtain $\pi/2<H_I/f_{a_1}$ and $H_I/f_{a_2}<\pi$, and consequently a lower bound $H_I>2.67\times10^{10}$ GeV from Eq.\,(\ref{a_nucleon03}), which could be comparable to the values in TABLE\,\ref{inf_result}.

%%%%%%%%%%%%%%%
%    Fig 2    %
%%%%%%%%%%%%%%%
\begin{figure}[b]
%\begin{tabular}{c}
\includegraphics[width=11.0cm]{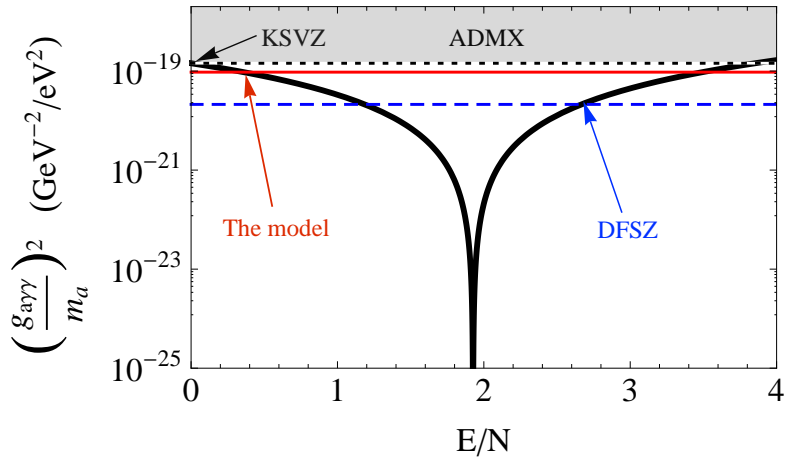}
%\end{tabular}
\caption{\label{Fig2} Plot of $(g_{a\gamma\gamma}/m_{a})^2$ versus $E/N$ for $z=0.56$. The gray-band represents the experimentally excluded  bound $(g_{a\gamma\gamma}/m_{a})^2\leq1.44\times10^{-19}\,{\rm GeV}^{-2}\,{\rm eV}^{-2}$ from ADMX\,\cite{Asztalos:2003px}. Here the dotted-black, dashed-blue, and solid-red lines stand for $(g_{a\gamma\gamma}/m_{a})^2=1.435\times10^{-19}\,{\rm GeV}^{-2}\,{\rm eV}^{-2}$ for $E/N=0$, $2.120\times10^{-20}\,{\rm GeV}^{-2}\,{\rm eV}^{-2}$ for $E/N=8/3$, and $9.010\times10^{-20}\,{\rm GeV}^{-2}\,{\rm eV}^{-2}$ for $E/N=14/39$, respectively. See more various supersymmetric and no-supersymmetric KSVZ and DFSZ-type models varying the parameter $E/N$ in Ref.\,\cite{Ahn:2015pia}.}
\end{figure}
%%%%%%%%%%%%%%%%%%%%%%%%%%%%%%%%%%%%%%%%%%%%%%%%%%%%%%%%%%%%%%%%%%%%%%%%%%%%%%%%%%%%%%%%%%
\subsubsection{Axion interactions with photon}
\noindent After integrating out the heavy $\pi^{0}$ and $\eta$ at low energies, there is an effective low energy Lagrangian with an axion-photon coupling $g_{a\gamma\gamma}$:
 \begin{eqnarray}
{\cal L}_{a\gamma\gamma}= \frac{1}{4}g_{a\gamma\gamma}\,a_{\rm phys}\,F^{\mu\nu}\tilde{F}_{\mu\nu}=-g_{a\gamma\gamma}\,a_{\rm phys}\,\vec{E}\cdot\vec{B}\,,
 \end{eqnarray}
where $\vec{E}$ and $\vec{B}$ are the electromagnetic field components.
And the axion-photon coupling can be expressed in terms of the axion mass, pion mass, pion decay constant, $z$ and $w$:
 \begin{eqnarray}
 g_{a\gamma\gamma}=\frac{\alpha_{\rm em}}{2\pi}\frac{m_a}{f_{\pi}m_{\pi^0}}\frac{1}{\sqrt{F(z,w)}}\left(\frac{E}{N}-\frac{2}{3}\,\frac{4+z+w}{1+z+w}\right)\,.
   \label{gagg}
 \end{eqnarray}
The upper bound on the axion-photon coupling is derived from the recent analysis of the horizontal branch (HB) stars in galactic globular clusters (GCs)\,\cite{Ayala:2014pea}, which translates into the lower bound of decay constant through Eq.\,(\ref{axiMass2}),  as
  \begin{eqnarray}
 |g_{a\gamma\gamma}|<6.6\times10^{-11}\,{\rm GeV}^{-1}\,(95\%\,{\rm CL})\,\Leftrightarrow F_A\gtrsim\left\{
       \begin{array}{lll}
         2.57\times10^{7}\,{\rm GeV}\, & \hbox{FA} \\
         3.20\times10^{7}\,{\rm GeV}, & \hbox{KSVZ}\\
         1.50\times10^{7}\,{\rm GeV}, & \hbox{DFSZ}
       \end{array}
     \right.
   \label{gagg_const}
 \end{eqnarray}
where in the right side $z=0.56$, $\omega=0.315\,z$, and $E/N=14/39,0,8/3$, FA, KSVZ, and DFSZ, respectively, are used. Subsequently, the bounden Eq.\,(\ref{gagg}) translates into the upper bound of axion mass through Eq.\,(\ref{gagg}) as $m_a<0.22$ eV, $<0.18$ eV, and $<0.38$ eV for FA, KSVZ, and DFSZ, respectively.  
%%%%%%%%%%%%%%%
%    Fig 3    %
%%%%%%%%%%%%%%%
\begin{figure}[t]
%\begin{tabular}{c}
\includegraphics[width=11.0cm]{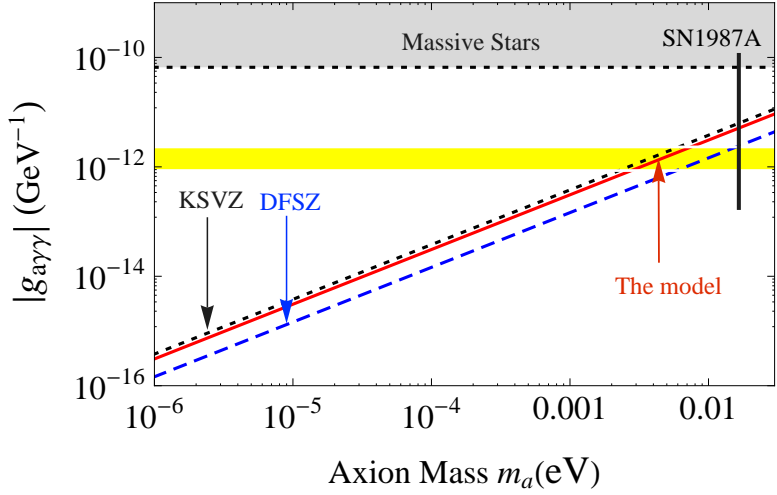}
%\end{tabular}
\caption{\label{Fig3} Plot of $|g_{a\gamma\gamma}|$ versus $m_{a}$ for KSVZ (black dotted line), DFSZ (blue dashed line) and our model (red solid line) in terms of $E/N=$ $0$, $8/3$ and $14/39$, respectively. Here the horizontal dotted line stands for the upper bound $|g_{a\gamma\gamma}|\lesssim6.6\times10^{-11}$ GeV$^{-1}$ which is from globular-cluster stars\,\cite{PDG}. And the black bar corresponding to $m_a\lesssim16$ meV is the constraint derived from the measured duration of the neutrino signal of the supernova SN1987A\,\cite{PDG}. Especially, in the model, for $F_A=1.30^{+0.66}_{-0.54}\times10^{9}$ GeV we obtain $m_{a}=4.34^{+3.37}_{-1.49}\,\text{meV}$ and $|g_{a\gamma\gamma}|=1.30^{+1.01}_{-0.45}\times10^{-12}\,\text{GeV}^{-1}$, which corresponds to the yellow band.}
\end{figure}
From Eq.\,(\ref{axiMass4}) and Eq.\,(\ref{a_nucleon03}) we predict the axion mass and its corresponding axion-photon coupling
  \begin{eqnarray}
  m_a=4.34^{+3.37}_{-1.49}\,\text{meV}\,\Leftrightarrow\,|g_{a\gamma\gamma}|=1.30^{+1.01}_{-0.45}\times10^{-12}\,\text{GeV}^{-1}\,.
  \label{Amass_pre}
 \end{eqnarray}
The corresponding Compton wavelength of axion oscillations is $\lambda_a=(2{\pi\!\!\not\!h}/m_a)c$ with $c\simeq2.997\times10^{8}\,{\rm m/s}$ and ${\!\!\not\!h}\simeq1.055\times10^{-34}\,{\rm J}\cdot{\rm s}$:
 \begin{eqnarray}
  \lambda_a=2.86^{+1.50}_{-1.25}\times10^{-2}\,{\rm cm}\,.
 \end{eqnarray}
 The axion to two-photon decay width is
 \begin{eqnarray}
 \Gamma_{a\rightarrow\gamma\gamma}=\frac{g^2_{a\gamma\gamma}m^3_a}{64\pi}\simeq1.66\times10^{-38}s^{-1}\left(\frac{g_{a\gamma\gamma}}{1.30\times10^{-12}{\rm GeV}^{-1}}\right)^2\left(\frac{m_a}{4.34\,{\rm meV}}\right)^3\,.
 \end{eqnarray}
So the axions decay much slower than the age of the Universe of $4.35\times10^{17}$ s.
The axion coupling to photon $g_{a\gamma\gamma}$ divided by the axion mass $m_{a}$ is dependent on $E/N$. Fig.\,\ref{Fig2} shows the $E/N$ dependence of $(g_{a\gamma\gamma}/m_{a})^2$ so that the experimental limit is independent of the axion mass $m_{a}$\,\cite{Ahn:2014gva}: the value of $(g_{a\gamma\gamma}/m_{a})^2$ of our model is located just a bit lower than that of the conventional axion model, {\it i.e.} KSVZ model. For the Weinberg value $z=0.56$, the anomaly value $E/N=14/39$ predicts $(g_{a\gamma\gamma}/m_{a})^2=9.010\times10^{-20}\,{\rm GeV}^{-2}\,{\rm eV}^{-2}$ which is lower than the ADMX (Axion Dark Matter eXperiment) bound\,\cite{Asztalos:2003px}, $(g_{a\gamma\gamma}/m_{a})^2_{\rm ADMX}\leq1.44\times10^{-19}\,{\rm GeV}^{-2}\,{\rm eV}^{-2}$.
Fig.\,\ref{Fig3} shows the plot for the axion-photon coupling $|g_{a\gamma\gamma}|$ as a function of the axion mass $m_{a}$ in terms of anomaly values $E/N=0, 8/3, 14/39$ which correspond to the KSVZ, DFSZ and FA model, respectively. The model will be tested in the very near future through the experiment such as CAPP (Center for Axion and Precision Physics research)\,\cite{CAPP}.

%%%%%%%%%%%%%%%%%%%%%%%%%%%%%%%%%%%%%%%%%%%%%%%%%%%%%%%%%%%%%%%%
\section{Neutrinos in flavor and astro-particle physics}
\label{low_nut}
Let us investigate how neutrino oscillations at low energies could be connected to new oscillations available on high energy neutrinos. Similar to the quark sector, in order to eliminate the NG modes $A_{1,2}$ from the Yukawa Lagrangian of the neutrinos in Eq.\,(\ref{AxionLag2}) we transform the neutrino fields by chiral rotations 
 \begin{eqnarray}
 U(1)_{X}: N\rightarrow e^{-iX_1\frac{A_1}{f_{a1}}\frac{\gamma_5}{2}}\,N,~~ S\rightarrow e^{i(-X_1\frac{A_1}{f_{a1}}+50\,X_2\frac{A_2}{f_{a2}})\frac{\gamma_5}{2}}\,S,~~
  \nu\rightarrow e^{i(-X_1\frac{A_1}{f_{a1}}+18X_2\frac{A_2}{f_{a2}})\frac{\gamma_5}{2}}\,\nu\,.
 \end{eqnarray}
Since the masses of Majorana neutrino $N_R$ are much larger than those of Dirac and light Majorana ones, after integrating out the heavy Majorana neutrinos, we obtain the following effective Lagrangian for neutrinos
 \begin{eqnarray}
  -{\cal L}^{a-\nu}_{W} &\simeq&\frac{1}{2} \begin{pmatrix} \overline{\nu^{c}_L} & \overline{S_R} \end{pmatrix} {\cal M}_{\nu} \begin{pmatrix} \nu_L \\ S^{c}_R \end{pmatrix}+\frac{1}{2}\overline{N_R}\,M_R\,N^c_R +\frac{g}{\sqrt{2}}W^-_\mu\overline{\ell_{L}}\gamma^\mu\,\nu_{L}+\text{h.c.}\nonumber\\
  &-&\frac{X_{1}}{2}\,\frac{A_1}{f_{a1}}\,M_i\overline{N}_i\,i\gamma_{5}\,N^c_i-\left\{\frac{X_1}{2}\frac{A_1}{f_{a1}}-25\,X_2\frac{A_2}{f_{a2}}\right\}\,m_{s_i}\overline{S}_i\,i\gamma_{5}\,S_i
  \nonumber\\
  &-&\left\{\frac{X_1}{2}\frac{A_1}{f_{a1}}-9X_2\frac{A_2}{f_{a2}}\right\}\,m_{\nu_i}\overline{\nu}_i\,i\gamma_{5}\,\nu_i
  -\frac{1}{2}\,\overline{N}_i\,i\! \! \not\!\partial\,N_i-\frac{1}{2}\,\overline{S}_i\,i\! \! \not\!\partial\,S_i-\frac{1}{2}\,\overline{\nu}_i\,i\! \! \not\!\partial\,\nu_i\,,
\label{Axion_nu_La}\\
\text{with}&&~~{\cal M}_{\nu}= \begin{pmatrix} -m^T_DM^{-1}_Rm_D & m^T_{DS}  \\ m_{DS} &  M_{S}   \end{pmatrix}\,,
  \label{neut1}
 \end{eqnarray}
where the mass matrices $M_R, m_D, m_{DS}$ and $M_S$ have off-diagonal components, and $M_i$ ($m_{s_i}$) and $m_{\nu_i}$ are mass eigenvalues of the heavy (very light) Majorana neutrinos and active neutrinos, respectively. Here we used four-component Majorana spinors, ($N^c=N, S^c=S$, and $\nu^c=\nu$).

According to the simple basis rotation by Lim and Kobayashi\,\cite{lim}, we perform basis rotations from weak to mass eigenstates in the leptonic sector,
 \begin{eqnarray}
  \begin{pmatrix} \nu_L \\ S^{c}_R \end{pmatrix}\longrightarrow W^\dag_\nu\begin{pmatrix} \nu_L \\ S^{c}_R \end{pmatrix}=\xi_L\,.
 \label{}
 \end{eqnarray}
Here the transformation matrix $W_\nu$ is unitary, which is given by 
 \begin{eqnarray}
  W_\nu={\left(\begin{array}{cc}
 U_L &  0  \\
 0 &  U_R 
 \end{array}\right)}{\left(\begin{array}{cc}
 V_1 &  iV_1  \\
 V_2 &  -iV_2 
 \end{array}\right)}Z\,,\qquad\text{with}\quad Z={\left(\begin{array}{cc}
 e^{i\frac{\pi}{4}}\cos\theta &  -e^{i\frac{\pi}{4}}\sin\theta  \\
 e^{-i\frac{\pi}{4}}\sin\theta &  e^{-i\frac{\pi}{4}}\cos\theta 
 \end{array}\right)}
 \label{Wnu}
 \end{eqnarray}
where the $3\times3$ matrix $U_L$ participates in the leptonic mixing matrix, the $3\times3$ matrix $U_R$ is an unknown unitary matrix and $V_1$ and $V_2$ are the diagonal matrices, $V_1={\rm diag}(1,1,1)/\sqrt{2}$ and $V_2={\rm diag}(e^{i\phi_1}, e^{i\phi_2}, e^{i\phi_3})/\sqrt{2}$ with $\phi_i$ being arbitrary phases.
Then the $6\times6$ light neutrino mass matrix in Eq.\,(\ref{neut1}) is diagonalized as 
 \begin{eqnarray}
  W^T_\nu{\cal M}_\nu W_\nu=Z^T{\left(\begin{array}{cc}
 \hat{M}_{\nu\nu} &  \hat{M}  \\
 \hat{M} &  \hat{M}_S 
 \end{array}\right)}Z
 \equiv{\rm diag}(m_{\nu_1},m_{\nu_2},m_{\nu_3},m_{s_1},m_{s_2},m_{s_3})
  \label{diag1}
 \end{eqnarray}
with 
 \begin{eqnarray}
  &\hat{M}_{\nu\nu}=U^T_LM_{\nu\nu}U_L\,,\qquad\qquad\hat{M}_{S}=U^T_RM_{S}U_R\,,\nonumber\\
  &\hat{M}=U^T_R\,m_{DS}\,U_L\equiv{\rm diag}(m_{1},m_{2},m_{3})\,.
 \end{eqnarray}
As can be seen in Eqs.\,(\ref{YDS1}-\ref{MR1}),
it is important to notice that the low energy effective light neutrinos become pseudo-Dirac particles since $\hat{M}$ (or $m_{DS}$) is dominant over $\hat{M}_{\nu\nu}$ and $\hat{M}_S$, that is, $|\hat{M}|\gg|\hat{M}_{\nu\nu}|\,,|\hat{M}_{S}|$ due to Eqs.\,(\ref{YDS1}-\ref{MR1}) and (\ref{neut1}). This is an important point because the masses of the low energy active neutrinos are determined by the Dirac neutrino mass term $m_{DS}$ (or $\hat{M}$) which is from the operators $y^s_i\,L_\alpha S^c_\alpha H_u$ in Eq.\,(\ref{lagrangian2}).
After some algebra if we name $\delta$ ``pseudo-Dirac mass splitting" 
 \begin{eqnarray}
   \delta\equiv\hat{M}_{\nu\nu}+\hat{M}^\dag_S\,,\qquad \text{with}\,\tan2\theta=\frac{|\hat{M}_{\nu\nu}|^2-|\hat{M}_S|^2}{2\hat{M}|\delta|}\,,
  \label{}
 \end{eqnarray}
due to $|\hat{M}_{\nu\nu}|\gg|\hat{M}_S|$ one obtains
 \begin{eqnarray}
   \delta\simeq\hat{M}_{\nu\nu}\,,
  \label{m_split0}
 \end{eqnarray}
 leading to $\tan2\theta\simeq|\delta|/2\hat{M}\ll1$. 
Keeping terms up to the first order in heavy Majorana mass, in the mass eigenstates $\nu_1, \nu_2, \nu_3$, $S^c_1,S^c_2,S^c_3$  basis the Hermitian matrix ${\cal M}_\nu{\cal M}^\dag_\nu$ can be diagonalized as a real and positive $6\times6$ squared mass matrix by the unitary transformation $W_\nu$ in Eq.\,(\ref{Wnu})
 \begin{eqnarray}
   W^T_\nu\,{\cal M}_\nu{\cal M}^\dag_\nu\,W^\ast_\nu&=& {\left(\begin{array}{cc}
 |\hat{M}|^2+|\hat{M}||\delta| &  0  \\
 0 &  |\hat{M}|^2-|\hat{M}||\delta| 
 \end{array}\right)}\nonumber\\
 &\equiv&{\rm diag}(m^2_{\nu_1}, m^2_{\nu_2}, m^2_{\nu_3}, m^2_{s_1},m^2_{s_2} m^2_{s_3})\,.
  \label{eff_nu_mass}
 \end{eqnarray}
As is well-known, because of the observed hierarchy $|\Delta m^{2}_{\rm Atm}|= |m^{2}_{\nu_3}-(m^{2}_{\nu_1}+m^{2}_{\nu_2})/2|\gg\Delta m^{2}_{\rm Sol}\equiv m^{2}_{\nu_2}-m^{2}_{\nu_1}>0$, and the requirement of a Mikheyev-Smirnov-Wolfenstein resonance for solar neutrinos, there are two possible neutrino mass spectra: (i) the normal mass ordering (NO) $m^2_{\nu_1}<m^2_{\nu_2}<m^2_{\nu_3},~m^2_{s_1}<m^2_{s_2}<m^2_{s_3}$, and (ii) the inverted mass ordering (IO) $m^2_{\nu_3}<m^2_{\nu_1}<m^2_{\nu_2},~m^2_{s_3}<m^2_{s_1}<m^2_{s_2}$, in which the mass-squared differences in the $k$-th pair $\Delta m^2_{k}\equiv m^2_{\nu_k}-m^2_{s_k}$ are enough small that the same mass ordering applies for the both eigenmasses, that is, 
 \begin{eqnarray}
   \Delta m^2_{k}=2m_k|\delta_k|\ll m^2_{\nu_k}
  \label{msd}
 \end{eqnarray}
 for all $k=1,2,3$. 
It is anticipated that $\Delta m^2_k\ll\Delta m^2_{\rm Sol}, |\Delta m^2_{\rm Atm}|$,
otherwise the effects of the pseudo-Dirac neutrinos should have been detected.
But in the limit that $\Delta m^2_k=0$, it is hard to discern the pseudo-Dirac nature of neutrinos. 
The pseudo-Dirac mass splittings will manifest themselves through very long wavelength oscillations characterized by the $\Delta m^2_{k}$. (Hereafter, shortly, we call $\Delta m^2_{k}$ mass splitting.) 
The pseudo-Dirac mass splittings could be limited by the following four constraints (i) the active neutrino mass hierarchy: for NO, $m^{2}_{3}\gtrsim\Delta m^2_{\rm Atm}\simeq2.5\times10^{-3}\,{\rm eV}^2$ and $m^{2}_{2}\gtrsim\Delta m^2_{\rm Sol}\simeq7.5\times10^{-5}\,{\rm eV}^2$, while for IO $m^{2}_{2}>m^{2}_{1}\gtrsim2.5\times10^{-3}\,{\rm eV}^2$, which gives the upper bounds for the values of $\delta_k$
 \begin{eqnarray}
   &|\delta_1|\ll3.8\times10^{-5}\,\text{eV}^2/m_1\,,~|\delta_2|\ll4.3\times10^{-3}\,\text{eV}\,,~|\delta_3|\ll7.5\times10^{-4}\,\text{eV},\quad\text{for NO} \nonumber\\
   &|\delta_{1,2}|\ll7.5\times10^{-4}\,\text{eV}\,,~|\delta_3|\ll3.8\times10^{-5}\,\text{eV}^2/m_3\,,\quad \text{for IO}\,,
  \label{pD_bound}
 \end{eqnarray}
(ii) the BBN constraints on the effective  number of species of light particles during nucleosynthesis\,\footnote{ If the effective number of neutrinos $N^{\rm eff}_{\nu}$ is larger than the SM prediction of $N^{\rm eff}_{\nu}=3.046$\,\cite{PDG} at the BBN era, the relativistic degree of freedom, and, consequently, the Hubble expansion rate, will also be larger, causing weak interactions to become ineffective earlier. This will lead to a larger neutron-to-proton ratio and will change the standard BBN predictions for light element abundances. However, the latest number combining Planck and BAO is $N^{\rm eff}_{\nu}=3.04\pm0.18$, spot on $3.046$ expected from the SM neutrinos\,\cite{Planck2014}.}; by requiring sterile neutrinos do not equilibrium at that time through large angle oscillations to active one\,\cite{BBN}, which implies 
 \begin{eqnarray}
   \Delta m^2_k\leq10^{-9}\,\text{eV}^2\,,
  \label{D_bound0}
 \end{eqnarray}
(iii) the solar neutrino oscillations; such $\Delta m^2_{1,2}$ can modify the LMA (large mixing angle) solution and detailed fits in case of pseudo-Dirac neutrinos imply a bound\,\cite{deGouvea:2009fp} 
 \begin{eqnarray}
   \Delta m^2_{1,2}<1.8\times10^{-12}\,\text{eV}^2\,\,\text{at}\,3\sigma\,.
  \label{D_bound}
 \end{eqnarray}
And (iv) the inflationary and leptogenesis scenarios in Ref.\,\cite{Ahn:2016hbn} in the context of our model gives a lower bound on the values of $\delta_k$
 \begin{eqnarray}
   \delta_k\gtrsim2.95\times10^{-14}\,{\rm eV}\,,
  \label{D_lbound}
 \end{eqnarray}
when the Hubble scale during inflation is $H_I\simeq10^{10}$ GeV. From the above constraints (i)-(iv) we roughly estimate a bound for tiny mass splittings
 \begin{eqnarray}
   6\times10^{-16}\lesssim\Delta m^2_k/{\rm eV}^2\lesssim1.8\times10^{-12}\,,
  \label{D_lbound1}
 \end{eqnarray}
where $m_{\nu_i}\sim0.01$ eV is assumed\,\footnote{In the present model the lightest effective neutrino mass could not be extremely small because the values of $\delta_k$ through the relation Eq.\,(\ref{msd}), are constrained by the $\mu-\tau$ powered mass matrix in Eq.\,(\ref{mass matrix}).} in the lower bound.

From the basis rotations of weak to mass eigenstates, one of Majorana neutrino mass matrices, $M_{\nu\nu}=-m^T_DM^{-1}_Rm_D$ in Eq.\,(\ref{neut1}), can be diagonalized as
 \begin{eqnarray}
   \hat{M}_{\nu\nu}=U^T_LM_{\nu\nu}U_L=-U^T_L\,m^T_DM^{-1}_Rm_D\,U_L\,,
  \label{sesa}
 \end{eqnarray}
as noticed in Eq.\,(\ref{diag1}). The three neutrino active states emitted by weak interactions are described in terms of the six mass eigenstates as
 \begin{eqnarray}
   \nu_\alpha=U_{\alpha k}\,\xi_k\,\quad\text{with}~\,\xi_{k}=\frac{1}{\sqrt{2}}\left(\begin{array}{cc}
 1 & i \end{array}\right)\left(\begin{array}{c}
 \nu_{k} \\
 S^c_k \end{array}\right)\,,
  \label{nu_mass_eigen}
 \end{eqnarray}
in which the redefinition of the fields $\nu_k\rightarrow e^{i\frac{\pi}{4}}\nu_k$ and $S^c_k\rightarrow e^{-i\frac{\pi}{4}}S^c_k$ is used. 
Since the active neutrinos are massive and mixed, the weak eigenstates $\nu_{\alpha}$ (with flavor $\alpha=e,\mu,\tau$) produced in a weak gauge interaction are linear combinations of the mass eigenstates with definite masses, given by $|\nu_{\alpha}\rangle=\sum^{N_{\nu}}_kW^{\ast}_{\alpha k}|\xi_{k}\rangle$ where $W_{\alpha k}$ are the matrix elements of the explicit form of the matrix $W_{\nu}$. 
Note that even the number $N_\nu$ of massive neutrinos can be larger than three, in the present model the light fermions $S_\alpha$ do not take part in the standard weak interaction and thus are not excluded by LEP results according to which the number of active neutrinos are coupled with the $W^{\pm}$ and $Z$ bosons is $N_{\nu}=2.984\pm0.008$\,\cite{Beringer:1900zz}. The charged gauge interaction in Eq.\,(\ref{neut1}) for the neutrino flavor production and detection is written in the charged lepton basis as
\begin{align}
-{\cal L}_{\rm c.c.} =\frac{g}{\sqrt{2}}W^{-}_{\mu}\overline{\ell_{\alpha}}\,\frac{1+\gamma_5}{2}\gamma^{\mu}\,U_{\alpha k}\,\xi_{k}+ {\rm h.c.}\,,
\label{WK1}
\end{align}
where $g$ is the SU(2) coupling constant, and $U\equiv U_L$ is the $3\times3$ Pontecorvo-Maki-Nakagawa-Sakata (PMNS) mixing matrix $U_{\rm PMNS}$. Thus
in the mass eigenstate basis the PMNS leptonic mixing matrix\,\cite{PDG} at low energies is visualized in the charged weak interaction, which is expressed in terms of three mixing angles, $\theta_{12}, \theta_{13}, \theta_{23}$, and three \cp-odd phases (one $\delta_{CP}$ for the Dirac neutrino and two $\varphi_{1,2}$ for the Majorana neutrino) as
 \begin{eqnarray}
  U_{\rm PMNS}=
  {\left(\begin{array}{ccc}
   c_{13}c_{12} & c_{13}s_{12} & s_{13}e^{-i\delta_{CP}} \\
   -c_{23}s_{12}-s_{23}c_{12}s_{13}e^{i\delta_{CP}} & c_{23}c_{12}-s_{23}s_{12}s_{13}e^{i\delta_{CP}} & s_{23}c_{13}  \\
   s_{23}s_{12}-c_{23}c_{12}s_{13}e^{i\delta_{CP}} & -s_{23}c_{12}-c_{23}s_{12}s_{13}e^{i\delta_{CP}} & c_{23}c_{13}
   \end{array}\right)}P_{\nu}~,
 \label{PMNS}
 \end{eqnarray}
where $s_{ij}\equiv \sin\theta_{ij}$, $c_{ij}\equiv \cos\theta_{ij}$ and $P_{\nu}$ is a diagonal phase matrix what is that particles are Majorana ones.

%%%%%%%%%%%%%%%%%%%%%%%%%%%%%%%%%%%%%%%%%%%%
\subsection{A bridge between Low and High energy Neutrinos}

Now there are four interesting features in the neutrino sector.
%%%%%%%%
\subsubsection{The active neutrino mixing angles and the pseudo-Dirac mass splittings responsible for new wavelength oscillations come from seesaw}
The first one is that the active neutrino mixing angles $(\theta_{12}, \theta_{13}, \theta_{23}, \delta_{CP})$ and the pseudo-Dirac mass splittings $\delta_k$ responsible for new wavelength oscillations characterized by the $\Delta m^2_k$ could be obtained from the mass matrix $M_{\nu\nu}$ formed by seesawing. Recalling that the $3\times3$ mixing matrix $U_L=U_{\rm PMNS}$ diagonalizing the mass matrix $M_{\nu\nu}$ participates in the charged weak interaction.  From Eqs.\,(\ref{AxionLag2}) and (\ref{WK1}), by redefining the light neutrino field $\nu_{L}$ as $P_{\nu}\nu_{L}$ and transforming $\ell_{L}\rightarrow P_{\nu}\ell_{L}$, $\ell_{R}\rightarrow P_{\nu}\ell_{R}$, $S_{R}\rightarrow P_{s}S_{R}$, one can always make the Yukawa couplings $\hat{y}^{\nu}_{1}, y_{2}, y_{3}$ in Eq.\,(\ref{Ynu1}) and $\hat{y}^s_1,\hat{y}^s_2,\hat{y}^s_3$ in Eq.\,(\ref{YDS1}) real and positive. Then, from Eqs.\,(\ref{Ynu1}) and (\ref{MR1})  we obtain the $\mu-\tau$ powered mass matrix as in Refs.\,\cite{Ahn:2012cg,Ahn:2014gva}
 \begin{eqnarray}
  M_{\nu\nu}
   &=& m_{0}\,e^{i\pi}
   {\left(\begin{array}{ccc}
   1+2F & (1-F)\,y_{2} & (1-F)\,y_{3} \\
   (1-F)\,y_{2} & (1+\frac{F+3\,G}{2})\,y^{2}_{2} & (1+\frac{F-3\,G}{2})\,y_{2}\,y_{3}  \\
   (1-F)\,y_{3} & (1+\frac{F-3\,G}{2})\,y_{2}\,y_{3} & (1+\frac{F+3\,G}{2})\,y^2_{3}
   \end{array}\right)}\nonumber\\
   &=&U^\ast_{\rm PMNS}\hat{M}_{\nu\nu}U^\dag_{\rm PMNS}\,,
\label{mass matrix}
 \end{eqnarray}
where
 \begin{eqnarray}
 m_{0}\equiv \left|\frac{\hat{y}^{\nu2}_{1}\upsilon^{2}_{u}}{3M}\right|\left(\frac{v_{T}}{\sqrt{2}\Lambda}\right)^2\left(\frac{v_{\Psi}}{\sqrt{2}\Lambda}\right)^{18},\quad F=\left(\tilde{\kappa}\,e^{i\phi}+1\right)^{-1},\quad G=\left(\tilde{\kappa}\,e^{i\phi}-1\right)^{-1}.
 \label{Numass1}
 \end{eqnarray}
In the limit  $y^\nu_1=y^\nu_2=y^\nu_3$ ($y_{2}, y_{3}\rightarrow1$), the  mass matrix\,(\ref{mass matrix}) gives the TBM angles\,\cite{Harrison:2002er} and their corresponding mass eigenvalues $|\delta_k|$ which are equivalent to $|(\hat{M}_{\nu\nu})_k|=\Delta m^2_k/2m_k$
 \begin{eqnarray}
 &&\sin^{2}\theta_{12}=\frac{1}{3}\,,\,\quad\qquad\sin^{2}\theta_{23}=\frac{1}{2}\,,~\,\qquad\sin\theta_{13}=0\,,\nonumber\\
 &&|\delta_1|=3\,m_{0}\,|F|~,\qquad |\delta_2|=3\,m_{0}~,\,\qquad |\delta_3|= 3\,m_{0}\,|G|~.
 \label{TBM1}
 \end{eqnarray}
These pseudo-Dirac mass splittings $|\delta_k|$, which is closely correlated with an axion decay constant 
(see Eq.\,(\ref{fA1}), the $U(1)_{X_1}$ symmetry breaking scale), are disconnected from the TBM mixing angles. It is in general expected that deviations of $y_2, y_3$ from unity, leading to the non-zero reactor mixing angle, {\it i.e.} $\theta_{13}\neq0$, and in turn opening a possibility to search for CP violation in neutrino oscillation experiments.
These deviations generate relations between mixing angles and eigenvalues $|\delta_k|$.
Therefore Eq.\,(\ref{mass matrix}) directly indicates that there could be deviations from the exact TBM if the Dirac neutrino Yukawa couplings in $m_D$ of Eq.\,(\ref{Ynu1}) do not have the same magnitude, and the pseudo-Dirac mass splittings are all of the same order
 \begin{eqnarray}
 |\delta_1|\simeq |\delta_2|\simeq |\delta_3|\simeq {\cal O}(m_{0})~.
 \label{m_split}
 \end{eqnarray}
The large values of the solar ($\theta_{12}$) and atmospheric ($\theta_{23}$) mixings as well as the non-zero but relatively large reactor mixing angle ($\theta_{13}$), as indicated in TABLE\,\ref{exp}, are consequences of a nontrivial structure of the $\mu-\tau$ powered mass matrix $M_{\nu\nu}$ in Eq.\,(\ref{mass matrix}) in the charged lepton basis. Let us consider the constraints on the $X$-symmetry (or PQ symmetry) breaking scale implied by the fermion mass scales in the model as well as the interactions between SM fermions and axions. In turn, this astro-particle constraint plays a crucial role in cosmology, as shown in the leptogenesis scenario of Ref.\,\cite{Ahn:2016hbn}. From the overall scale of the mass matrix in Eq.\,(\ref{Numass1}) the pseudo-Dirac mass splitting, $\delta_2$, is expected to be
\begin{eqnarray}
  |\delta_2|\simeq2.94\times10^{-11}\left(\frac{4.24\times10^{9}{\rm GeV}}{M}\right)\left|\hat{y}^{\nu}_{1}\frac{v_{T}}{\sqrt{2}\Lambda}\right|^{2}\sin^2\beta~{\rm eV}~,
 \label{scaleLambda}
\end{eqnarray}
in which the scale of $M$ can be estimated from Eqs.\,(\ref{MR2}) and (\ref{a_nucleon03}) through the astrophysical constrainsts as
\begin{eqnarray}
  M=|\hat{y}_\Theta|\times2.75^{+1.50}_{-1.25}\times10^{9}\,{\rm GeV}~.
 \label{scaleM}
\end{eqnarray}
Note that the scale of the heavy neutrino, $M$, is connected to the PQ symmetry breaking scale via the axion decay constant in Eq.\,(\ref{fA1}).
As shown in Eq.\,(\ref{vev_ph}), the scale of $M$ is expected as ${\cal O}(v_{\Theta})\sim{\cal O}(v_{S})\sim{\cal O}(M)$. And Eq.\,(\ref{scaleLambda}) shows that the value of $\delta_2$ depends on the magnitude $\hat{y}^{\nu}_{1}v_{T}/\Lambda$ since $M$ is constrained by the astrophysical constraints in Eq.\,(\ref{a_nucleon03}): the smaller the ratio $v_{T}/\Lambda$, the smaller becomes $|\delta_k|$ responsible for the pseudo-Dirac mass splittings\,\footnote{Moreover, the overall scale of the heavy neutrino mass $M$ is closely related with a successful leptogenesis in Ref.\,\cite{Ahn:2016hbn}, constraints of the mass splittings in Eq.\,(\ref{msd}), and the CKM mixing parameters, therefore it is very important to fit the parameters $v_{T}/\Lambda$ and $M$.}. However, the value of $|\delta_k|$ is constrained from Eqs.\,(\ref{D_lbound}) and (\ref{D_lbound1}); for example, using $\tan\beta=2$ and $v_T/\Lambda\simeq\lambda^2/\sqrt{2}$ we obtain
\begin{eqnarray}
  |\delta_2|\simeq1.50\times10^{-14}\,|\hat{y}^\nu_1|^2\,{\rm eV}~.
 \label{scaleLambda2}
\end{eqnarray}
The value of $v_{T}/\Lambda$ is also related to the $\mu$-term in Eq.\,(\ref{muterm}): when soft SUSY breaking terms are included into the flavon potential, the driving fields attain VEVs, and in turn the magnitude of $\mu$-term is expected to be $200~{\rm GeV}\lesssim\mu_{\rm eff}\lesssim1$ TeV for $m_S\sim{\cal O}(10)$ TeV and $v_{T}/\Lambda\sim0.04$.
Since the values of $v_{T}/\Lambda$ and $v_{S}/\Lambda$ are closely associated with the CKM mixing matrix and the down-type quark masses, respectively, their values should lie in the ranges
\begin{eqnarray}
  \frac{v_{T}}{\Lambda}\sim{\cal O}(0.1)\,,\qquad \frac{v_{S}}{\Lambda}\lesssim\frac{v_{\Theta}}{\Lambda}\sim\lambda^2<\frac{v_{\Psi}}{\Lambda}=\lambda<1\,.
 \label{MassRangge}
\end{eqnarray}
Here the first term is derived from the requirement that the term should fit its size down to generate the correct CKM matrix in Ref.\,\cite{Ahn:2014gva} as well as the $\mu$-term in Eq.\,(\ref{muterm}), and the second one comes from Eq.\,(\ref{Cabbibo}) and  $v_{\Theta}=v_{\Psi}\delta^{\rm G}_1/\delta^{\rm G}_2\sqrt{1+\kappa^2}$ with $\delta^{\rm G}_1=6$, $\delta^{\rm G}_2=13$ and $\kappa\equiv v_{S}/v_{\Theta}$ (see also its related parameter $\tilde{\kappa}$ in Eq.\,(\ref{MR2})), as shown in Eq.\,(\ref{AhnMass}).

Naively speaking, the charged-lepton superpotential in (\ref{lagrangian2}) does not contribute to the PMNS matrix due to the diagonal form of mass matrix. However, the neutrino superpotential in\,(\ref{lagrangian2}) has totally 22 parameters (except for $\hat{y}_{\tilde{\Theta}}$), which means the mass matrix in Eq.\,(\ref{neut1}) has 22 parameters. Since the transform matrix $W_\nu$ in Eq.\,(\ref{Wnu}) has 16 parameters: $U_L$ contains 6, $U_R$ contains 6, $V_2$ contains 3, and $Z$ contains 1, instead of using Eq.\,(\ref{diag1}) due to ambiguity of phases if we look at the equation\,(\ref{eff_nu_mass}) we see that there are 6 real mass squared eigenvalues. From Eqs.\,(\ref{eff_nu_mass}) and (\ref{mass matrix}) we see that there are 8 physical degree of freedoms, {\it i.e.}, $m_0, y_2, y_3, \tilde{\kappa}, \phi$, and $\Delta m^2_k$ with $k=1,2,3$. One can reduce the physical degree of freedoms more: once the three pseudo-Dirac mass splittings $\Delta m^2_k$ are fixed by high energy very long wave experiments, such as IceCube, there are only 5 physical degree of freedoms left in neutrino sector; among nine observables the five measured quantities ($\theta_{12}, \theta_{23}, \theta_{13}, \Delta m^2_{\rm Sol}$, and $\Delta m^2_{\rm Atm}$) are used as constraints, and four quantities could be predicted, see Sec.\,\ref{Nuannu}.

%%%%%%%%
\subsubsection{The sum of active neutrino masses constrained from cosmology}
The second interesting feature is that the masses  of the active neutrinos $m_{\nu_i}$ are determined in a completely independent way that the neutrino mixing angles are obtained through the seesaw formula in Eq.\,(\ref{mass matrix}); but they are tied to each other by the tiny mass splittings in Eq.\,(\ref{eff_nu_mass}). Thus the sum of light neutrino masses given by
 \begin{eqnarray}
 \sum_{i}m_{\nu_i}=\frac{1}{2}\left(\frac{\Delta m^2_1}{\delta_1}+\frac{\Delta m^2_2}{\delta_2}+\frac{\Delta m^2_3}{\delta_3}\right)
  \label{nu_sum_c}
 \end{eqnarray}
could be controlled by the $\mu-\tau$ powered mass matrix in Eq.\,(\ref{mass matrix}). And a bound on the sum of the light neutrino masses can be extracted as
 \begin{eqnarray}
 0.06\lesssim\sum_{i}m_{\nu_i}/\text{eV}<0.194\,;
  \label{nu_sum_b}
 \end{eqnarray}
a lower limit for the sum of the neutrino masses, $\sum_{i=1}^{3} m_{\nu_i}\gtrsim0.06$ eV could be provided by the neutrino oscillation measurements;
a upper limit\,\footnote{Massive neutrinos could leave distinct signatures on the CMB and large-scale structure (LSS) at different epochs of the Universe's evolution\,\cite{Abazajian:2008wr}. To a large extent, these signatures could be extracted from the available cosmological observations, from which the total neutrino mass could be constrained. } is given by Planck Collaboration\,\cite{Planck2014} which is subject to the cosmological bounds $\sum_{i}m_{\nu_i}<0.194$ eV at $95\%$ CL (the CMB temperature and polarization power spectrum from Planck 2015
in combination with the baryon acoustic oscillations (BAO) data, assuming a standard $\Lambda$CDM cosmological model). 
And another interesting quantity related to our leptogenesis scenario in Ref.\,\cite{Ahn:2016hbn} could be extracted as 
 \begin{eqnarray}
 \sum_{i}\frac{m_{\nu_i}}{\Delta m^2_i}=\frac{1}{2}\left(\frac{1}{\delta_1}+\frac{1}{\delta_2}+\frac{1}{\delta_3}\right)\lesssim0.5\times10^{14}\,\text{eV}^{-1}\,,
  \label{nu_sum_d}
 \end{eqnarray}
 where the upper limit is derived from a lower bound on $\delta_i$ in Eq.\,(\ref{D_lbound}); see, the leptogenesis scenario in Ref.\,\cite{Ahn:2016hbn}.
It is expected that, once the tiny mass splittings $\Delta m^2_k$ are fixed through new oscillation experiments, the above quantities in Eqs.\,(\ref{nu_sum_c}) and (\ref{nu_sum_d}) has a dependence on $\theta_{23}$ along with the $\mu-\tau$ powered mass matrix in Eq.\,(\ref{mass matrix}).
Also remark that the tritium beta decay experiment KATRIN\,\cite{KATRIN} will be sensitive to an effective electron neutrino mass  $m_{\beta}=\sqrt{\sum_{i}|U_{ei}|^{2}\,m^2_{\nu_i}}$\,\cite{beta} down to about $0.2$ eV. 

%%%%%%%%
\subsubsection{The active neutrino mixing parameters constrained from astronomical-scale baseline neutrino oscillations.}
The third interesting feature is that, once very tiny mass splittings are determined by performing astronomical-scale baseline experiments to uncover the oscillation effects of very tiny
mass splitting $\Delta m^2_k$, the active neutrino mixing parameters ($\theta_{12}, \theta_{23}, \theta_{13}, \delta_{CP}$ and $m_{\nu_1}, m_{\nu_2}, m_{\nu_3}$) are predicted in the model due to Eqs.\,(\ref{msd}) and (\ref{sesa}).
Thus we can possibly connect the pseudo-Dirac neutrino oscillations with the low energy neutrino properties as well as a successful leptogenesis in Ref.\,\cite{Ahn:2016hbn}.
With the help of the mixing matrix Eq.\,(\ref{Wnu}), the flavor conversion probability between the active neutrinos follows from the time evolution of the state $\xi_k$ as,
\begin{eqnarray}
 P_{\nu_{\alpha}\rightarrow\nu_{\beta}}(W_{\nu},L,E)=\left|\left(W^{\ast}_{\nu}e^{-i\frac{\hat{{\cal M}}^{2}_{\nu}}{2E}L}W^T_{\nu}\right)_{\alpha\beta}\right|^2=\frac{1}{4}\left|\sum^3_{k=1}U_{\beta k}\left\{e^{i\frac{m^2_{\nu k}L}{2E}}+e^{i\frac{m^2_{S k}L}{2E}}\right\}U^{\ast}_{\alpha k}\right|^2\,,
\end{eqnarray}
in which $L=$ flight length, $E=$ neutrino energy, and $\hat{\cal M}_\nu\equiv W^T_\nu\,{\cal M}_\nu\,W_\nu$, see Eq.\,(\ref{diag1}).
For the baseline, $4\pi E/\Delta m^{2}_{\rm Sol, Atm}\ll L$, the probability of  neutrino flavor conversion reads
\begin{eqnarray}
 P_{\nu_{\alpha}\rightarrow\nu_{\beta}}\equiv P_{\alpha\beta}=\sum^3_{k=1}|U_{\alpha k}|^2|U_{\beta k}|^2\cos^2\left(\frac{\Delta m^2_{k}L}{4E}\right)\,,
\label{proba1}
\end{eqnarray}
where the oscillatory terms involving the atmospheric and solar mass-squared differences are averaged out over these long distances.
Such new oscillation lengths far beyond the earth-sun distance will be provided by astrophysical neutrinos, which fly galactic and extra galactic distances with very high energy neutrinos. It has been shown\,\cite{Lunardini:2000swa} that inside the Gamma Ray Burst (GRB) sources $\int V_{C,N}dt\ll1$ where the effective potentials due to the matter effects are $V_C=\sqrt{2}G_Fn_e$ with $n_e$ being the electron number density in matter and $V_N=-\sqrt{2}G_Fn_n/2$ with $n_n$ being the neutron number density in matter, so the matter effects inside the source are not relevant for neutrino oscillation, while inside the earth for $V_{C,N}\gg\Delta m^2_{k}/2E$ again the matter effect will not be significant because of the very tiny effective mixing angle.
So, we only consider neutrino oscillation in vacuum for astrophysical neutrinos.
Neutrinos arriving at neutrino telescopes from astrophysical sources such as GRBs\,\cite{GRBs}, active galactic nuclei\,\cite{Becker:2007sv}, and type Ib/c supernova\,\cite{Kappes:2006fg} travel large distances over $\sim100$ Mpc.
Neutrino telescope, such as IceCube\,\footnote{IceCube\,\cite{ice-cube} is a powerful neutrino telescope but also a huge muon detector that registers more than 100 billion muons per year, produced by the interaction of cosmic rays in the Earth�s atmosphere.}, observes neutrinos from extragalactic sources 
located far away from the earth and with
neutrino energy $10^5\,{\rm GeV}\lesssim E\lesssim10^7$ GeV.
Given neutrino trajectory $L$ and energy $E$, the oscillation effects  become prominent
when  $\Delta m^2_k\sim E/4\pi L$, 
where $L\equiv L(z)$ is a distance-measure with redshift $z$, which is different from comoving or luminosity distance,
given by
\begin{eqnarray}
 L(z)\equiv D_H\int^z_0\frac{dz'}{(1+z')^2\sqrt{\Omega_m(1+z')^3+\Omega_\Lambda}}\,,
 \label{}
\end{eqnarray}
where the Hubble length $D_H=c/H_0\simeq4.42$ Gpc with the results of the Planck Collaboration\,\cite{Ade:2015xua}: 
 \begin{eqnarray}
 \Omega_\Lambda=0.6911\pm0.0062\,,\quad\Omega_m=0.3089\pm0.0062\,,\quad  H_0=67.74\pm0.46\,{\rm km}\,{\rm s}^{-1}{\rm Mpc}^{-1}\,,
 \label{cosmo_const0}
 \end{eqnarray}
in which $\Omega_\Lambda$, $\Omega_m$, and $H_0$ stand for the dark energy density of the Universe, the matter density of the Universe, and the present Hubble expansion rate, respectively.
The asymptotic value of $L(z)$ is about $2.1$ Gpc achieved by large value of $z$, which means that  the smallest $\Delta m^2_k$ that can
be probed with astrophysical neutrinos with $E$ is $10^{-17}\,\mbox{eV}^2 \,(E/\rm PeV)$\,\cite{redshift}.
If this is case, in order to observe the oscillation effects the oscillation lengths should not be  much larger than the flight length
before arriving at neutrino telescopes in earth for given tiny mass splittings, that is,
\begin{eqnarray}
 L^k_{\rm osc}\simeq\left(\frac{5\times10^{-15}\,{\rm eV}^2}{\Delta m^2_{k}}\right)\left(\frac{E}{5\times10^5{\rm GeV}}\right)8\,\text{Mpc}\lesssim8\,\text{Mpc}
\label{osc_length}
\end{eqnarray}
which means that astrophysical neutrinos with  $L\simeq8$ Mpc (the flight length) and energy $E\simeq0.5\,{\rm PeV}$ would be useful
to probe the pseudo-Dirac property of neutrinos with the very tiny mass splitting $\Delta m^2_{k}\simeq5\times10^{-15}\,{\rm eV}^2$.
From Eq.\,(\ref{osc_length}), we see that given the tiny mass splittings $\Delta m^2_k=10^{-14 -15}{\rm eV}^2$ with the energies around 100 TeV--1 PeV,
 a new oscillation curve at neutrino trajectory $\lesssim{\cal O}(10)$ Mpc is naively expected to occur.
In Refs.\,\cite{Palladino:2015zua,Ahn:2016hhq} the track-to-shower ratio for the number of shower $N_S$ and track events $N_T$  in the IceCube detector
is expressed in terms of tiny mass splittings $\Delta m^2_{k}$, flight length $L$, neutrino mixing angles and CP phase ($\theta_{12},\theta_{23},\theta_{13},\delta_{CP}$), and initial flavor composition $\phi^0_{\beta}$
\begin{eqnarray}
 \frac{N_T}{N_S}&=& \frac{a_\mu\,p_T\,\tilde{F}_\mu}{a_e\,\tilde{F}_e+a_\mu\,(1-p_T)\,\tilde{F}_\mu+a_\tau\,\tilde{F}_\tau}\,,
 \label{}
\end{eqnarray}
where
\begin{eqnarray}
 &\tilde{F}_\alpha=\sum_{\beta k}|U_{\alpha k}|^2|U_{\beta k}|^2\,\phi^0_{\beta}\,,\nonumber\\
 &a_\alpha=4\pi\int dE\cos^2\left(\frac{\Delta m^2_{k}L}{4E}\right)E^{-\omega}A_\alpha(E)\,,
 \label{}
\end{eqnarray}
with a spectral index $\omega$.
Here $p_T$ is the probability that an observed event produced by a muon neutrino is a track event, which is mildly dependent on energy and approximately equals to $0.8$\,\cite{Aartsen:2013jdh}.
Then above equation can be simplified to
\begin{eqnarray}
 \frac{N_T}{N_S}=\frac{\phi_\mu}{\frac{a_e}{a_\mu\,p_T}+\left(\frac{a_\tau}{a_\mu\,p_T}-\frac{a_e}{a_\mu\,p_T}\right)\,\phi_\tau+\left(\frac{1-p_T}{p_T}-\frac{a_e}{a_\mu\,p_T}\right)\,\phi_{\mu}},
\label{NTS}
\end{eqnarray}
where $\phi_{e}=1-\phi_\mu-\phi_\tau$ with $\phi_\ell\equiv\tilde{F}_\ell/(\tilde{F}_e+\tilde{F}_\mu+\tilde{F}_\tau)$ is assumed.

%%%%%%%%
\subsubsection{No observable $0\nu\beta\beta$-decay rate.}
The fourth important feature is that, since the two mass eigenstates in each pseudo-Dirac pair have opposite CP parity, no observable $0\nu\beta\beta$-decay rate is expected.
In the model the $0\nu\beta\beta$-decay rate effectively measures the absolute value of the $ee$-component of the effective neutrino mass matrix $\mathcal{M}_{\nu}$ in Eq.\,(\ref{neut1}) in the basis where the charged lepton mass matrix is real and diagonal, which can be expressed as 
 $|m_{ee}|=\big|\sum^3_{k=1}\big(U_{ek}/\sqrt{2}\big)^2(m_{\nu_k}-m_{s_k})\big|$, which in turn
 is roughly re-expressed  in terms of the pseudo-Dirac mass splittings as
 \begin{eqnarray}
 |m_{ee}|\simeq\Big|\sum^3_{k=1}\left(\frac{U_{ek}}{\sqrt{2}}\right)^2\delta_k\Big|\sim{\cal O}(10^{-14-15})\,{\rm eV}\,,
 \label{mee1}
 \end{eqnarray}
where the last equality is deduced from the numerical analysis in Sec.\,\ref{Nuannu}. This clearly indicates that the $0\nu\beta\beta$-decay would be highly suppressed due to the constraints in Eqs.\,(\ref{pD_bound}-\ref{D_lbound1}).
 The pseudo-Dirac neutrinos (Majorana neutrinos) are almost Dirac particles and the lepton number is only slightly violated by their Majorana masses and $M_{\nu\nu},M_{S}\ll m_{DS}$.
Therefore, the discovery of $0\nu\beta\beta$-decay in the on-gong or future $0\nu\beta\beta$ experiments\,\cite{nuBB}, with sensitivities $0.01<|m_{ee}|/{\rm eV}<0.1$, will rule out the present model. Current $0\nu\beta\beta$-decay experimental upper limits and the reach of near-future experiments are collected for example in Ref.\,\cite{Schwingenheuer:2012zs}.

%%%%%%%%%%%%%%%%%%%%%%%%%%%%%%%%%%%%%%%%%%%%%%%%%%%%%%
\subsection{Numerical analysis}
\label{Nuannu}
After the relatively large reactor angle $\theta_{13}$ measured in Daya Bay\,\cite{An:2012eh} and RENO\,\cite{Ahn:2012nd} including Double Chooz, T2K and MINOS experiments\,\cite{Other}, the recent analysis based on global fits\,\cite{global_nu,Gonzalez-Garcia:2015qrr} of the neutrino oscillations enters into a new phase of precise determination of mixing angles and mass squared differences, indicating that the TBM\,\cite{Harrison:2002er} for three flavors should be corrected in the lepton sector: especially, in the most recent analysis\,\cite{Gonzalez-Garcia:2015qrr} their allowed ranges at $1\sigma$ best-fit $(3\sigma)$ from global fits are given by TABLE\,\ref{exp}.
\begin{table}[h]
%\begin{widetext}
%\begin{center}
\caption{\label{exp} The global fit of three-flavor oscillation parameters at the best-fit (BF) and $3\sigma$ level\,\cite{Gonzalez-Garcia:2015qrr}. NO = normal neutrino mass ordering; IO = inverted mass ordering. And $\Delta m^{2}_{\rm Sol}\equiv m^{2}_{\nu_2}-m^{2}_{\nu_1}$, $\Delta m^{2}_{\rm Atm}\equiv m^{2}_{\nu_3}-m^{2}_{\nu_1}$ for NO, and  $\Delta m^{2}_{\rm Atm}\equiv m^{2}_{\nu_2}-m^{2}_{\nu_3}$ for IO.}
\begin{ruledtabular}
\begin{tabular}{cccccccccccc} &$\theta_{13}[^{\circ}]$&$\delta_{CP}[^{\circ}]$&$\theta_{12}[^{\circ}]$&$\theta_{23}[^{\circ}]$&$\Delta m^{2}_{\rm Sol}[10^{-5}{\rm eV}^{2}]$&$\Delta m^{2}_{\rm Atm}[10^{-3}{\rm eV}^{2}]$\\
\hline
BF $\begin{array}{ll}
\hbox{NO}\\
\hbox{IO}
\end{array}$&$\begin{array}{ll}
8.50 \\
8.51
\end{array}$&$\begin{array}{ll}
306 \\
254
\end{array}$&$34.63$&$\begin{array}{ll}
42.3 \\
49.5
\end{array}$&$7.50$
 &$\begin{array}{ll}
2.457 \\
2.449
\end{array}$ \\
\hline
$3\,\sigma$$\begin{array}{ll}
\hbox{NO}\\
\hbox{IO}
\end{array}$&$\begin{array}{ll}
7.85\rightarrow9.10 \\
7.87\rightarrow9.11
\end{array}$&$0\rightarrow360$&~$31.29\rightarrow35.91$&$\begin{array}{ll}
38.2\rightarrow53.3 \\
38.6\rightarrow53.3
\end{array}$
 &$7.02\rightarrow8.09$&$ \begin{array}{ll}
                           2.317\rightarrow2.607 \\
                           2.307\rightarrow2.590
                          \end{array}$\\
\end{tabular}
\end{ruledtabular}
%\end{center}
%\end{widetext}
\end{table}
In addition, recently the high energy neutrino events observed by IceCube\,\cite{Aartsen:2013jdh} are analyzed in Refs.\,\cite{Palladino:2015zua}, aiming to probe the initial flavor of cosmic neutrinos; 
the bound on the track-to-shower ratio of a cosmic neutrino\,\footnote{ We note that much larger detectors than the present IceCube would be required to get fully meaningful result for the test of our model in detail.} is extracted as
\begin{eqnarray}
 \frac{N_T}{N_S}&=& 0.18^{+0.13}_{-0.05}\,.
 \label{NTS_bound}
\end{eqnarray}
First, in order to obtain low energy neutrino data we perform a numerical analysis using the linear algebra tools of Ref.\,\cite{Antusch:2005gp}.
The seesaw formula in Eq.\,(\ref{mass matrix}) for obtaining neutrino mixing angles and pseudo-Dirac mass splittings contains seven parameters : $y_{1}(\equiv\hat{y}^{\nu}_{1}\frac{v_T}{\sqrt{2}\Lambda}(\frac{v_\Psi}{\sqrt{2}\Lambda})^{9}),v_{u}, \,M,\,y_{2},\,y_{3},\,\tilde{\kappa},\,\phi$. The first three ($y_{1}$, $M,$ and $v_{u}$) lead to the overall scale parameter $m_{0}$, which is closely related to the $U(1)_{X_1}$ breaking scale, see Eq.\,(\ref{a_nucleon03}). The next four ($y_2,\,y_3,\,\tilde{\kappa},\,\phi$) give rise to the deviations from TBM as well as the CP phases and corrections to the pseudo-Dirac mass splittings (see Eq.\,(\ref{TBM1})).
In our numerical analysis, we take $M=4.24\times10^{9}$ GeV and\,\footnote{From Eqs.\,(\ref{MR2}) and (\ref{a_nucleon03}) we simply square the axion decay constant $f_{a_1}$ with the scale $M$. As noticed in Eq.\,(\ref{YukawaW}), in our model small values of $\tan\beta=v_{u}/v_{d}$ are preferred.} $\tan\beta=2$ (see Eq.\,(\ref{scaleM}) and below Eq.\,(\ref{YukawaW})), for simplicity, as inputs. Recalling that all the hat Yukawa couplings are of order unity, {\it i.e.}, $1/\sqrt{10}\lesssim|\hat{y}|\lesssim\sqrt{10}$. 
Then the effective mass matrix in Eq.\,(\ref{mass matrix}) contains only the five parameters $m_{0},y_{2},y_{3},\tilde{\kappa},\phi$, which can be determined from the experimental results of three mixing angles, $\theta_{12},\theta_{13},\theta_{23}$, and the
three tiny mass splittings, $\Delta m^{2}_{k}=2m_k\,|\delta_k|$, if they are fixed by high energy very long wavelength experiments, such as IceCube. In addition, the individual neutrino masses $m_{\nu_i}=m_i$ and the CP phases $\delta_{CP},\varphi_{1,2}$ can be predicted after determining the model parameters. Scanning all the parameter spaces by putting the experimental constraints in TABLE\,\ref{exp} with the above input parameters,
we obtain for the normal mass ordering (NO) with $\Delta m^2_1=\Delta m^2_2=2.7\times10^{-15}\,{\rm eV}^2$, $\Delta m^2_3=5\times10^{-15}\,{\rm eV}^2$
 \begin{align}
  &\tilde{\kappa} \in [0.15,0.66] ,
  & \phi \in [92^{\circ},112^{\circ}]\cup[248^{\circ},268^{\circ}]\,,
  \nonumber\\
  & \hat{y}^\nu_{1} \in [1.28,1.98],
  & y_{2} \in [0.81,1.29],\qquad\qquad\quad
  & y_{3} \in [0.82,1.31],
  \label{input1}
 \end{align}
 leading to $\hat{y}^s_1 \in [0.93,2.10]$, $\hat{y}^s_2 \in [0.98,2.12]$, and $ \hat{y}^s_3 \in [2.11,2.89]$;
 for the inverted mass ordering (IO) with $\Delta m^2_1=\Delta m^2_2=10^{-14}\,{\rm eV}^2$, $\Delta m^2_3=5.5\times10^{-15}\,{\rm eV}^2$, we obtain
 \begin{align}
  &\tilde{\kappa} \in [0.10,0.66] ,
  & \phi \in [90^{\circ},112^{\circ}]\cup[248^{\circ},269^{\circ}]\,,
  \nonumber\\
  & \hat{y}^\nu_{1} \in [2.15,2.48],
  & y_{2} \in [0.82,1.22],\qquad\qquad\quad
  & y_{3} \in [0.80,1.22],
  \label{input2}
 \end{align}
 leading to $\hat{y}^s_1 \in [2.07,2.86]$, $\hat{y}^s_2 \in [2.10,2.88]$, and $\hat{y}^s_3 \in [0.85,2.12]$.
%%%%%%%%%%%%%%%
%   Fig A-3   %
%%%%%%%%%%%%%%%
\begin{figure}[h]
%\vspace*{-5.0cm}
%\hspace*{-1cm}
\begin{minipage}[h]{7.3cm}
\epsfig{figure=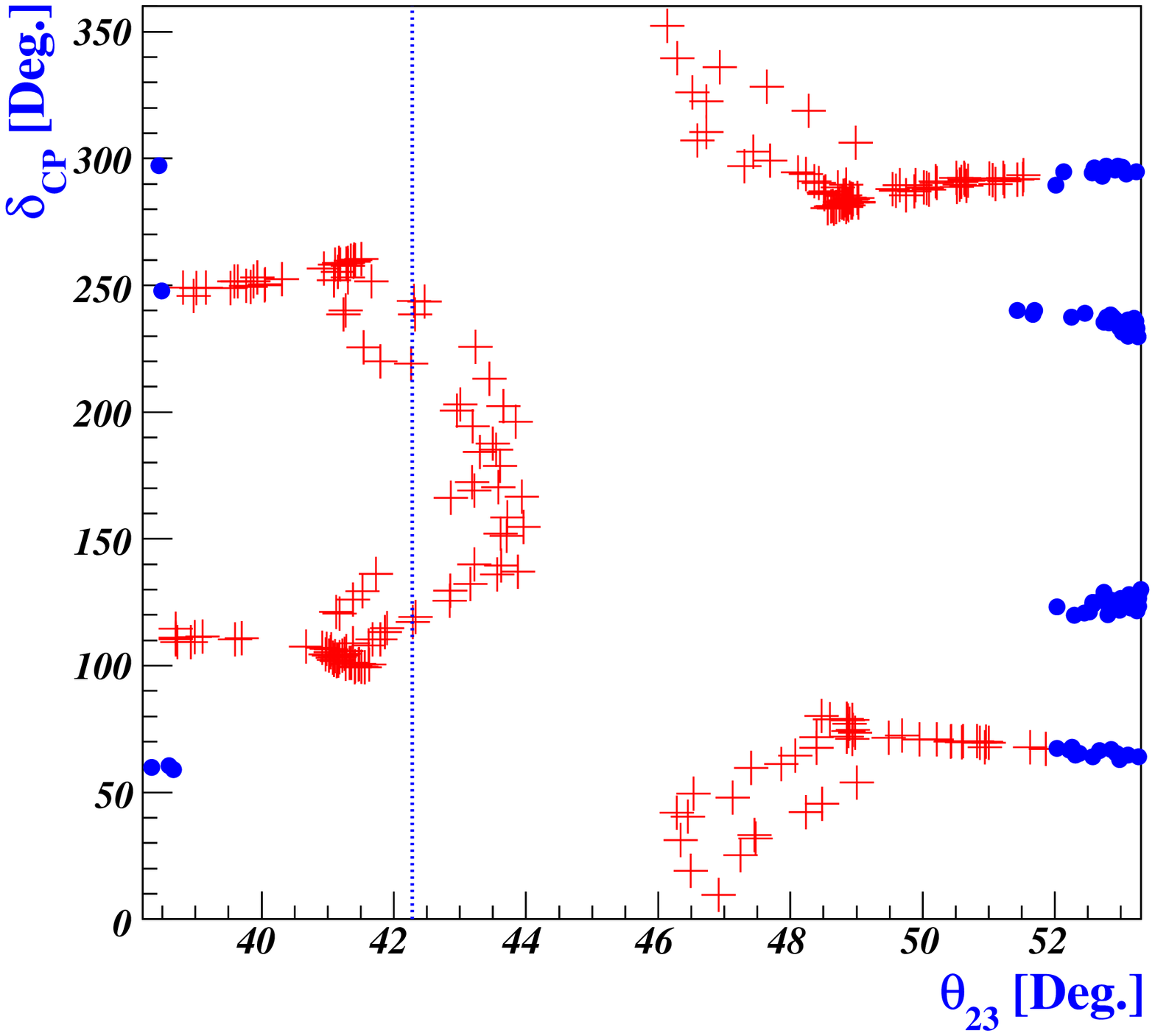,width=7.3cm,angle=0}
\end{minipage}
\hspace*{1.0cm}
\begin{minipage}[h]{7.3cm}
\epsfig{figure=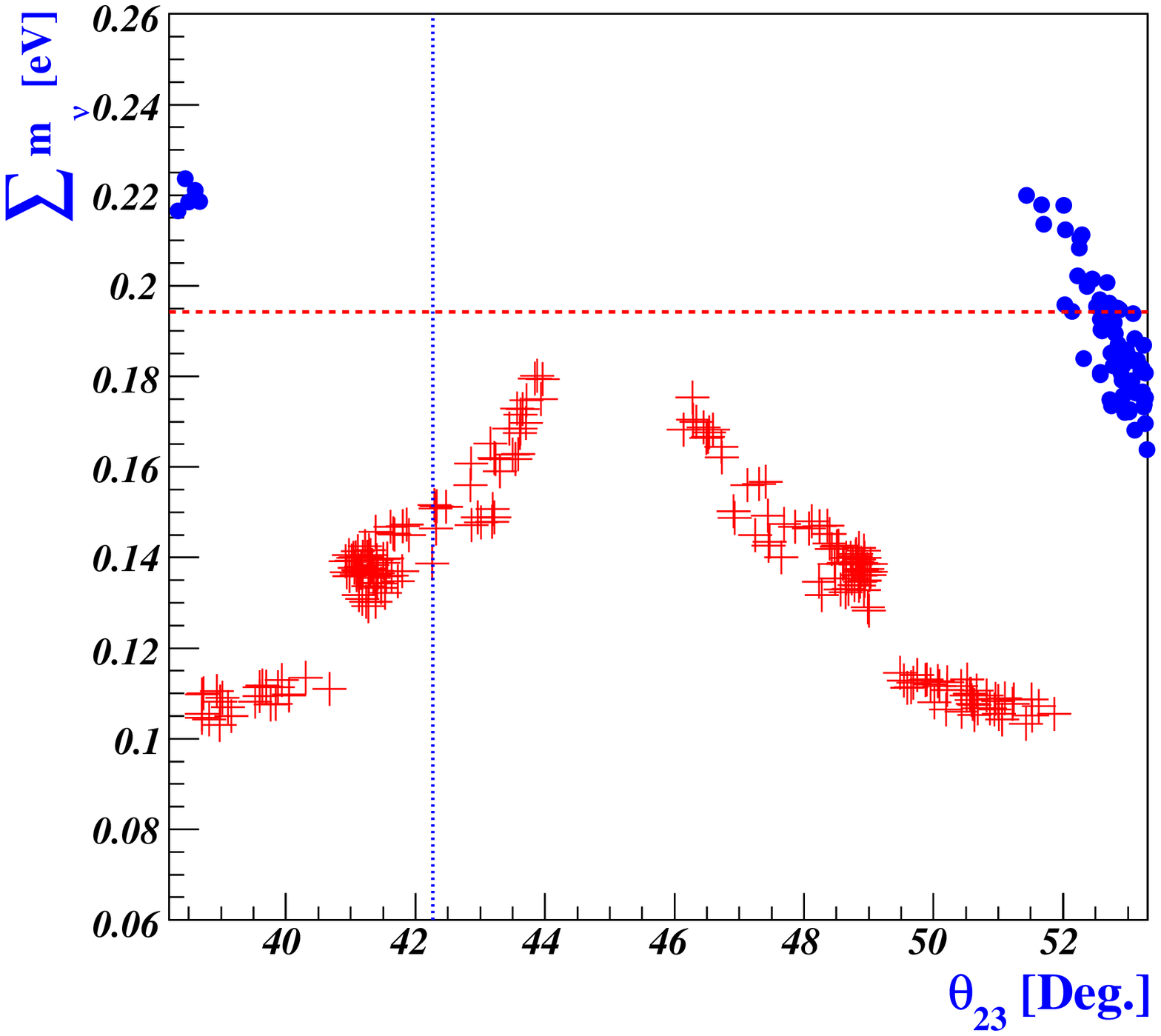,width=7.3cm,angle=0}
\end{minipage}
%\vspace*{-5.5cm}
\caption{\label{FigA3} For NO, the left plot shows predictions of $\delta_{CP}$ as a function of $\theta_{23}$, while the right plot shows predictions of $\sum_{i} m_{\nu_i}$ in terms of $\theta_{23}$. Here, the red crosses and blue dots correspond to $\Delta m^{2}_{1}=\Delta m^{2}_{2}=2.7\times10^{-15}\,{\rm eV}^2<\Delta m^{2}_{3}=5\times10^{-15}\,{\rm eV}^2$ and $\Delta m^{2}_{1}=\Delta m^{2}_{2}=\Delta m^{2}_{3}=5\times10^{-15}\,{\rm eV}^2$, respectively. In both plots the vertical dotted lines indicate the best-fit value for NO, and in the right plot the horizontal dotted line shows the cosmological bounds $\sum_i m_{\nu_i}<0.194$ eV at $95\%$ CL\,\cite{Planck2014}.}
\end{figure}

On the other hand, in case of degenerate mass splittings $\Delta m^2_i=5\times10^{-15}\,{\rm eV}^2$ ($i=1,2,3$), 
we obtain for the NO
 \begin{align}
   &\tilde{\kappa} \in [0.37,0.62] ,
   & \phi \in [99^{\circ},105^{\circ}]\cup[255^{\circ},262^{\circ}]\,,
  \nonumber\\
   & \hat{y}^\nu_{1} \in [1.48,1.88],
   & y_{2} \in [0.79,1.14],\qquad\qquad\quad
   & y_{3} \in [0.84,1.18],
  \label{input1a}
 \end{align}
 leading to $\hat{y}^s_1 \in [1.92,3.01]$, $\hat{y}^s_2 \in [1.95,3.02]$, and $\hat{y}^s_3 \in [2.72,3.33]$;
 for the IO we obtain
 \begin{align}
  &\tilde{\kappa} \in [0.41,0.68] ,
  & \phi \in [102^{\circ},112^{\circ}]\cup[249^{\circ},256^{\circ}]\,,
  \nonumber\\
  & \hat{y}^\nu_{1} \in [1.35,1.54],
  & y_{2} \in [1.08,1.20], \qquad\qquad\quad 
  & y_{3} \in [0.80,1.20],
  \label{input2a}
 \end{align}
 leading to $\hat{y}^s_1 \in [2.75,3.32]$, $\hat{y}^s_2 \in [2.77,3.33]$, and $\hat{y}^s_3 \in [2.03,2.74]$.
%%%%%%%%%%%%%%%
%   Fig A-4   %
%%%%%%%%%%%%%%%
\begin{figure}[h]
%\vspace*{-5.0cm}
%\hspace*{-1cm}
\begin{minipage}[h]{7.3cm}
\epsfig{figure=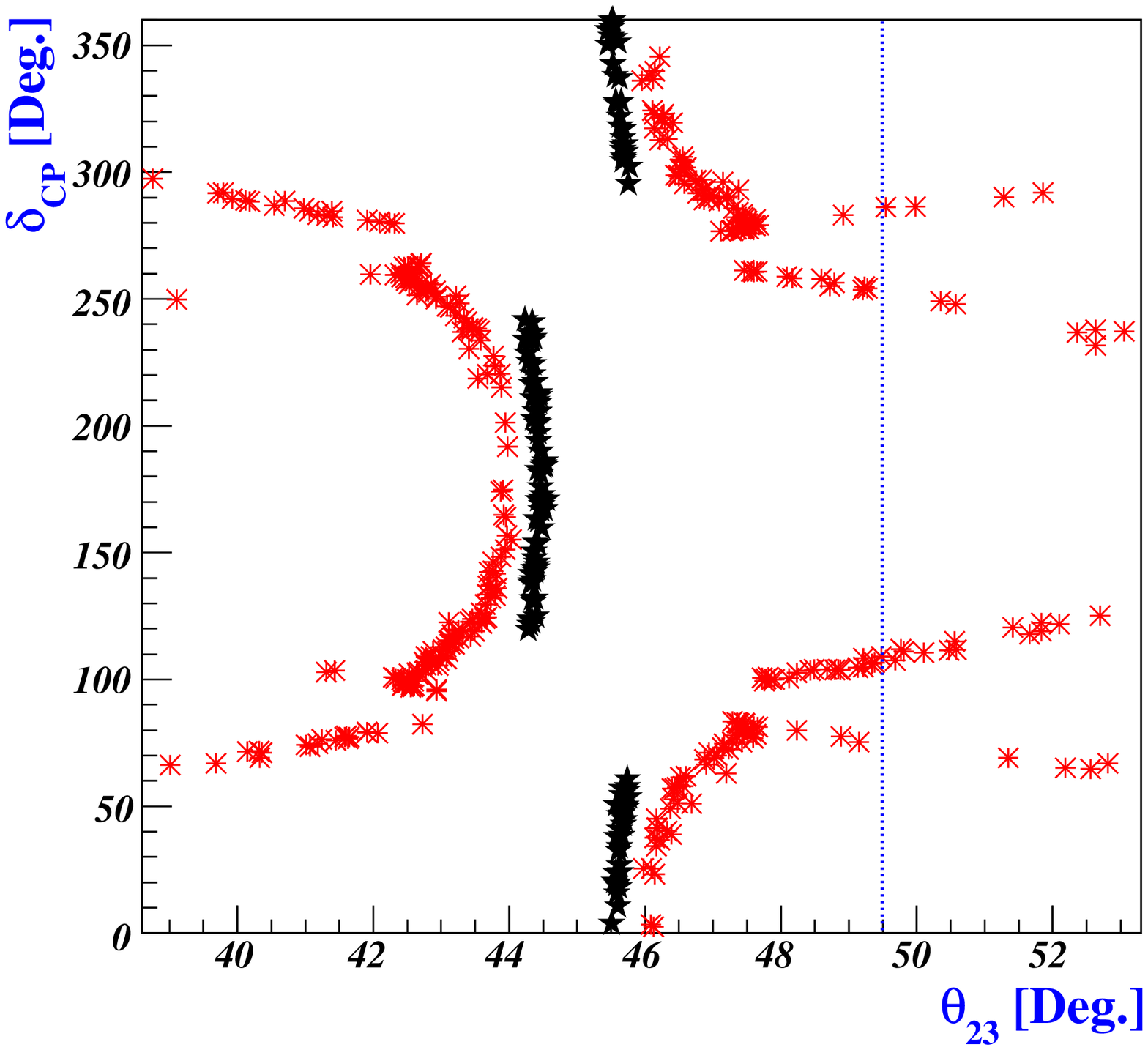,width=7.3cm,angle=0}
\end{minipage}
\hspace*{1.0cm}
\begin{minipage}[h]{7.3cm}
\epsfig{figure=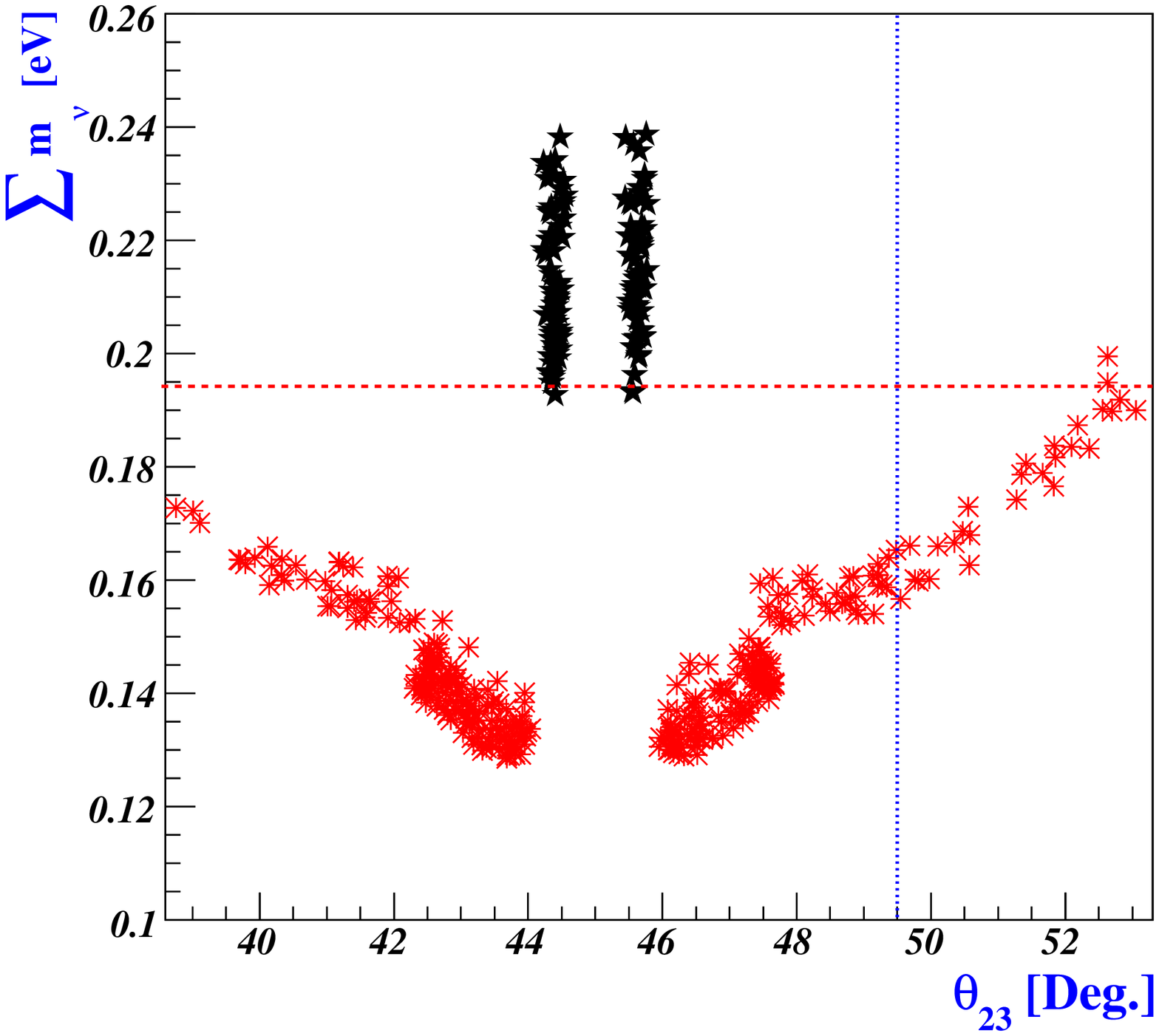,width=7.3cm,angle=0}
\end{minipage}
%\vspace*{-5.5cm}
\caption{\label{FigA4} For IO, the left plot shows predictions of $\delta_{CP}$ as a function of $\theta_{23}$, while the right plot shows predictions of $\sum_{i} m_{\nu_i}$ in terms of $\theta_{23}$. Here, the red asters and black stars correspond to $\Delta m^{2}_{1}=\Delta m^{2}_{2}=10^{-14}\,{\rm eV}^2>\Delta m^{2}_{3}=5.5\times10^{-15}\,{\rm eV}^2$ and $\Delta m^{2}_{1}=\Delta m^{2}_{2}=\Delta m^{2}_{3}=5\times10^{-15}\,{\rm eV}^2$, respectively. In both plots the vertical dotted lines indicate the best-fit value for IO, and in the right plot the horizontal dotted line shows the cosmological bounds $\sum_i m_{\nu_i}<0.194$ eV at $95\%$ CL\,\cite{Planck2014}.}
\end{figure}
The active neutrino oscillation experiments are now on a new step to confirm the CP violation in the lepton sector. Actually, the T2K and NO$\nu$A experiments indicate a finite CP phase\,\cite{finite_CP}. As can be seen in the left side figures of Figs.\,\ref{FigA3} and \ref{FigA4} there is a remarkable behavior correlated between $\delta_{CP}$ and $\theta_{23}$. Thus, accurate measurements of $\theta_{23}$ are crucial for a test of our model.

Figs.\,\ref{FigA3} and \ref{FigA4} show predictions of $\delta_{CP}$ (left plot) and $\sum_i m_{\nu_i}$ (right plot) as a function of the atmospheric mixing angle $\theta_{23}$.
For the hierarchical mass splittings $\Delta m^{2}_{1}=\Delta m^{2}_{2}=2.7\times10^{-15}\,{\rm eV}^2<\Delta m^{2}_{3}=5\times10^{-15}\,{\rm eV}^2$ for NO (red crosses in Fig.\,\ref{FigA3}) and $\Delta m^{2}_{3}=5.5\times10^{-15}\,{\rm eV}^2<\Delta m^{2}_{1}=\Delta m^{2}_{2}=10^{-14}\,{\rm eV}^2$ for IO (red asters in Fig.\,\ref{FigA4}),  the value of $\theta_{23}$ would lie on $|\theta_{23}-45^{\circ}|\sim1^{\circ}-8^{\circ}$, while the values of Dirac CP phase have predictive but wide ranges for both NO and IO. The left plots in Fig.\,\ref{FigA3} and Fig.\,\ref{FigA4} on $\delta_{CP}$ as a function of $\theta_{23}$ predict $\delta_{CP}=220^{\circ}-240^{\circ}$, $120^{\circ}-140^{\circ}$ on the global best-fit $\theta_{23}=42.3^{\circ}$ for NO, and $\delta_{CP}=283^{\circ},250^{\circ},100^{\circ},70^{\circ}$ on $\theta_{23}=49.5^{\circ}$ for IO.
 For the degenerate mass splittings $\Delta m^{2}_{1}=\Delta m^{2}_{2}=\Delta m^{2}_{3}=5\times10^{-15}\,{\rm eV}^2$ the value of $\theta_{23}$ would lie on $|\theta_{23}-45^{\circ}|\sim1^{\circ}$ for IO (black stars in Fig.\,\ref{FigA4}), while $|\theta_{23}-45^{\circ}|\sim7^{\circ}-8^{\circ}$ for NO (blue dots in Fig.\,\ref{FigA3}). Due to the relation $\Delta m^2_k=2m_k\,|\delta_k|$, as the value of $\Delta m^2_k$ decreases up to the bound in Eq.\,(\ref{D_lbound1}) the sum of the light neutrino masses could become lower than the bounds from Planck Collaboration\,\cite{Planck2014}. Hence, future precise measurement on the atmospheric mixing angle $\theta_{23}$ is of importance in order to distinguish between hierarchy and degeneracy of the mass splittings $\Delta m^2_k$ in the model. 
The magnitude of the CP-violating effects is determined by the invariant $J_{CP}$ associated with the Dirac CP-violating phase
 \begin{eqnarray}
J_{CP}\equiv-{\rm Im}[U^{\ast}_{e1}U_{e3}U_{\tau1}U^{\ast}_{\tau3}]=\frac{1}{8}\sin2\theta_{12}\sin2\theta_{13}\sin2\theta_{23}\cos\theta_{13}\sin\delta_{CP}.
 \end{eqnarray}
Here $U_{\alpha j}$ is an element of the PMNS matrix in Eq.\,(\ref{PMNS}), with $\alpha=e,\mu,\tau$
corresponding to the lepton flavors and $j=1,2,3$ corresponding to the light neutrino mass eigenstates.
Due to the precise measurement of $\theta_{13}$, which is relatively large, it may now be possible to put constraints on the Dirac phase $\delta_{CP}$ which will be obtained in the long baseline  neutrino oscillation experiments T2K, NO$\nu$A, etc. (see, Ref.\,\cite{PDG}). However, the current large uncertainty on $\theta_{23}$ is at present limiting the information that can be extracted from the $\nu_{e}$ appearance measurements. Precise measurements of all the mixing angles, especially $\theta_{23}$, are needed to maximize the sensitivity to the leptonic CP violation.

%%%%%%%%%%%%%%%
%   Fig A-6   %
%%%%%%%%%%%%%%%
\begin{figure}[t]
%\vspace*{-5.0cm}
%\hspace*{-1cm}
\begin{minipage}[h]{7.5cm}
\epsfig{figure=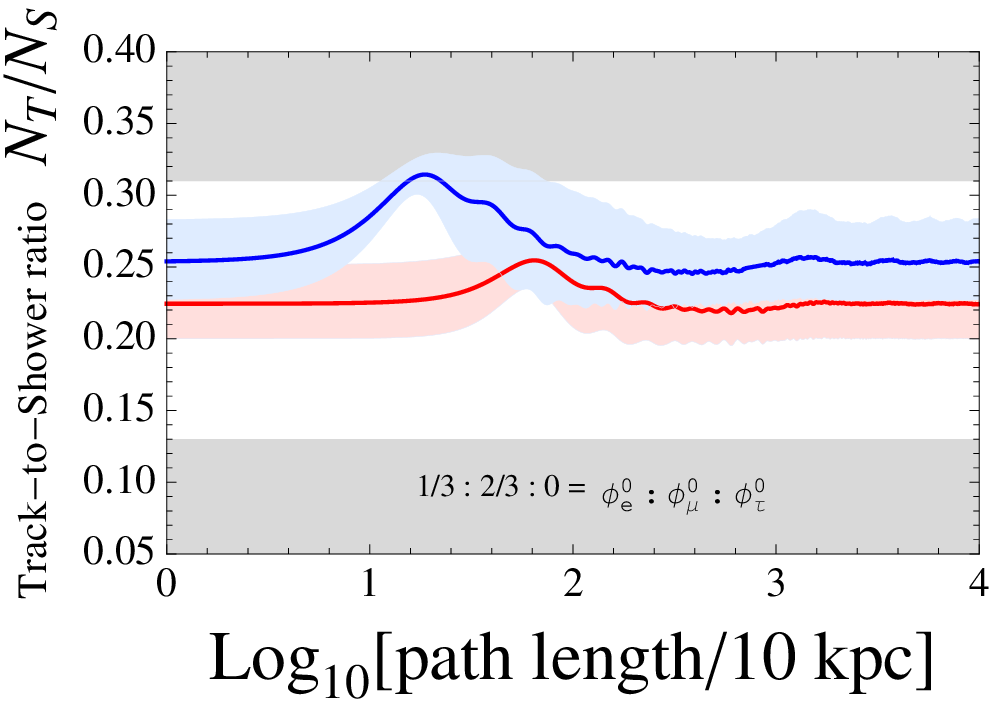,width=8.5cm,angle=0}
\end{minipage}
\hspace*{1.0cm}
\begin{minipage}[h]{7.5cm}
\epsfig{figure=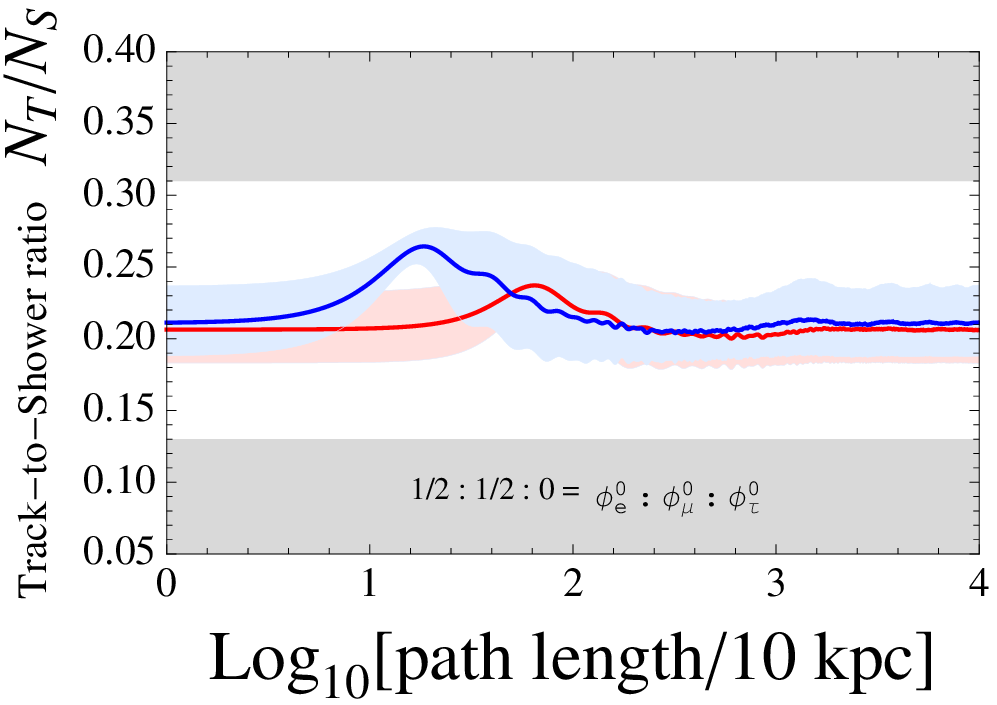,width=8.5cm,angle=0}
\end{minipage}\\
\begin{minipage}[h]{7.5cm}
\epsfig{figure=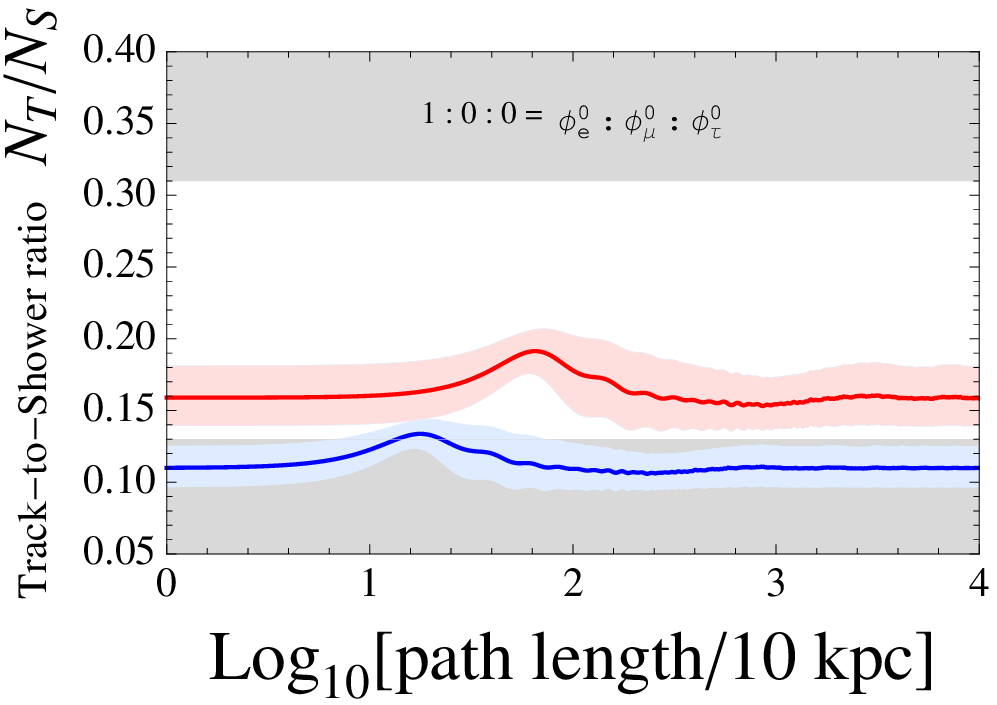,width=8.5cm,angle=0}
\end{minipage}
\hspace*{1.0cm}
\begin{minipage}[h]{7.5cm}
\epsfig{figure=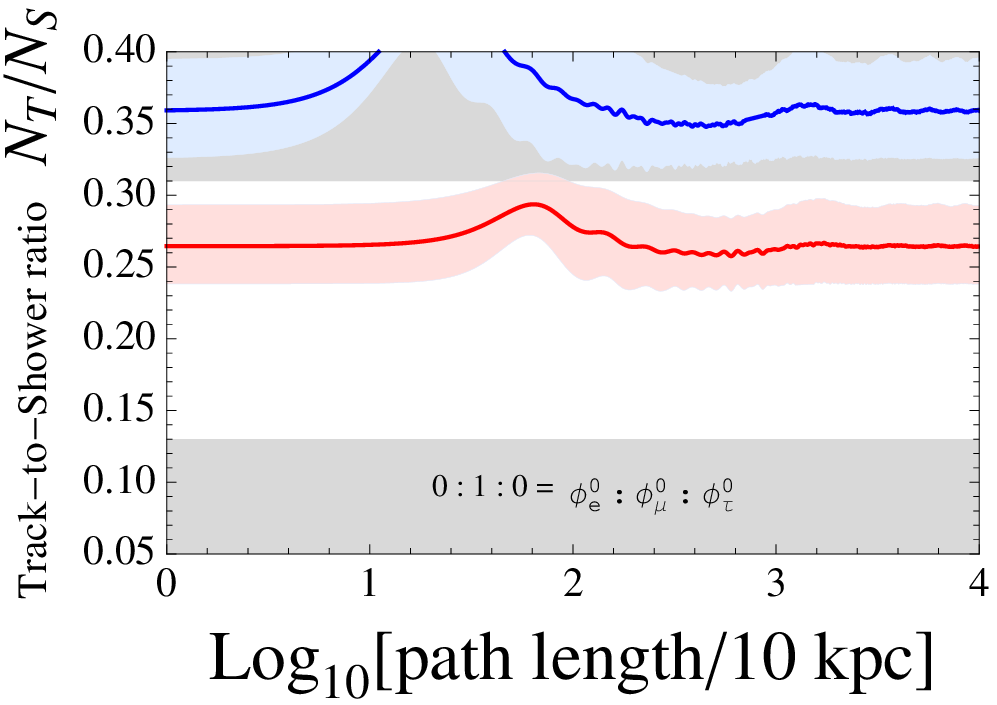,width=8.5cm,angle=0}
\end{minipage}
\caption{\label{FigA6} Plots of the track-to-shower ratio $N_{T}/N_S$ as a function of $L\,(\log_{10}[{\rm path~length/10\,kpc}])$ for NO with $\Delta m^{2}_{1}=\Delta m^{2}_{2}=2.7\times10^{-15}\,{\rm eV}^2<\Delta m^{2}_{3}=5.0\times10^{-15}\,{\rm eV}^2$, and for IO with $\Delta m^{2}_{3}=5.5\times10^{-15}\,{\rm eV}^2<\Delta m^{2}_{1}=\Delta m^{2}_{2}=10^{-14}\,{\rm eV}^2$.
Each panel corresponds to the specific initial flavor composition ($\phi^{0}_e: \phi^0_\mu: \phi^0_\tau$) at the source.
  For three neutrino mixing angles and Dirac-type CP phase, we take the global fit results at $1\sigma$ \,\cite{Gonzalez-Garcia:2015qrr}.
Red and blue curved lines correspond to normal and inverted neutrino mass orderings, respectively, for $\omega=2.2$, whereas light red and light blue regions represent the corresponding results for $\omega=1.8-2.6$.
Gray shaded regions represent the forbidden bound from $N_T/N_S=0.18^{+0.13}_{-0.05}$ in Ref.\,\cite{Palladino:2015zua}.}
\end{figure}

Second, to investigate how large the value of $N_T/N_S$ can be deviated by the oscillatory terms as in Ref.\,\cite{Ahn:2016hhq}, we perform numerical
analysis by taking the values of the neutrino mixing angles and CP phase from the above low energy neutrino oscillation data consistent with the global fit results at $1\sigma$ level\,\cite{Gonzalez-Garcia:2015qrr} as shown in TABLE\,\ref{exp}. Since we are interested in the data consistent with the global fit results at $1\sigma$ level\,\cite{Gonzalez-Garcia:2015qrr}, we take the hierarchical mass splittings in Eqs.\,(\ref{input1}) and  (\ref{input2}) for NO and IO, respectively. Future precise measurement on the atmospheric mixing angle $\theta_{23}$ is of importance in order to distinguish between hierarchy and degeneracy of the mass splittings $\Delta m^2_k$ in the model.   In the limit of large or null mass splitting $\Delta m^2_{k}$, there is no oscillation effects, and thus it is expected that the value of $N_T/N_S$ becomes constant for a given data set of neutrino mixing angles and CP phase.
By using the high energy neutrino events in the IceCube detector which lie in energies between $60$ TeV and $3$ PeV\,\cite{Palladino:2015zua}, Eq.\,(\ref{NTS}) shows directly that track-to-shower ratio $N_T/N_S$ can give a new oscillation curve as a signal dependent on neutrino flight length if the neutrino mixing angles and CP phase, initial flavor composition, and tiny mass splittings are given as inputs.
Our numerical results depend on the initial flavor composition $\phi^0_e:\phi^0_\mu:\phi^0_\tau$ at the source which are relevant for the interpretation of observational data.
We consider the well-known four production mechanisms for high energy neutrinos from which the flavor compositions are given as : (i) $(\frac{1}{3}: \frac{2}{3}: 0)$ for $\pi$ decay, (ii) $(\frac{1}{2}: \frac{1}{2}: 0)$ for charmed mesons decay, (iii) $(1: 0 : 0)$ for $\beta$ decay of neutrons, and (iv) $(0 : 1 : 0)$ for $\pi$ decay with damped muons.
The tiny mass splittings $\Delta m^2_k$ can be searched for, looking at high energy cosmic neutrinos by measuring the track-to-shower ratio $N_T/N_S$ as the function of $L\,(\log_{10}[{\rm path~length}/{\rm 10\,kpc}])$ in Eq.\,(\ref{NTS}).
In the numerical analysis shown by Fig.\,\ref{FigA6}, we use the spectral index given by $\omega=2.2\pm0.4$\,\cite{Aartsen:2013jdh} and the best-fit values for NO (IO) in TABLE\,\ref{exp}.
Fig.\,\ref{FigA6} shows plots of the track-to-shower ratio $N_{T}/N_S$ as a function of $L\,(\log_{10}[{\rm path~length/10\,kpc}])$ with the neutrino energy $60\,{\rm TeV}\lesssim E_\nu\lesssim3\,{\rm PeV}$ studied in Ref.\,\cite{Palladino:2015zua}.
According to four specific assumptions at each panel for the flavor compositions at the source ($\phi^{0}_e: \phi^0_\mu: \phi^0_\tau$), for $\omega=2.2$ the normal mass ordering is presented as the red curved line (for $\Delta m^{2}_{1}=\Delta m^{2}_{2}=2.7\times10^{-15}\,{\rm eV}^2<\Delta m^{2}_{3}=5\times10^{-15}\,{\rm eV}^2$), and the inverted one as the blue curved line (for $\Delta m^{2}_{3}=5.5\times10^{-15}\,{\rm eV}^2<\Delta m^{2}_{1}=\Delta m^{2}_{2}=10^{-14}\,{\rm eV}^2$), respectively, whereas light red and light blue regions represent the corresponding results for $\omega=1.8-2.6$.
Clearly, Fig.\,\ref{FigA6} shows the oscillation peaks occur at distances of 0.65 Mpc and 0.18 Mpc for NO and IO, respectively. In order for the track-to-shower ratio $N_{T}/N_S$ to have the ability to distinguish between NO and IO, much larger detectors than the present IceCube would be required\,\cite{Aartsen:2014njl}. See also similar study in Ref.\,\cite{Ahn:2016hhq}.

%%%%%%%%%%%%%%%%%%%%%%%%%%%%%%%%%%%%%%%%%%%%%%%%%%%%%%%%%%%%%%%%%%%%%%%%%%%%%%%%%%%%%%%%%%
\section{Conclusion}

We have constructed an explicit model for rather recent but fast growing issues of astro-particle physics, encompassing several main issues which are connected to each other: leptonic mixings and CP violation in neutrino oscillation, high-energy neutrinos, QCD axion, and axion cooling of stars. 
The model based on the $SM\times U(1)_X\times A_4$ symmetry has effective physical degree of freedoms: (i) The up-type quark and charged-lepton superpotentials in (\ref{lagrangian1}) and (\ref{lagrangian2}), respectively, does not contribute to the CKM and PMNS mixing matrices due to their diagonal form of mass matrices.  While (ii) the down-type quark superpotential (\ref{lagrangian1}) having six physical parameters including next-to-leading order corrections could explain the four CKM parameters and three down-type quark masses.  And (iii) in neutrino sector there are eight physical degree of freedoms, {\it i.e.}, $m_0, y_2, y_3, \tilde{\kappa}, \phi$, and $\Delta m^2_k$ with $k=1,2,3$. One can reduce the physical degree of freedoms more: once the three pseudo-Dirac mass splittings $\Delta m^2_k$ are fixed by high energy very long wave experiments, such as IceCube, there are only five physical degree of freedoms left in neutrino sector; among nine observables the five measured quantities ($\theta_{12}, \theta_{23}, \theta_{13}, \Delta m^2_{\rm Sol}$, and $\Delta m^2_{\rm Atm}$) are used as constraints, and four quantities could be predicted, see Sec.\,\ref{Nuannu}.  Finally, (iv) in string moduli sector there are eight physical degree of freedoms, {\it i.e.}, three moduli plus two gauge bosons. In the context of supersymmetric moduli stabilization three size moduli and one axionic partner with positive masses are stabilized while leaving two axions massless. Two massive gauge bosons corresponding to gauged $U(1)_{X_i}$ with $i=1,2$ eat the gauged flat two axionic degree of freedoms, leaving behind low energy global $U(1)_{X_i}$ symmetries so that the two axionic directions survive to low energies as the flavored PQ axions.

The model has the following desirable features, in that such flavored-PQ supersymmetric model can be testable in the very near future through on-going experiments for neutrino oscillation, $0\nu\beta\beta$ decay, axion, and IceCube searches for neutrinos:\\
{\bf (i)} The anomalous global $U(1)_X$, which originates from the broken gauged $U(1)_X$ symmetry, see Sec.\,\ref{qla1}, is introduced as a fundamental symmetry in nature in a way that the mixed $U(1)_X$-$[{gravity}]^2$ anomaly is free. Especially, such anomaly free condition together with the observed mass hierarchies of the SM charged fermions demands additional sterile neutrinos; the $U(1)_X$ quantum numbers of the SM quarks are assigned in a way that no axionic domain-wall problem occurs, implying that flavor structure of the SM may be correlated to axionic domain-wall. Such additional sterile neutrinos play the role of a realization of baryogenesis via a new Affleck-Dine leptogenesis\,\cite{Ahn:2016hbn}.
The spontaneous breakdown of the automatic flavored $U(1)_X$ symmetry together with the GS mechanism produces\,\footnote{Here ``automatic" means that the quantum number assignment of $U(1)_X$ is determined by hierarchies of the SM fermions in a way that no axionic domain-wall problem occurs if the $X$-symmetry breakdown occurs after inflation.} NG modes, $A_{1,2}$, (and QCD axion $A$) whose decay constants are fixed by several astrophysical constraints\,\cite{Keller:2012yr, wd_recent2, Raffelt:2011ft, WD01, Ayala:2014pea, Sedrakian:2015krq, Leinson:2014ioa, Bertolami:2014wua, Raffelt:1985nj, Viaux:2013lha, Redondo:2013wwa}. %enough to solve the strong CP problem as well as the flavor problems associated with the mass hierarchies of quarks and leptons including their peculiar mixing patterns.
Then the flavored PQ symmetry $U(1)_X$ embedded in the non-Abelian $A_4$ finite group could economically explain the mass hierarchies of quarks and leptons including their peculiar mixing patterns as well as provide a neat solution to the strong CP problem and its resulting axion.
 Such flavored PQ symmetry breakdown leads to two Majorana neutrino mass scales of order much larger and smaller than the electroweak symmetry breaking scale according to the $A_4\times U(1)_X$ symmetry. And the NG modes couple very weakly to both active and sterile neutrinos, so that they are not in thermal equilibrium with the neutrinos during nucleosynthesis.
Interestingly enough, since the NG mode $A_2$ interacts with electrons at tree level, indeed, the NG mode $A_2$ emitted from the dense interior of WDs play a crucial role in direct searches as fundamental particles, see Eq.\,(\ref{wd_bound}); the bound of the QCD axion mass could be inferred from such astrophysical considerations on star coolings since the QCD axion emission causes energy loss $\sim1/F_A$ affecting crucial stellar evolution, see Eq.\,(\ref{a_nucleon01}). Interestingly, we found that the QCD axion decay constant is shown to be located at $F_A=1.30^{+0.66}_{-0.54}\times10^{9}$ GeV. Consequently, we have shown model predictions on the axion mass $m_a=4.34^{+3.37}_{-1.49}$ meV and  the axion coupling to photon $|g_{a\gamma\gamma}|=1.30^{+1.01}_{-0.45}\times10^{-12}\,{\rm GeV}^{-1}$. In turn, the square of the ratio between them is shown to be located just a bit lower than that of the conventional KSVZ model as shown in Fig.\,\ref{Fig2}.

{\bf (ii)} We have shown that, after the symmetry breakdown, the active neutrino masses are achieved by pseudo-Dirac mass scheme, and which are determined in a completely independent way that the active neutrino mixing angles are obtained through the seesaw framework. But they are linked each other through astronomical mass splittings responsible for new wavelength oscillations characterized by the mass squared differences of the light neutrino pairs, $\Delta m^2$. So in this framework which leads to pairs of almost degenerate neutrinos, the pseudo-Dirac mass splittings $|\delta_k|$ as eigenvalues of the seesaw formula are much smaller than the active neutrino masses. Such mass splittings $\Delta m^2$ are well constrained by the low energy neutrino oscillation data, the BBN constraints on the effective number of species of light particles during nucleosynthesis, and a leptogenesis scenario Ref.\,\cite{Ahn:2016hbn}. Since the mass eigenstates in each pseudo-Dirac pair have opposite CP parity, $0\nu\beta\beta$-decay rate is expected to be $\lesssim{\cal O}(\delta_k)$ eV, which might not be observable in the near future. 
Once the mass splittings $\Delta m^2$ are fixed by astronomical-scale baseline experiments, such as IceCube\,\cite{ice-cube}, the active neutrino mixing angles ($\theta_{12}, \theta_{13}, \theta_{23}, \delta_{CP}$)\,\cite{PDG} and the pseudo-Dirac mass splittings $|\delta_k|$ are well constrained since stars are employed to place constraints on the decay constant of the NG mode $A_2$ (QCD axion $A$) through the $A_2$ interaction to electron (the $A$ interaction to photon and neutron). We have investigated how neutrino oscillations at low energies could be connected to new oscillations available on high energy neutrinos, connected by a new Affleck-Dine leptogenesis scenario in Ref.\,\cite{Ahn:2016hbn}.

On phenomenological examples, taking specific but realistic mass splittings $\Delta m^2_k$ for normal mass ordering (NO) and inverted one (IO), we have examined leptonic CP violation and the sum of the light neutrino masses as a function of the atmospheric mixing angle $\theta_{23}$: Figs.\,\ref{FigA3} and \ref{FigA4} show the main results. Future precise measurement on the atmospheric mixing angle $\theta_{23}$ is of importance in order to distinguish between hierarchy and degeneracy of the mass splittings $\Delta m^2_k$ in the model. 
For the hierarchical mass splittings $\Delta m^{2}_{1}=\Delta m^{2}_{2}=2.7\times10^{-15}\,{\rm eV}^2<\Delta m^{2}_{3}=5\times10^{-15}\,{\rm eV}^2$ for NO (red crosses in Fig.\,\ref{FigA3}) and $\Delta m^{2}_{3}=5.5\times10^{-15}\,{\rm eV}^2<\Delta m^{2}_{1}=\Delta m^{2}_{2}=10^{-14}\,{\rm eV}^2$ for IO (red asters in Fig.\,\ref{FigA4}),  the value of $\theta_{23}$ would lie on $|\theta_{23}-45^{\circ}|\sim1^{\circ}-8^{\circ}$, while the values of Dirac CP phase have predictive but wide ranges for both NO and IO. Especially, the left plots in Fig.\,\ref{FigA3} and Fig.\,\ref{FigA4} on $\delta_{CP}$ as a function of $\theta_{23}$ predict $\delta_{CP}=220^{\circ}-240^{\circ}$, $120^{\circ}-140^{\circ}$ on the global best-fit $\theta_{23}=42.3^{\circ}$ for NO, and $\delta_{CP}=283^{\circ},250^{\circ},100^{\circ},70^{\circ}$ on $\theta_{23}=49.5^{\circ}$ for IO.
 For the degenerate mass splittings $\Delta m^{2}_{1}=\Delta m^{2}_{2}=\Delta m^{2}_{3}=5\times10^{-15}\,{\rm eV}^2$ the value of $\theta_{23}$ would lie on $|\theta_{23}-45^{\circ}|\sim1^{\circ}$ for IO (black stars in Fig.\,\ref{FigA4}), while $|\theta_{23}-45^{\circ}|\sim7^{\circ}-8^{\circ}$ for NO (blue dots in Fig.\,\ref{FigA3}). Due to the relation $\Delta m^2_k=2m_k\,|\delta_k|$, as the value of $\Delta m^2_k$ decreases up to the bound in Eq.\,(\ref{D_lbound1}) the sum of the light neutrino masses could become lower than the bounds from Planck Collaboration\,\cite{Planck2014}. Hence, future precise measurement on the atmospheric mixing angle $\theta_{23}$ is of importance in order to distinguish between hierarchy and degeneracy of the mass splittings $\Delta m^2_k$ in the model.
Moreover, by using the high energy neutrino events in the IceCube detector which lie in energies between 60 TeV and 3 PeV, we have plotted the track-to-shower ratio $N_{T}/N_S$ as a function of flight length $L\,(\log_{10}[{\rm path~length/10\,kpc}])$ and found the new oscillation peaks as signals at distances of 0.65 Mpc and 0.18 Mpc for NO and IO, respectively, when the best-fit values for NO(IO) in TABLE\,\ref{exp} and the given hierarchical mass splittings for NO(IO) are given as inputs. In order for the track-to-shower ratio $N_{T}/N_S$ to have the ability to distinguish between NO and IO, much larger detectors than the present IceCube would be required\,\cite{Aartsen:2014njl}.
Although it is a little bit hard to confirm the tiny pseudo-Dirac mass splittings, it can be tested indirectly. 
A crucial observation here is that such tiny mass splittings together with the sum of neutrino masses obtained from cosmological constraints suggest a high predictability of very long wavelength oscillations as well as the non-observational $0\nu\beta\beta$ decay rate.

{\bf (iii)} Under the gauged $U(1)_X\equiv U(1)_{X_1}\times U(1)_{X_2}$ symmetry, the string theoretic axions, vector fields, and Kahler moduli participate in the four-dimensional GS mechanism. The string theoretic QCD axions originate from antisymmetric tensor gauge fields in compactified string theory, with the string theoretic axion decay constants depending on the Kahler metric. Since the three moduli all appear in the Kahler potential, the three size moduli and one axionic partner with positive masses are stabilized, while leaving two axions massless, through non-perturbative superpotentials\,\cite{Ahn:2016typ}. 
The two gauged anomalous $U(1)$ symmetries have the mixed $U(1)_{X}$-$[SU(3)_C]^2$, $U(1)_{X}$-$[SU(2)]^2$, $U(1)_{X}$-$[U(1)_Y]^2$, and $U(1)_{Y}$-$[U(1)_X]^2$ anomalies which are cancelled by the GS mechanism, where the gauged anomalous $U(1)_X$ mixes with the axionic moduli and which in turn couples to a multiple of the QCD instanton density.  
The two axionic directions are gauged by the $U(1)$ gauge interactions associated with $D$-branes, and the gauged flat directions of the $F$-term potential are removed through the Stuckelberg mechanism. Below the mass scale of heavy gauge boson the gauge bosons decouple, leaving behind low energy symmetries which are anomalous global $U(1)_{X}$. In such a way, the QCD axion decay constant could be much lower than the scale of moduli stabilization when the matter fields charged under the global anomalous $U(1)_X$ get VEVs induced by tachyonic SUSY breaking masses. One linear combination of the global $U(1)_{X_i}$ is broken explicitly by instantons, and such would-be QCD axions play crucial role in evolution of stars and solving the strong CP problem.

%\newpage

\appendix
%%%%%%%%%%%%%%%%%%%%%%%%%%%%%%%%%%%%%%%%%%%%%%%%%%%%%%%%%%%%%%%%%%%%%%%%%%%%%%%%%%%%%%%%%%
\section{The $A_{4}$ Group}
 \label{A4group}
The  group $A_{4}$ is the symmetry group of the
tetrahedron, isomorphic to the finite group of the even permutations of four
objects. The group $A_{4}$ has two generators, denoted $S$ and $T$, satisfying the relations $S^{2}=T^{3}=(ST)^{3}=1$. In the three-dimensional complex representation, $S$ and $T$ are given by
 \begin{eqnarray}
 S=\frac{1}{3} \, {\left(\begin{array}{ccc}
 -1 &  2 &  2 \\
 2 &  -1 & 2 \\
 2 &  2 &  -1
 \end{array}\right)}~,\qquad T={\left(\begin{array}{ccc}
 1 &  0 &  0 \\
 0 &  \omega &  0 \\
 0 &  0 &  \omega^2
 \end{array}\right)}~.
 \label{generator}
 \end{eqnarray}
$A_{4}$ has four irreducible representations: one triplet ${\bf
3}$ and three singlets ${\bf 1}, {\bf 1}', {\bf 1}''$.
An $A_4$ singlet $a$ is invariant under the action of $S$ ($Sa=a$), while the action of $T$ produces $Ta=a$ for ${\bf 1}$, $Ta=\omega a$ for ${\bf 1}'$, and $Ta=\omega^2 a$ for ${\bf 1}''$, where $\omega=e^{i2\pi/3}=-1/2+i\sqrt{3}/2$ is a complex cubic-root of unity.
Products of two $A_4$ representations decompose into irreducible representations according to the following multiplication rules: ${\bf 3}\otimes{\bf 3}={\bf 3}_{s}\oplus{\bf
3}_{a}\oplus{\bf 1}\oplus{\bf 1}'\oplus{\bf 1}''$, ${\bf
1}'\otimes{\bf 1}''={\bf 1}$, ${\bf 1}'\otimes{\bf 1}'={\bf 1}''$
and ${\bf 1}''\otimes{\bf 1}''={\bf 1}'$. Explicitly, if $(a_{1},
a_{2}, a_{3})$ and $(b_{1}, b_{2}, b_{3})$ denote two $A_4$ triplets, then we have Eq.\,(\ref{A4reps}).
% \begin{eqnarray}
%  (a\otimes b)_{{\bf 3}_{\rm s}} &=& \frac{1}{\sqrt{3}}(2a_{1}b_{1}-a_{2}b_{3}-a_{3}b_{2}, 2a_{3}b_{3}-a_{2}b_{1}-a_{1}b_{2}, 2a_{2}b_{2}-a_{3}b_{1}-a_{1}b_{3})~,\nonumber\\
%  (a\otimes b)_{{\bf 3}_{\rm a}} &=& i\,(a_{3}b_{2}-a_{2}b_{3}, a_{2}b_{1}-a_{1}b_{2}, a_{1}b_{3}-a_{3}b_{1})~,\nonumber\\
%  (a\otimes b)_{{\bf 1}} &=& a_{1}b_{1}+a_{2}b_{3}+a_{3}b_{2}~,\nonumber\\
%  (a\otimes b)_{{\bf 1}'} &=& a_{1}b_{2}+a_{2}b_{1}+a_{3}b_{3}~,\nonumber\\
%  (a\otimes b)_{{\bf 1}''} &=& a_{1}b_{3}+a_{2}b_{2}+a_{3}b_{1}~.
% \end{eqnarray}

To make the presentation of our model physically more transparent, we define the $T$-flavor quantum number $T_f$ through the eigenvalues of the operator $T$, for which $T^3=1$. In detail, we say that a field $f$ has $T$-flavor $T_f=0$, +1, or -1 when it is an eigenfield of the $T$ operator with eigenvalue $1$, $\omega$, $\omega^2$, respectively (in short, with eigenvalue $\omega^{T_f}$ for $T$-flavor $T_f$, considering the cyclical properties of the cubic root of unity $\omega$). The $T$-flavor is an additive quantum number modulo 3. We also define the $S$-flavor-parity through the eigenvalues of the operator $S$, which are +1 and -1 since $S^2=1$, and we speak of $S$-flavor-even and $S$-flavor-odd fields.
For $A_4$-singlets, which are all $S$-flavor-even, the $\mathbf{1}$ representation has no $T$-flavor ($T_f=0$), the $\mathbf{1}'$ representation has $T$-flavor $T_f=+1$, and the $\mathbf{1}''$ representation has $T$-flavor $T_f=-1$. Since for $A_4$-triplets, the operators $S$ and $T$ do not commute, $A_4$-triplet fields cannot simultaneously have a definite $T$-flavor and a definite $S$-flavor-parity.

The real representation, in which $S$ is diagonal, is obtained through the unitary transformation
\begin{align}
A \to A'=U_{\omega}\,A\,U^{\dag}_{\omega},
\end{align}
where $A$ is any $A_4$ matrix in the real representation and
\begin{align}
U_{\omega}=\frac{1}{\sqrt{3}}{\left(\begin{array}{ccc}
 1 &  1 &  1 \\
 1 & \omega & \omega^{2} \\
 1 & \omega^{2} & \omega
 \end{array}\right)}.
 \label{eq:Uomega}
\end{align}
We have
 \begin{eqnarray}
 S'={\left(\begin{array}{ccc}
 1 &  0 &  0 \\
 0 &  -1 & 0 \\
 0 &  0 &  -1
 \end{array}\right)}~,\qquad T'={\left(\begin{array}{ccc}
 0 &  1 &  0 \\
 0 &  0 &  1 \\
 1 &  0 &  0
 \end{array}\right)}~.
 \label{generator2}
 \end{eqnarray}
For reference, an $A_4$ triplet field with $T$-flavor eigenfields $(a_1,a_2,a_3)$ in the complex representation can be expressed in terms of components $(a_{R1},a_{R2},a_{R3})$ as
\begin{align}
 a_{1R} = \frac{a_{1}+a_{2}+a_{3}}{\sqrt{3}} , \quad
 a_{2R} = \frac{a_{1}+\omega\,a_{2}+\omega^2 a_{3}}{\sqrt{3}} , \quad
 a_{3R} = \frac{a_{1}+\omega^2 a_{2}+\omega\,a_{3}}{\sqrt{3}} .
 \label{eq:Ua1}
\end{align}
Inversely,
 \begin{align}
 a_{1}  = \frac{a_{1R}+a_{2R}+a_{3R}}{\sqrt{3}} , \quad
 a_{2}  = \frac{a_{1R}+\omega^2 a_{2R}+\omega \,a_{3R}}{\sqrt{3}} , \quad
 a_{3}  = \frac{a_{1R}+\omega \,a_{2R}+\omega^2 a_{3R}}{\sqrt{3}} .
 \label{eq:Ua2}
 \end{align}
Now, in the $S$ diagonal basis the product rules of two triplets $(a_{R1},a_{R2},a_{R3})$ and $(b_{R1},b_{R2},b_{R3})$ according to ${\bf 3}\otimes{\bf 3}={\bf 3}_{s}\oplus{\bf
3}_{a}\oplus{\bf 1}\oplus{\bf 1}'\oplus{\bf 1}''$ are as follows
 \begin{eqnarray}
  (a_R\otimes b_R)_{{\bf 3}_{\rm s}} &=& (a_{2R}\,b_{3R}+a_{3R}\,b_{2R}, \,a_{3R}\,b_{1R}+a_{1R}\,b_{3R}, \,a_{1R}\,b_{2R}+a_{2R}\,b_{1R})~,\nonumber\\
  (a_R\otimes b_R)_{{\bf 3}_{\rm a}} &=& (a_{2R}\,b_{3R}-a_{3R}\,b_{2R}, \,a_{3R}\,b_{1R}-a_{1R}\,b_{3R}, \,a_{1R}\,b_{2R}-a_{2R}\,b_{1R})~,\nonumber\\
  (a_R\otimes b_R)_{{\bf 1}} &=& a_{1R}\,b_{1R}+a_{2R}\,b_{2R}+a_{3R}\,b_{3R}~,\nonumber\\
  (a_R\otimes b_R)_{{\bf 1}'} &=& a_{1R}\,b_{1R}+\omega^{2} a_{2R}\,b_{2R}+\omega\, a_{3R}\,b_{3R}~,\nonumber\\
  (a_R\otimes b_R)_{{\bf 1}''} &=& a_{1R}\,b_{1R}+\omega\,a_{2R}\,b_{2R}+\omega^{2} a_{3R}\,b_{3R}~.
 \end{eqnarray}

%%%%%%%%%%%%%%%%%%%%%%%%%%%%%%%%%%%%%%%%%%%%%%%%%%%%%%%%%%%%%%%%%%%%%%%%%%%%%%%%%%%%%%%%%%
\section{Vacuum configuration}
\subsection{Vacuum configuration for the flavon fields}
 \label{flavon}
We review the vacuum configuration shown in Ref.\,\cite{Ahn:2014gva}. Indeed, the VEV pattern of the flavons is determined dynamically, in which the vacuum alignment problem can be solved by the supersymmetric driving field method in Ref.\,\cite{Altarelli:2005yp}\,\footnote{There is another generic way for the vacuum alignment problem by extending the model with a spacial extra dimension\,\cite{Altarelli:2005yp}.}. In order to make a non-trivial scalar potential in the SUSY breaking sector, we introduce driving fields $\Phi^{T}_{0},\Phi^{S}_{0},\Theta_{0},\Psi_{0}$ whose have the representation of $A_{4}\times U(1)_{X}$ as in TABLE\,\ref{DrivingRef}.
The leading order superpotential dependent on the driving fields, which is invariant under the flavor symmetry $A_{4}\times U(1)_{X}$, is given by the superpotential\,(\ref{potential}).
In the SUSY limit, the vacuum configuration is obtained by the $F$-terms of all fields being required to vanish. The vacuum alignment of the flavon $\Phi_{T}$ is determined by
 \begin{eqnarray}
 \frac{\partial W_{v}}{\partial\Phi^{T}_{01}}&=&\tilde{\mu}\,\Phi_{T1}+\frac{2\tilde{g}}{\sqrt{3}}\left(\Phi^{2}_{T1}-\Phi_{T2}\Phi_{T3}\right)=0~,\nonumber\\
 \frac{\partial W_{v}}{\partial\Phi^{T}_{02}}&=&\tilde{\mu}\,\Phi_{T3}+\frac{2\tilde{g}}{\sqrt{3}}\left(\Phi^{2}_{T2}-\Phi_{T1}\Phi_{T3}\right)=0~,\nonumber\\
 \frac{\partial W_{v}}{\partial\Phi^{T}_{03}}&=&\tilde{\mu}\,\Phi_{T2}+\frac{2\tilde{g}}{\sqrt{3}}\left(\Phi^{2}_{T3}-\Phi_{T1}\Phi_{T2}\right)=0~.
 \label{potential1}
 \end{eqnarray}
From this set of three equations, we can obtain the supersymmetric vacuum for $\Phi_T$,
 \begin{eqnarray}
  \langle\Phi_T\rangle=\frac{v_T}{\sqrt{2}}(1, 0, 0)\qquad\text{with}~v_T=-\frac{\tilde{\mu}}{\tilde{g}}\sqrt{\frac{3}{2}}\,,
 \end{eqnarray}
where $\tilde{g}$ is a dimensionless coupling.
And the minimization equations for the vacuum configuration of $\Phi_{S}$ and $(\Theta,\tilde{\Theta})$ are given by
 \begin{eqnarray}
 \frac{\partial W_{v}}{\partial\Phi^{S}_{01}}&=&\frac{2g_{1}}{\sqrt{3}}\left(\Phi_{S1}\Phi_{S1}-\Phi_{S2}\Phi_{S3}\right)+g_{2}\Phi_{S1}\tilde{\Theta}=0~,\nonumber\\
 \frac{\partial W_{v}}{\partial\Phi^{S}_{02}}&=&\frac{2g_{1}}{\sqrt{3}}\left(\Phi_{S2}\Phi_{S2}-\Phi_{S1}\Phi_{S3}\right)+g_{2}\Phi_{S3}\tilde{\Theta}=0~,\nonumber\\
 \frac{\partial W_{v}}{\partial\Phi^{S}_{03}}&=&\frac{2g_{1}}{\sqrt{3}}\left(\Phi_{S3}\Phi_{S3}-\Phi_{1}\Phi_{S2}\right)+g_{2}\Phi_{S2}\tilde{\Theta}=0~,\nonumber\\
 \frac{\partial W_{v}}{\partial\Theta_{0}}&=&g_{3}\left(\Phi_{S1}\Phi_{S1}+2\Phi_{S2}\Phi_{S3}\right)+g_{4}\Theta^{2}+g_{5}\Theta\tilde{\Theta}+g_{6}\tilde{\Theta}^{2}=0~.
 \label{potential2}
 \end{eqnarray}
From the above four equations, we can get the supersymmetric vacua for the fields $\Phi_S$, $\Theta$, $\tilde{\Theta}$,
\begin{eqnarray}
 \langle\Phi_S\rangle=\frac{v_S}{\sqrt{2}}(1, 1, 1)\,,\quad\langle\Theta\rangle=\frac{v_\Theta}{\sqrt{2}}\,,\quad\langle\tilde{\Theta}\rangle=0\,,\qquad\text{with}~v_\Theta=v_S\sqrt{-3\frac{g_3}{g_4}}\,,
\end{eqnarray}
where $v_\Theta$ is undetermined, and the VEVs $v_\Theta$ and $v_S$ are naturally of the same order of magnitude (here, the dimensionless parameters $g_3$ and $g_4$ are the same order of magnitude.).
Finally, the minimization equation for the vacuum configuration of $\Psi$ ($\tilde{\Psi}$) is given by
 \begin{eqnarray}
 \frac{\partial W_{v}}{\partial\Psi_{0}}&=&g_{7}(\Psi\tilde{\Psi}-\mu^{2}_{\Psi})=0~,
 \label{potential3}
 \end{eqnarray}
where $\mu_{\Psi}$ is the $U(1)_{X}$ breaking scale and $g_{7}$ is a dimensionless coupling. From the above equation we obtain the supersymmetric vacua for the fields $\Psi$ and $\tilde{\Psi}$
\begin{eqnarray}
 \langle\Psi\rangle=\langle\tilde{\Psi}\rangle=\frac{v_\Psi}{\sqrt{2}}\,,\qquad\text{with}~v_\Psi=\mu_\Psi\sqrt{2}\,.
\end{eqnarray}

%%%%%%%%%%%%%%%%%%%%%%%%%%%%%%%%%%%%%%%%%%%%%%
\subsection{Vacuum configuration for the driving fields}
 \label{driving}
From the vanishing of the F-terms associated to the flavons, the vacuum configuration of the driving fields $\Phi^{T}_{0},\Phi^{S}_{0},\Theta_{0},\Psi_{0}$ are determined by
 \begin{eqnarray}
  \frac{\partial W_{v}}{\partial\Phi_{T1}}&=&\frac{2\tilde{g}}{\sqrt{3}}\left(2\Phi_{T1}\Phi^{T}_{01}-\Phi_{T2}\Phi^{T}_{03}-\Phi_{T3}\Phi^{T}_{02}\right)+\tilde{\mu}\Phi^{T}_{01}=0\,,\nonumber\\
  \frac{\partial W_{v}}{\partial\Phi_{T2}}&=&\frac{2\tilde{g}}{\sqrt{3}}\left(2\Phi_{T2}\Phi^{T}_{02}-\Phi_{T3}\Phi^{T}_{01}-\Phi_{T1}\Phi^{T}_{03}\right)+\tilde{\mu}\Phi^{T}_{03}=0\,,\nonumber\\
  \frac{\partial W_{v}}{\partial\Phi_{T3}}&=&\frac{2\tilde{g}}{\sqrt{3}}\left(2\Phi_{T3}\Phi^{T}_{03}-\Phi_{T2}\Phi^{T}_{01}-\Phi_{T1}\Phi^{T}_{02}\right)+\tilde{\mu}\Phi^{T}_{02}=0\,,
 \end{eqnarray}
 \begin{eqnarray}
  \frac{\partial W_{v}}{\partial\Phi_{S1}}&=&\frac{2g_{1}}{\sqrt{3}}\left(2\Phi_{S1}\Phi^{S}_{01}-\Phi_{S2}\Phi^{S}_{03}-\Phi_{S3}\Phi^{S}_{02}\right)+g_{2}\Phi^{S}_{01}\tilde{\Theta}+2g_{3}\Phi_{S1}\Theta_{0}=0\,,\nonumber\\
  \frac{\partial W_{v}}{\partial\Phi_{S2}}&=&\frac{2g_{1}}{\sqrt{3}}\left(2\Phi_{S2}\Phi^{S}_{02}-\Phi_{S3}\Phi^{S}_{01}-\Phi_{S1}\Phi^{S}_{03}\right)+g_{2}\Phi^{S}_{03}\tilde{\Theta}+2g_{3}\Phi_{S3}\Theta_{0}=0\,,\nonumber\\
  \frac{\partial W_{v}}{\partial\Phi_{S3}}&=&\frac{2g_{1}}{\sqrt{3}}\left(2\Phi_{S3}\Phi^{S}_{03}-\Phi_{S1}\Phi^{S}_{02}-\Phi_{S2}\Phi^{S}_{01}\right)+g_{2}\Phi^{S}_{02}\tilde{\Theta}+2g_{3}\Phi_{S2}\Theta_{0}=0\,,
 \end{eqnarray}
 \begin{eqnarray}
  \frac{\partial W_{v}}{\partial\Theta}&=&\Theta_{0}\left(2g_{4}\Theta+g_{5}\tilde{\Theta}\right)=0\,,\nonumber\\
  \frac{\partial W_{v}}{\partial\tilde{\Theta}}&=&\Theta_{0}\left(g_{5}\Theta+2g_{6}\tilde{\Theta}\right)+g_{2}\left(\Phi_{S1}\Phi^{S}_{01}+\Phi_{S2}\Phi^{S}_{03}+\Phi_{S3}\Phi^{S}_{02}\right)=0\,,\nonumber\\
  \frac{\partial W_{v}}{\partial\Psi}&=&g_{7}\Psi_{0}\tilde{\Psi}=0\,,\nonumber\\
  \frac{\partial W_{v}}{\partial\tilde{\Psi}}&=&g_{7}\Psi_{0}\Psi=0\,.
  \label{drivingF-term}
 \end{eqnarray}
From this set of ten equations, we obtain
 \begin{eqnarray}
\langle\Phi^{T}_{0}\rangle=(0,0,0)\,,\qquad\langle\Phi^{S}_{0}\rangle=(0,0,0)\,,\qquad\langle\Theta_{0}\rangle=0\,,\qquad\langle\Psi_{0}\rangle=0\,,
 \label{Drdirection1}
 \end{eqnarray}
which are valid to all orders.

%%%%%%%%%%%%%%%%%%%%%%%%%%%%%%%%%%%%%%%%%%%%%%%%%%%%%%%%%%%%%%%%%%%%%%%%%%%%%%%%%%%%%%%%%%
\section{Mixing between Axion and meson}
 \label{ECL}

The mass terms reads
 \begin{eqnarray}
  {\cal L}_{\rm mass} &=& \mu m_{u}\Big\{\frac{f^{2}_{\pi}}{2(1+z+w)F^{2}_{A}}a^{2}+\frac{1+z}{2z}\pi^{2}_{0}+\frac{w+4z+zw}{6zw}\eta^{2}\nonumber\\
  &-&\frac{1-z}{2\sqrt{3}z}\pi_{0}\eta+\left(\frac{z+w}{zw}\right)\bar{K}^{0}K^{0}+\frac{1+w}{w}K^{+}\bar{K}^{-}+\frac{1+z}{z}\pi^{+}\pi^{-}\Big\}\,.
  \label{neut-A}
 \end{eqnarray}
As for the axion-photon coupling, both the $\pi^{0}$ and $\eta$ couple to photons through triangle anomalies.
%\begin{eqnarray}
%  \frac{1}{4}G_{\pi\gamma\gamma}\pi^0\,F\tilde{F}+\frac{1}{4}G_{\eta\gamma\gamma}\eta\,F\tilde{F}=\frac{e^2}{32\pi^2}(2)\frac{\pi^0}{f_\pi}\,F\tilde{F}+\frac{e^2}{32\pi^2}\left(\frac{2}{\sqrt{3}}\right)\frac{\eta}{f_\pi}\,F\tilde{F}\,.
%  \label{}
% \end{eqnarray}
However, from Eq.\,(\ref{neut-A}) we see that there are no mixings with the axion and the heavy $\pi^0$ and $\eta$.
We explicitly show the mass squared terms in Eq.\,(\ref{neut-A}) and the boson-photon-photon couplings $G_{a\gamma\gamma}, G_{\pi\gamma\gamma}$, and $G_{\eta\gamma\gamma}$ for the axion, $\pi^{0}$ and $\eta$, respectively:
 \begin{eqnarray}
  \frac{1}{2}{\left(\begin{array}{ccc}
 a &  \pi^{0} &  \eta
 \end{array}\right)}{\cal M}^2{\left(\begin{array}{c}
 a\\
 \pi^{0} \\
 \eta
 \end{array}\right)}+\frac{1}{4}{\left(\begin{array}{ccc}
 a &  \pi^{0} &  \eta
 \end{array}\right)}{\left(\begin{array}{c}
 G_{a\gamma\gamma}\\
 G_{\pi\gamma\gamma} \\
 G_{\eta\gamma\gamma}
 \end{array}\right)}F\tilde{F}
  \label{neut-B}
 \end{eqnarray}
where
 \begin{eqnarray}
 {\cal M}^2={\left(\begin{array}{ccc}
 \mu m_{u}\frac{f^{2}_{\pi}}{F^{2}_{A}(1+z+w)} & 0 & 0 \\
 0 &  \mu m_{u}\frac{1+z}{z} &  \mu m_{u}\frac{z-1}{\sqrt{3}z} \\
 0 &  \mu m_{u}\frac{z-1}{\sqrt{3}z} &  \mu m_{u}\frac{w+4z+zw}{3zw}
 \end{array}\right)}\,.
  \label{neut-C}
 \end{eqnarray}
Diagonalization of the mass squared matrix ${\cal M}^2$ in a basis $a-\pi^{0}-\eta$ basis, one can find the physical masses for the axion $a$, $\pi^{0}$, and $\eta$. And, the physical masses for $\pi^0$ and $K^0$ mesons as well as the electromagnetic contributions to the physical $\pi^{\pm}$ and $K^{\pm}$ mesons are expressed as
 \begin{eqnarray}
 (m^{2}_{\pi^0})_{\rm phys}&=&2\mu m_{u}\left(\frac{z+w+zw-\sqrt{(z+w+zw)^2-3zw(1+z+w)}}{3zw}\right)\,,\nonumber\\
 \nonumber\\
 (m^{2}_{K^0})_{\rm phys}&=& \mu m_{u}\left(\frac{1}{z}+\frac{1}{w}\right)\,,~\,\qquad(m^{2}_{K^\pm}-m^{2}_{\pi^\pm})_{\rm phys}=\mu m_{u}\left(\frac{1}{w}-\frac{1}{z}\right)\,.
\label{physMeson}
 \end{eqnarray}

%%%%%%%%%%%%%%%%%%%%%%%%%%%%%%%%%%%%%%%%%%%%%%%%%%%%%%%%%%%%%%%%%%%%%%%%%%%%%%%%%%%%%%%%%%%%%%%%%%%%%%%%%%%%%%%%
\acknowledgments{This work was supported by IBS under the project code, IBS-R018-D1.
}

%%%%%%%%%%%%%%%%%%%%%%%%%%%%%%%%%%%%%%%%%%%%%%%%%%%%%%%%%%%%%%%%%%%%%%%%%%%%%%%%%%%%%%%%%%%%%%%%%%%%%%%%%%%

\end{document}